\documentclass[reprint,twocolumn,notitlepage,superscriptaddress,nofootinbib]{revtex4-2}

\usepackage[english]{babel}
\usepackage{xcolor}
\usepackage{braket}
\usepackage{amsmath}
\usepackage{amssymb}
\usepackage{mathrsfs}
\usepackage{amsthm}
\usepackage{mathtools}
\usepackage{comment}
\usepackage[hidelinks]{hyperref}
\usepackage{stackrel}
\usepackage{enumitem}

\usepackage{pdfpages} 
\usepackage{pgffor} 

\makeatletter
\AtBeginDocument{\let\LS@rot\@undefined}
\makeatother

\def\supplementfilename{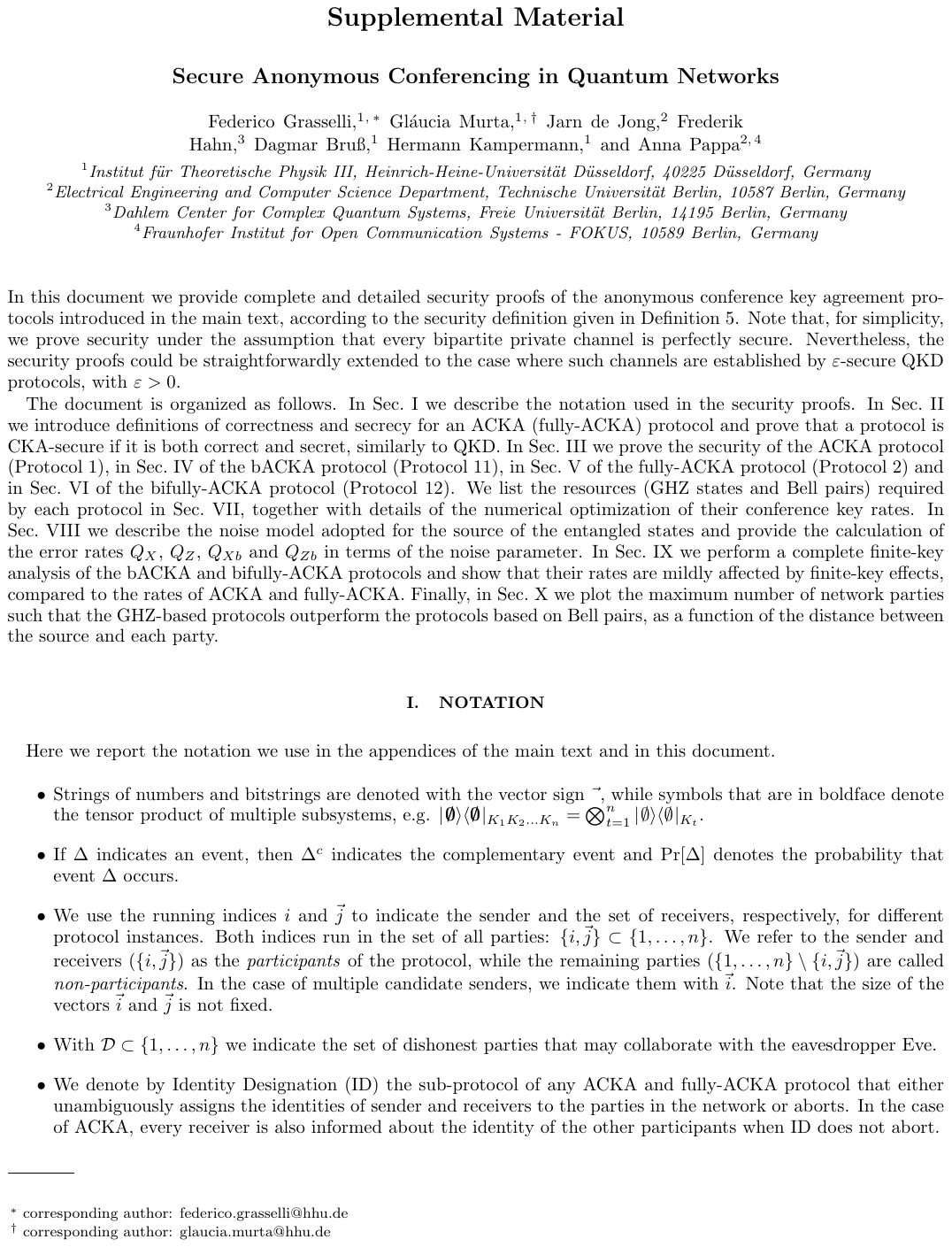}

\pdfximage{\supplementfilename}
\def\numbersupplementpages{\the\pdflastximagepages}

\newif\ifarXiv
\arXivtrue 

\setlist[enumerate]{label*=\arabic*.}

\usepackage{soul}

\usepackage{graphicx}
\usepackage{tikz}
\usepackage{framed}
\usepackage{float}

\addto\captionsenglish{}
\addto\captionsenglish{}

\usepackage{pifont}

\def\diracspacing{0.7pt}
\newcommand{\ketbra}[2]{| \hspace{\diracspacing} #1 \rangle \langle #2 \hspace{\diracspacing} |} 

\newcommand{\norm}[1]{\left\|#1\right\|_{\mathrm{tr}}}
\newcommand{\abs}[1]{\left|#1\right|}
\DeclareMathOperator{\Tr}{Tr}

\newcommand{\proj}[1]{\ketbra{#1}{#1}}

\theoremstyle{definition}
\newtheorem{Thm}{Theorem}

\theoremstyle{definition}
\newtheorem{Def}{Definition}

\newtheorem{Lmm}{Lemma}

\usepackage{algorithm}

\makeatletter
\newenvironment{protocol}
{
		\renewcommand{\ALG@name}{Protocol}
		\refstepcounter{algorithm}
		\hrule height.8pt depth0pt \kern2pt
		\renewcommand{\caption}[2][\relax]{
			{\raggedright\textbf{\fname@algorithm~\thealgorithm} ##2\par}%
			\ifx\relax##1\relax 
			\addcontentsline{loa}{algorithm}{\protect\numberline{\thealgorithm}##2}%
			\else 
			\addcontentsline{loa}{algorithm}{\protect\numberline{\thealgorithm}##1}%
			\fi
			\kern2pt\hrule\kern2pt
		}
	}{
	\kern2pt\hrule\relax
}
\makeatother

\usepackage{natbib}
\makeatletter
\let\cat@comma@active\@empty
\makeatother

\begin{document}
	
	\title{Secure Anonymous Conferencing in Quantum Networks}

	\author{Federico Grasselli}
	\email[corresponding author:]{ federico.grasselli@hhu.de}
	\affiliation{Institut f\"ur Theoretische Physik III, Heinrich-Heine-Universit\"at D\"usseldorf, 40225 D\"usseldorf, Germany}
	
	\author{Gl\'{a}ucia Murta}
	\email[corresponding author:]{ glaucia.murta@hhu.de}
	\affiliation{Institut f\"ur Theoretische Physik III, Heinrich-Heine-Universit\"at D\"usseldorf, 40225 D\"usseldorf, Germany}
	
	\author{Jarn de Jong}
	\affiliation{Electrical Engineering and Computer Science Department, Technische Universit\"at Berlin, 10587 Berlin, Germany}
	
	\author{Frederik Hahn}
	\affiliation{Dahlem Center for Complex Quantum Systems, Freie Universit\"at Berlin, 14195 Berlin, Germany}
	
	\author{Dagmar Bru\ss}
	\author{Hermann Kampermann}
	\affiliation{Institut f\"ur Theoretische Physik III, Heinrich-Heine-Universit\"at D\"usseldorf, 40225 D\"usseldorf, Germany}
	
	\author{Anna Pappa}
	\affiliation{Electrical Engineering and Computer Science Department, Technische Universit\"at Berlin, 10587 Berlin, Germany}
	\affiliation{Fraunhofer Institut for Open Communication Systems - FOKUS, 10589 Berlin, Germany}

	\begin{abstract}
		\noindent Users of quantum networks can securely communicate via so-called (quantum) conference key agreement --- making their identities publicly known. In certain circumstances, however, communicating users demand anonymity. Here, we introduce a security framework for anonymous conference key agreement with different levels of anonymity, which is inspired by the epsilon-security of quantum key distribution. We present efficient and noise-tolerant protocols exploiting multipartite Greenberger–Horne–Zeilinger (GHZ) states and prove their security in the finite-key regime. We analyze the performance of our protocols in noisy and lossy quantum networks and compare with protocols that only use bipartite entanglement to achieve the same functionalities. Our simulations show that GHZ-based protocols can outperform protocols based on bipartite entanglement and that the advantage increases for protocols with stronger anonymity requirements. Our results strongly advocate the use of multipartite entanglement for cryptographic tasks involving several users.
		
	\end{abstract}
	\maketitle

	\section{Introduction}\label{sec:intro}
	
	\noindent Building on the ``second quantum revolution'' \cite{second-quantum-rev1,second-quantum-rev2,QTEU}, quantum-cryptography technologies have recently seen a rapid development both in academia and industry, with quantum key distribution (QKD) \cite{BB84,Diamanti2016,pir2019advances} being a prominent example \cite{QKDmarkets}. The task of QKD has been generalized to multiple users with quantum conference key agreement (CKA) \cite{Epping,Grasselli_2018,WstateProtocol,FirstCVMDI,ZSG18,OLLP19,CKAreview,Chen-3CKA,Chen-3CKAasymm}, where $n$ parties establish a common secret key when linked by an insecure quantum network \cite{Das2019}. The key, called conference key, can subsequently be used for group-wise encryption among the $n$ parties \cite{CKAexperiment}.
	
	Besides achieving secure communication with encryption keys, an equally important privacy aspect that arises in a multipartite network is anonymity. Indeed, in numerous scenarios the communicating parties want to share a secret message and, at the same time, remain anonymous. That is, they want to keep their identities hidden from the other parties in the network, from the network manager, and even between themselves. In a world ever more demanding for privacy, many individuals would benefit from secure anonymous communication, including: voters, anonymous informants, investigative reporters, secret agents and users updating an online database (e.g.~medical records).
	
	Here, this cryptographic task is formalized as follows. Consider an $n$-party quantum network where, a priori, each party could be a participant, i.e.~the sender or one of the $m$ receivers ($m<n$) chosen by the sender. The task is to identify a set of participants and establish a secret conference key between them, such that they remain anonymous at least with respect to the other $n-m-1$ parties. The established conference key then enables secure anonymous conferencing among the participants.

	We define protocols that accomplish the described task with different anonymity requirements. In particular, an anonymous conference key agreement (ACKA) protocol reveals the participants' identities to each participant and provides them with the same conference key. Beyond that, in a fully-anonymous conference key agreement (fully-ACKA) protocol we require that only the sender knows the number and identities of the receivers, while each receiver is only aware of their role. In both ACKA and fully-ACKA protocols the participants' identities (and the conference key) are unknown to the remaining parties in the network and to a potential eavesdropper controlling the network and the quantum source.
	
	A fully-ACKA protocol could be used by a whistle-blower (the sender) within a company, who wants to expose an illicit activity to some of the company's managers (the receivers). The fully-ACKA protocol would ensure the anonymity of sender and receivers, thus protecting them from potential reprisals. Alternatively, an ACKA protocol could be employed by journalists to send reports from an area with limited freedom of press.
	
	We distinguish three features characterizing the security of ACKA and fully-ACKA protocols. Specifically, the protocol must be \emph{CKA-secure}, which means that the established conference key is identical for all participants, uniformly distributed and unknown to anybody else (as in standard CKA \cite{CKAreview,CKAbook}). The protocol must also be \emph{anonymous}, in the sense that the identity of each participant must be kept secret from a subset of parties, depending on the required level of anonymity. Finally, the protocol must be \emph{integrous}, that is, the identities of the sender and of the chosen receivers are correctly assigned and communicated.

	In this work, we introduce rigorous security definitions for ACKA and fully-ACKA, which encompass the above-mentioned security notions and are inspired by the composable security framework of QKD. Moreover, we design ACKA and fully-ACKA protocols based on multipartite entangled states, namely $n$-party GHZ states, distributed by an untrusted source and prove their security according to our definitions.
	
	To benchmark the performance of our GHZ-based protocols, we introduce two multiparty generalizations of the Anonymous Message Transmission protocol \cite{Broadbent2007}. The protocol in \cite{Broadbent2007} allows two parties to anonymously send classical messages in a network with dishonest parties, by employing bipartite private authenticated channels between every pair of parties. Such channels can be established beforehand by distributing Bell pairs and running pairwise QKD protocols. Our multiparty generalizations of the protocol in \cite{Broadbent2007}, named bACKA and bifully-ACKA, achieve the same functionalities of ACKA and fully-ACKA, respectively, while exclusively relying on bipartite entanglement (Bell pairs) shared between every pair of parties.
	
\begin{figure}[ht] 
		\centering
		\includegraphics[width=1\linewidth,keepaspectratio]{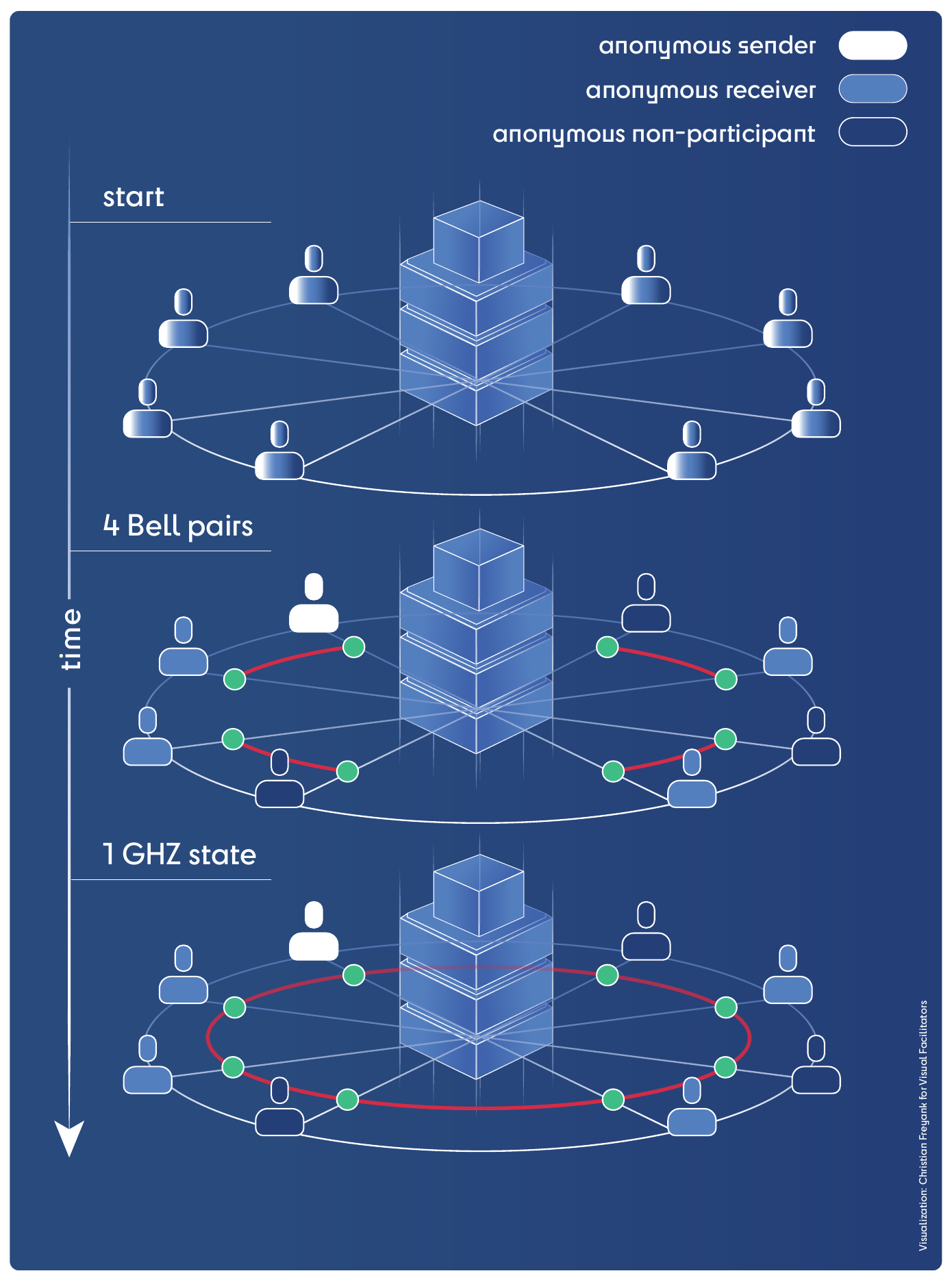}
		\caption{Our protocols are implemented on a quantum network consisting of $n$ parties ($n=8$ in the picture) linked to a central quantum server. The parties' roles are not predetermined and are instead assigned during the protocol's execution. The party designated as sender selects the desired receivers to anonymously establish a shared conference key. The quantum server distributes --in each network use-- either Bell pairs between distinct pairs of parties (e.g. four pairs when $n=8$) or one GHZ state shared between all parties. While the bACKA and bifully-ACKA protocols only require Bell pairs, our ACKA and fully-ACKA protocols mainly use GHZ states. The figure of merit to compare our protocols is the secret conference key rate, i.e., the number of secret conference key bits generated per network use (see Sec.~\ref{sec:results}~C).}
		\label{fig:network}
	\end{figure}

	Our simulations, run on a quantum network like the one in Fig.~\ref{fig:network}, show that the ACKA and fully-ACKA protocol based on GHZ states significantly outperform the corresponding protocol based on Bell pairs even in the finite-key regime. The advantage provided by GHZ states over Bell pairs increases with the anonymity requirements of the protocol. In the case of fully-ACKA, multipartite entanglement increases the secret conference key rate by a factor proportional to the square of the number of parties in the network ($n^2$). Our results clearly demonstrate the benefit of multipartite entanglement for cryptographic tasks involving more than two parties and reinforce previous results supporting this claim \cite{Epping,CKAexperiment,WalkCVQSS}.
	
	\subsection{Security assumptions} \label{sec:security-assumptions}
	\noindent The eavesdropper --called Eve-- can completely control the entanglement source and the quantum network. Eve may also listen to the public communication generated during the protocol and corrupt any subset of parties. Any party who misbehaves and willingly does not follow the prescriptions of the protocol is considered to be a corrupt party and is called \emph{dishonest}. Eve can collaborate with the dishonest parties, having access to their private classical inputs and outputs, but is not allowed to impersonate them (e.g. store their quantum systems). Analogously to QKD, Eve has unbounded quantum power and holds a perfect quantum memory.
	
	Conversely, the $n$ parties are equipped with trusted quantum-measurement devices and at most short-lived quantum memories. This assumption is motivated by the current state-of-the-art in the field of quantum memories \cite{qmemory,qmemory2} and by the fact that most end-nodes in quantum networks are simply measuring stations \cite{BristolNetwork,ChinaNetwork}.

	\subsection{Relation to previous works} 
	\label{sec:previous-work}
	
	\noindent Early anonymous communication protocols in quantum networks \cite{Wehner2005,lipinska_anonymous_2018} aim at transmitting classical as well as quantum information, going beyond the functionality achieved by the Anonymous Message Transmission protocol of~\cite{Broadbent2007}. However, such protocols \cite{Wehner2005,lipinska_anonymous_2018} require trusted sources of multipartite entangled states, do not address secrecy and neither tolerate dishonest network users nor noise.
	
	These drawbacks are partially resolved by following protocols \cite{Brassard2007,unnikrishnan_anonymity_2019,yang_examining_2020}, which allow for untrusted sources of entanglement and potentially dishonest parties by introducing source-verification schemes. However, such schemes happen to be highly inefficient, technologically demanding, and are still vulnerable to noise.
	
	Recently, the task of anonymously establishing a secret conference key, with the anonymity requirements identified as fully-ACKA in our work, was addressed in \cite{acka_old} and further explored in \cite{ACKA}. The protocols in \cite{acka_old,ACKA}, while requiring some level of trust in the source of multipartite entangled states, also build on the same source-verification scheme of \cite{unnikrishnan_anonymity_2019, pappa_multipartite_2012} and thus carry the same drawbacks.
	
	In contrast to previous protocols \cite{Brassard2007,unnikrishnan_anonymity_2019,yang_examining_2020,acka_old,ACKA} allowing for an untrusted albeit noiseless source, our protocols are robust against noise in the quantum network and in the untrusted source, thereby representing the first secure anonymous protocols with a \textit{noisy} untrusted source.
	
	Moreover, our protocols are remarkably more efficient and implementable with present-day technology, compared to previous proposals \cite{Wehner2005,lipinska_anonymous_2018,Brassard2007,unnikrishnan_anonymity_2019,yang_examining_2020,acka_old,ACKA}. Indeed, they verify the entangled states prepared by the source over a small fraction of states through local measurements, leaving the majority of the states for establishing the conference key. Previous protocols, instead, either assume a noiseless trusted source (which is a strong assumption) \cite{Wehner2005,lipinska_anonymous_2018}, or consume the majority of the entangled states for verification (resulting in asymptotically null key rates even in a noiseless scenario) \cite{unnikrishnan_anonymity_2019,yang_examining_2020,acka_old,ACKA}, or require multi-qubit gates and perfect quantum memories \cite{Brassard2007}, which are not yet technologically available.
	
	Regarding security, unlike any previous anonymous quantum protocol, we prove the security and anonymity of our protocols in the experimentally-relevant regime of finite keys, by extending the $\varepsilon$-security framework of QKD to include anonymity and integrity.
	
	Altogether, the efficiency, noise-tolerance and $\varepsilon$-security of our anonymous protocols allow us to compute non-zero conference key rates in noisy and lossy quantum networks, both in the asymptotic and finite-key regimes. We emphasize that such a result could not be obtained in any previous anonymous communication protocol based on multipartite entanglement for the reasons outlined above.

	\subsection{Use of the anonymous conference key} \label{sec:use-of-key}
	
	\noindent Contrary to standard QKD and CKA, the secret conference key established by ACKA and fully-ACKA protocols must be used in a non-trivial encryption scheme in order to protect the participants' identities. If the sender simply encrypts a message with the conference key and broadcasts it, every other party would identify the sender as the party who performed the broadcast.
	
	In the ACKA scenario, the solution is straightforward. Every party in the network is asked to broadcast a random string, while the sender broadcasts the encrypted message. Only the receivers are able to decrypt the sender's broadcast by using the conference key.
	
	In the fully-ACKA scenario the receivers do not know the sender's identity and the previous solution could reveal it. In this case, a possibility would be to employ classical anonymous broadcasting protocols \cite{anonymous-broadcast}.

	\section{Results} \label{sec:results}
	
	\subsection{Security definitions}
	
	\noindent Our approach to define the security of ACKA and fully-ACKA protocols starts from the identification of three properties that an ideal protocol is expected to satisfy. 
	
	The first property that we require is integrity. At the beginning of an ACKA (fully-ACKA) protocol, the participants are not yet determined. Therefore, the first step of such protocols consists in running an identity designation (ID) sub-protocol. The ID sub-protocol determines the sender, notifies the receivers of their roles and, in the case of ACKA, notifies the receivers of the other participants' roles. A (fully-)ACKA protocol is integrous if its ID sub-protocol works perfectly, i.e., if either it correctly communicates the roles of sender and receivers, or it aborts for every party in the network.
	
	Conditioned on the fact that the ID sub-protocol does not abort and correctly assigns the identities, we require the (fully-)ACKA protocol to be CKA-secure. That is, either it outputs the same random conference key for every participant, uncorrelated from any information held by the non-participants and Eve, or it aborts from the point of view of the participants.
	
	Finally, an ACKA (fully-ACKA) protocol should be anonymous, i.e., the identity of each participant must be kept secret from the non-participants and Eve --as well as from the other receivers in the case of fully-ACKA.
	
	We remark that dishonest participants may broadcast their identity --or the identity of all the participants in the case of ACKA. Hence, in the presence of dishonest participants we cannot impose the same anonymity requirements. Similarly, no CKA-security is required if some participants are dishonest, as they may publicly reveal the secret conference key.

	Inspired by the composable security framework of QKD \cite{PortmannRenner2014,Portmann2021}, we define the security of ACKA and fully-ACKA protocols from the output state of the protocol --i.e., the state of the classical and quantum registers of each party, including the eavesdropper, at the end of the protocol. More specifically, we introduce a security definition which quantifies, for every property, how close the output state is to a state with the required property. Informally, our security definition can be stated as following (the rigorous definition, Definition~\ref{def:suff-cond-security}, is given in Appendix~\ref{sec:security-defs}).
	
	\begin{Def}[Security (informal)]\label{def:security-informal}
		A (fully-)ACKA protocol is $\varepsilon$-secure, with $\varepsilon=\varepsilon_{\rm IN}+\varepsilon_{\rm CKA}+\varepsilon_{\rm AN}$, if it satisfies the following three conditions:
		\begin{itemize}
			
			\item $\varepsilon_{\rm IN}$-integrity: The ID sub-protocol correctly assigns the roles of sender and receivers or aborts for every party in the network, except for a probability smaller than $\varepsilon_{\rm IN}$. 
			
			\item $\varepsilon_{\rm CKA}$-CKA-security:
			In the case of honest participants, conditioned on the ID sub-protocol correctly assigning the participants' identities, the output state is $\varepsilon_{\rm CKA}$-close to the output state of a protocol that either delivers the same secret conference key to every participant, or aborts for every participant.
			
			\item $\varepsilon_{\rm AN}$-anonymity for ACKA:
			the output state of any subset of non-participants and Eve is $\varepsilon_{\rm AN}$-close to a state which is independent of the identity of the remaining parties.
			
			\item $\varepsilon_{\rm AN}$-anonymity for fully-ACKA:
			the output state of any subset of parties (except for the sender) and Eve is $\varepsilon_{\rm AN}$-close to a state which is independent of the identities of the other parties.
		\end{itemize}
	\end{Def}
	
	In Subsec.~\ref{sec:protocols} we introduce our ACKA and fully-ACKA protocols based on GHZ states shared by all parties. Motivated by the fact that our fully-ACKA protocol does not satisfy the (strong) anonymity condition of Definition~\ref{def:security-informal} but still retains important anonymity features, we provide a weaker anonymity definition satisfied by our fully-ACKA protocol.
	
	The reason for which the fully-ACKA protocol cannot satisfy the strong anonymity condition is that the receivers, by executing the protocol, gain access to information (e.g., whether the protocol aborts or not) that can depend on the identities of the other participants. Such dependence can occur if the untrusted source distributes asymmetric states which are noisier for some parties than for others, instead of the permutationally-symmetric GHZ states. Similar problems affecting anonymity were already mentioned in \cite{Brassard2007} regarding the protocol proposed in \cite{Bouda07}.
	
	However, we emphasize that honest-but-curious receivers may be able to deduce the identity of other participants only if they combine the identity-dependent information obtained from the protocol with a detailed knowledge of the asymmetric states distributed by the source --or with any other asymmetric specification of the protocol causing identity-dependent events. Therefore, the anonymity of the parties can be preserved if honest-but-curious receivers do not have access to the asymmetric specifications of the protocol's implementation (e.g. if they are secret or if the publicly available specifications are symmetric), even if the source actually distributes asymmetric states. Note, however, that we cannot prevent dishonest parties or Eve from broadcasting the actual, asymmetric, specifications of the protocol's implementation at any point in time, thus jeopardizing anonymity with respect to honest receivers. 
	
	This weaker version of anonymity is captured by the following definition, where $\rm w$ abbreviates ``weak'' and $\rm (n)p$ abbreviates ``(non-)participant''. The definition is satisfied by our fully-ACKA protocol with GHZ states (the formal definition is given in Appendix~\ref{sec:security-defs}).
	
	\begin{Def}[Weak-anonymity for fully-ACKA (informal)] \label{def:sweak-anonym}
		A fully-ACKA protocol is $\varepsilon_{\rm w AN}$-weak-anonymous, with $\varepsilon_{\rm wAN}=\varepsilon_{\rm npAN} + \varepsilon_{\rm pAN}$, if the following two conditions are satisfied:
		\begin{enumerate}
			\item (Anonymity with respect to non-participants and Eve) The output state of the protocol satisfies the $\varepsilon$-anonymity condition  for ACKA protocols with $\varepsilon=\varepsilon_{\rm npAN}$.
			\item (Anonymity with respect to honest-but-curious receivers) When the protocol's specifications known to the parties are symmetric --i.e., invariant under permutations of parties--, any subset of honest-but-curious receivers cannot guess the identity of other participants with a higher probability than the trivial guess, except for a small deviation $\varepsilon_{\rm pAN}$.
		\end{enumerate}
	\end{Def}

	\subsection{ACKA and fully-ACKA protocols}  \label{sec:protocols}
	
	\noindent Here we present an ACKA and a fully-ACKA protocol which rely on the multipartite correlations of GHZ states shared by all the parties in the network. However, we emphasize that the untrusted source could prepare completely arbitrary states and potentially be controlled by Eve. Similarly, any dishonest party can behave differently from the actions prescribed by the protocol. Yet, our ACKA and fully-ACKA protocols are secure in the sense of Definition~\ref{def:security-informal}.
	
	For convenience, we will identify the sender as Alice and the intended receivers as Bob$_l$, for $l\in\{1,\dots,m\}$, where the number of receivers $m$ is not predetermined and can be chosen by Alice during the protocol. Both our protocols require:
	\begin{enumerate}
		\item A shielded laboratory for each party, equipped with a trusted measurement device, a trusted post-processing unit and a private source of randomness.
		\item A bipartite private channel for each pair of the $n$ parties. We obtain it with a bipartite public authenticated channel and a shared secret key to encrypt the communication over the channel (with one-time pad). Hence, the communication over the bipartite private channel is secret (i.e., only known to the legitimate parties), whereas the identities of the parties using the channel are public.
		\item An authenticated broadcast channel.
		\item A public source of randomness that is not controlled by the adversary.
	\end{enumerate}
	
	The ACKA protocol also requires previously-shared conference keys among every subset of parties in the network (note that this requirement could be dropped by introducing a minor overhead in the number of bipartite private channel uses). Part of these keys are consumed during the execution of the ACKA protocol, which is thus a key-growing algorithm.
	
	In the following, we provide a high-level description of the steps of our ACKA and fully-ACKA protocols, whose core consists in measuring GHZ states to anonymously generate a shared conference key. The additional steps required to ensure integrity and CKA-security (namely the ID sub-protocol, error correction and privacy amplification) are summarized here and further detailed in Appendix~\ref{sec:GHZ-protocols}, together with more exhaustive protocol descriptions.
	
	In order to ensure that the classical communication required by ACKA and fully-ACKA is anonymous, we make use of classical sub-routines introduced in \cite{Broadbent2007} --specifically Parity, Veto and Collision Detection-- which are run on the bipartite private channels. Note that, in our simulations, we account for the generation of the secret keys --needed to implement the bipartite private channels-- by running pairwise QKD protocols over distributed Bell pairs. As the name suggests, the Parity protocol computes the parity of the input bits while preserving the anonymity of the parties.\\

	\begin{protocol} \label{prot:sACKA}
		\caption{Anonymous Conference Key Agreement (\textbf{ACKA})}
		\begin{enumerate}[label=\arabic*., ref=\arabic*]
			\item\label{sACKA:step1} The parties run the ACKA-ID sub-protocol (Protocol~\ref{prot:ID}), after which Alice is established to be the sender and Bob$_l$, for $l\in\{1,\dots,m\}$, the receivers. 
			\item\label{sACKA:step2} Alice and the Bobs recover a pre-established conference key.
			\item\label{sACKA:step3} Alice generates a random bitstring, called testing key, where 1 corresponds to a test round and 0 to a key generation round (each bit equals 1 with probability $p$). Alice broadcasts a compressed version of the testing key, encrypted (with one-time pad) with a portion of the pre-established conference key of step~\ref{sACKA:step2}, so that each Bob can decrypt it and recover the testing key. All the other parties broadcast a random string of the same length.
			\item\label{sACKA:step4} Repeat for $L$ rounds:
			\begin{enumerate}
				\item An untrusted source distributes a state to each of the $n$ parties. Ideally, the source prepares an $n$-party GHZ state.
				\item Alice and the Bobs measure their qubits according to the testing key. They measure in the (Pauli) $Z$ basis if the round is a key generation round, or in the (Pauli) $X$ basis if the round is a test round. All the other parties measure $X$. The outcomes $+1$ and $-1$ of each Pauli measurement are mapped to the binary values $0$ and $1$, respectively. The $Z$ outcomes of each participant form their raw conference key.
			\end{enumerate}
			\item\label{sACKA:step5} Once all the qubits have been measured (in case the parties hold short-lived quantum memories, we wait for a time longer than their coherence time), the testing key is anonymously revealed to all parties by iterating the Parity protocol (Protocol~\ref{prot:parity}). 
			\item\label{sACKA:step6} For every round labelled as a test round, the $n$ parties perform the Parity protocol with the following inputs: Every party, except for Alice, inputs the outcome of their $X$ measurement while Alice inputs a random bit. By combining the outputs of Parity with her test-round outputs, Alice computes $Q^{\mathrm{obs}}_X$, which is the fraction of test rounds where the $X$ outcomes of the $n$ parties have parity 1.
			\item\label{sACKA:step7} Verification of secrecy: Alice compares $Q^{\mathrm{obs}}_X$ with the predefined value $Q_X$. If $Q_X^{\mathrm{obs}}+ \gamma(Q_X^{\mathrm{obs}}) >Q_X + \gamma(Q_X)$, where $\gamma(Q_X)$ is the statistical fluctuation, Alice concludes that the verification failed.
			\item\label{sACKA:step8} Error correction (ACKA-EC, Protocol \ref{prot:err-corr-acka}): Alice anonymously broadcasts error correction information based on a predefined value $Q_Z$ for the pairwise error rate between the $Z$ outcomes of Alice and of each Bob. The Bobs use the information to correct their raw keys and verify that they match Alice's raw key. Alice's broadcast is encrypted with the pre-established conference key and only the Bobs can decrypt it. If the error correction or the verification of secrecy (step~\ref{sACKA:step7}) failed, the participants abort the protocol, but this information is encrypted and only available to them.
			\item\label{sACKA:step9} Privacy amplification (PA): The public randomness outputs a two-universal hash function.
			Alice and each Bob apply the two-universal hash function on their error-corrected keys and obtain the secret conference keys of length $\ell$.
		\end{enumerate}
	\end{protocol}\vspace{0.5cm}
	
	In the fully-ACKA scenario, the participants do not know each other's identity, hence they cannot use pre-established conference keys to share the testing key as in the ACKA protocol (Protocol~\ref{prot:sACKA}). Therefore, in the fully-ACKA protocol we introduce a sub-protocol called Testing Key Distribution (TKD) protocol (Protocol~\ref{prot:TKD} in Appendix~\ref{sec:GHZ-protocols}), which allows Alice to anonymously distribute the testing key to the Bobs.\vspace{0.5cm}
	
	\begin{protocol}\label{prot:sf-ACKA}
		\caption{Fully Anonymous Conference Key Agreement (\textbf{fully-ACKA})}
		\begin{enumerate}[label=\arabic*., ref=\arabic*]
			\item\label{sf-ACKA:step1} The parties perform the fully-ACKA-ID sub-protocol (Protocol~\ref{prot:f-ID}), after which Alice is established to be the sender and Bob$_l$, for $l\in\{1,\dots,m\}$, the receivers. 
			\item\label{sf-ACKA:step2} Alice generates a random bitstring, called testing key, where 1 corresponds to a test round and 0 to a key generation round (each bit equals 1 with probability $p$). Additionally, Alice generates the bitstrings $\vec{r}_l$ (for $l\in\{1,\dots,m\}$), which are later used to encrypt some communication between Alice and each Bob.
			\item\label{sf-ACKA:step3} The parties perform the TKD protocol (Protocol~\ref{prot:TKD}) in order for Alice to distribute the testing key and the string $\vec{r}_l$ to the corresponding Bob$_l$.
			\item\label{sf-ACKA:step4} Repeat for $L$ rounds:
			\begin{enumerate}
				\item An untrusted source distributes a state to each of the $n$ parties. Ideally, the source prepares an $n$-party GHZ state.
				\item Alice and the Bobs measure their qubits according to the testing key. They measure in the $Z$ basis if the round is a key generation round, or in the $X$ basis if the round is a test round. All the other parties measure $X$. The $Z$ outcomes of each participant form their raw conference key.
			\end{enumerate}
			\item\label{sf-ACKA:step5} Once all the qubits have been measured, the testing key is anonymously revealed by iterating the Parity protocol (Protocol~\ref{prot:parity}).
			\item\label{sf-ACKA:step6} For every round labelled as a test round, the $n$ parties perform the Parity protocol with the following inputs: Every party, except for Alice, inputs the outcome of their $X$ measurement while Alice inputs a random bit. By combining the outputs of Parity with her test-round outputs, Alice computes $Q^{\mathrm{obs}}_X$, which is the fraction of test rounds where the $X$ outcomes of the $n$ parties have parity 1.
			\item\label{sf-ACKA:step7} Verification of secrecy: Alice compares $Q^{\mathrm{obs}}_X$ with the predefined value $Q_X$. If $Q_X^{\mathrm{obs}}+ \gamma(Q_X^{\mathrm{obs}}) >Q_X + \gamma(Q_X)$, where $\gamma(Q_X)$ is the statistical fluctuation, Alice concludes that the verification failed. If the verification of secrecy failed, or if a party detected any malfunctioning in the TKD protocol (step~\ref{sf-ACKA:step3}), the protocol aborts for every party.
			\item\label{sf-ACKA:step8} Error correction (fully-ACKA-EC, Protocol \ref{prot:err-corr}): Alice anonymously broadcasts error correction information based on a predefined value $Q_Z$ for the pairwise error rate between the $Z$ outcomes of Alice and of each Bob. The Bobs use the information to correct their raw keys and verify that they match Alice's raw key. If the error correction fails, the protocol aborts but this information is encrypted with the strings $\vec{r}_l$ and thus only available to Alice and the Bobs.
			\item\label{sf-ACKA:step9} Privacy amplification (PA): The public randomness outputs a two-universal hash function.
			Alice and each Bob apply the two-universal hash function on their error-corrected keys and obtain the secret conference keys of length  $\ell$. 
		\end{enumerate}
	\end{protocol}\vspace{1em}
	
	In the Supplemental Material \cite{supplmat} we prove the security of Protocols \ref{prot:sACKA} and \ref{prot:sf-ACKA} according to the formal statement of Definition~\ref{def:security-informal}, provided in Appendix~\ref{sec:security-defs} (Definition~\ref{def:suff-cond-security}). The security claims of the protocols are reported in the following theorem.
	
	\begin{Thm}[Security]\label{thm:security-acka-and-f-acka}
		The ACKA protocol based on GHZ states (Protocol~\ref{prot:sACKA}) yields a secret conference key of net length
		\begin{align}
			\ell_{\rm net}=  &L (1-p) \left[1-h\left(Q_X + \gamma(Q_X) \right)-h(Q_Z)\right] \nonumber\\
			&-\log_2 \frac{2(n-1)}{\varepsilon_{\rm EC}}-2\log_2 \frac{1}{2\varepsilon_{\mathrm{PA}}} - Lh(p) -n \label{skeylength-acka},
		\end{align}
		and is $\varepsilon_{\rm tot}$-secure according to Definition~\ref{def:security-informal}, with $\varepsilon_{\mathrm{tot}}=2^{-r_V}+(n-1)\,\varepsilon_{\rm enc} + 2\varepsilon_x + \varepsilon_{\rm EC} + \varepsilon_{\rm PA}$ and where $h(x)=-x\log_2 x -(1-x)\log_2 (1-x)$ is the binary entropy function.
		
		The fully-ACKA protocol based on GHZ states (Protocol~\ref{prot:sf-ACKA}) yields a secret conference key of length
		\begin{align}
			\ell =  &L (1-p) \left[1-h\left(Q_X + \gamma(Q_X) \right) -h(Q_Z)\right] \nonumber\\
			&-\log_2 \frac{2(n-1)}{\varepsilon_{\mathrm{EC}}}-2\log_2 \frac{1}{2\varepsilon_{\mathrm{PA}}}  \label{skeylength},
		\end{align}
		and is $\varepsilon_{\rm tot}$-secure according to Definition~\ref{def:security-informal} but with the anonymity condition replaced by Definition~\ref{def:sweak-anonym}, with $\varepsilon_{\mathrm{tot}}=2^{-(r_V-2)}+(n-1)(6\varepsilon_{\rm enc} + 2^{-(r_N-1)})+\varepsilon_{\mathrm{EC}} +6\varepsilon_x+3\varepsilon_{\mathrm{PA}}$.
	\end{Thm}
	
	Since Protocol~\ref{prot:sACKA}, differently from Protocol~\ref{prot:sf-ACKA}, is a key-growing algorithm, in \eqref{skeylength-acka} we reported the net key length after one run of the protocol, which is obtained by subtracting from $\ell$ the number of consumed bits of pre-established conference keys. The protocols' parameters appearing in the conference key length and in the security parameter $\varepsilon_{\rm tot}$ are specified in Table~\ref{tab:parameters} of Appendix~\ref{sec:GHZ-protocols}.

	\subsection{Performance comparison}
	\noindent In order to assess the performance of Protocols~\ref{prot:sACKA} and \ref{prot:sf-ACKA} and the benefit of GHZ states in anonymously establishing a conference key, we design protocols achieving the same tasks without resorting to multipartite entanglement. The protocols, which only use bipartite private channels hence named bACKA and bifully-ACKA, are generalizations of the Anonymous Message Transmission protocol \cite{Broadbent2007} to more than two parties. In Appendix~\ref{sec:Bell-protocols} we provide a detailed description of bACKA (Protocol~\ref{prot:bACKA}) and bifully-ACKA (Protocol~\ref{prot:bif-ACKA}) along with their security claims according to Definition~\ref{def:security-informal}.
	
	All the protocols are run on the same quantum network (see Fig.~\ref{fig:network}). We model the network of $n$ parties as a star-shaped network where every party is linked to an untrusted quantum server --potentially operated by Eve-- by an equally-lossy quantum channel of transmittance $\eta$.
	
	The quantum server is programmed to distribute either the $n$-party GHZ state
	$\ket{\mathrm{GHZ}_n} = (\ket{0}^{\otimes n} + \ket{1}^{\otimes n})/\sqrt{2}$, which is used to extract the conference key in ACKA and fully-ACKA, or the Bell state $\ket{\mathrm{GHZ}_2}$ to every pair of parties in order to implement the bipartite private channels by running BB84 protocols \cite{BB84}.
	
	We assume that both states are encoded in some binary degree of freedom of single photons (e.g. polarization), such that the probability of detecting a (potentially noisy) Bell pair and GHZ state is $\eta^2$ and $\eta^n$, respectively. We allow for noisy states due to a faulty state preparation by the quantum server, which prepares the states by applying CNOT gates with failure probability $f_G$. This leads to non-zero error rates $Q_X$ and $Q_Z$ of the GHZ state and $Q_{Xb}$ and $Q_{Zb}$ of the Bell pairs. We refer the reader to the Supplemental Material \cite{supplmat} for the exact relation between the error rates and $f_G$.

	\onecolumngrid

	\begin{figure}[hb] 
		\centering
		\begin{minipage}{0.495\textwidth}
			\centering
			\textbf{ACKA vs bACKA}\par\medskip
			\includegraphics[width=1\linewidth,keepaspectratio]{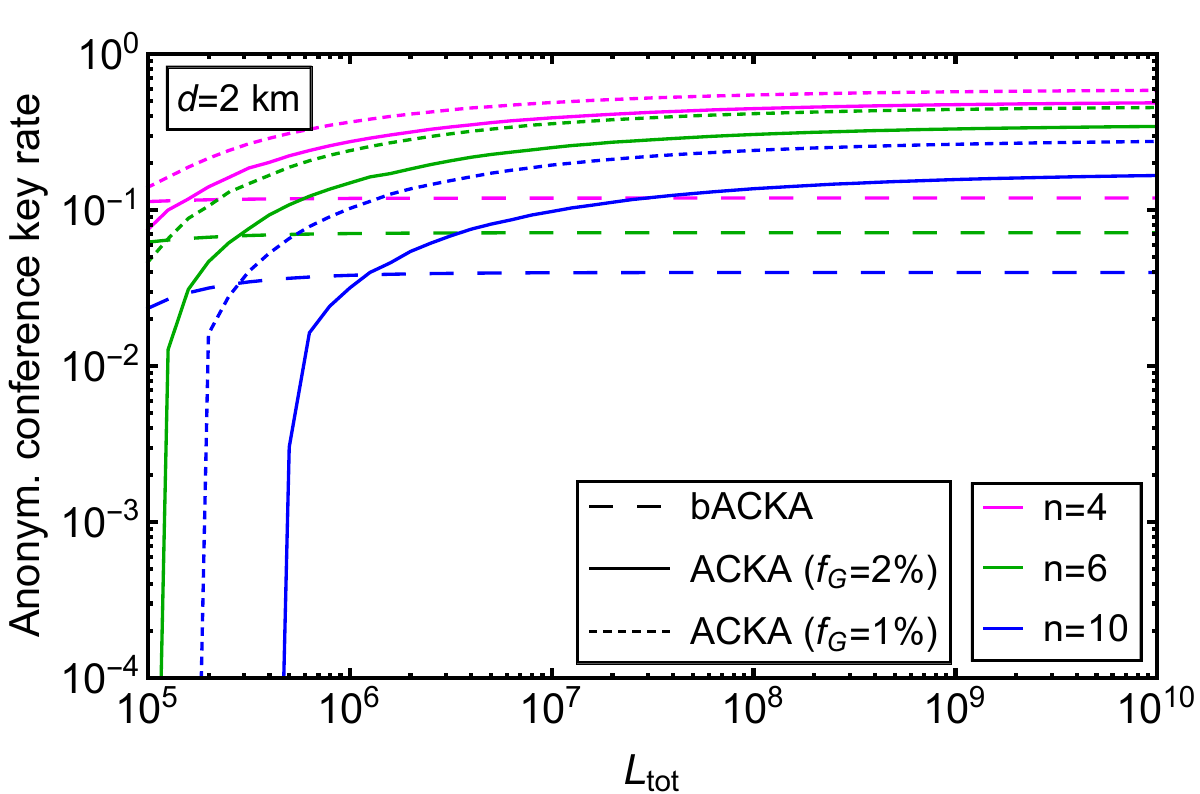}
			\includegraphics[width=1\linewidth,keepaspectratio]{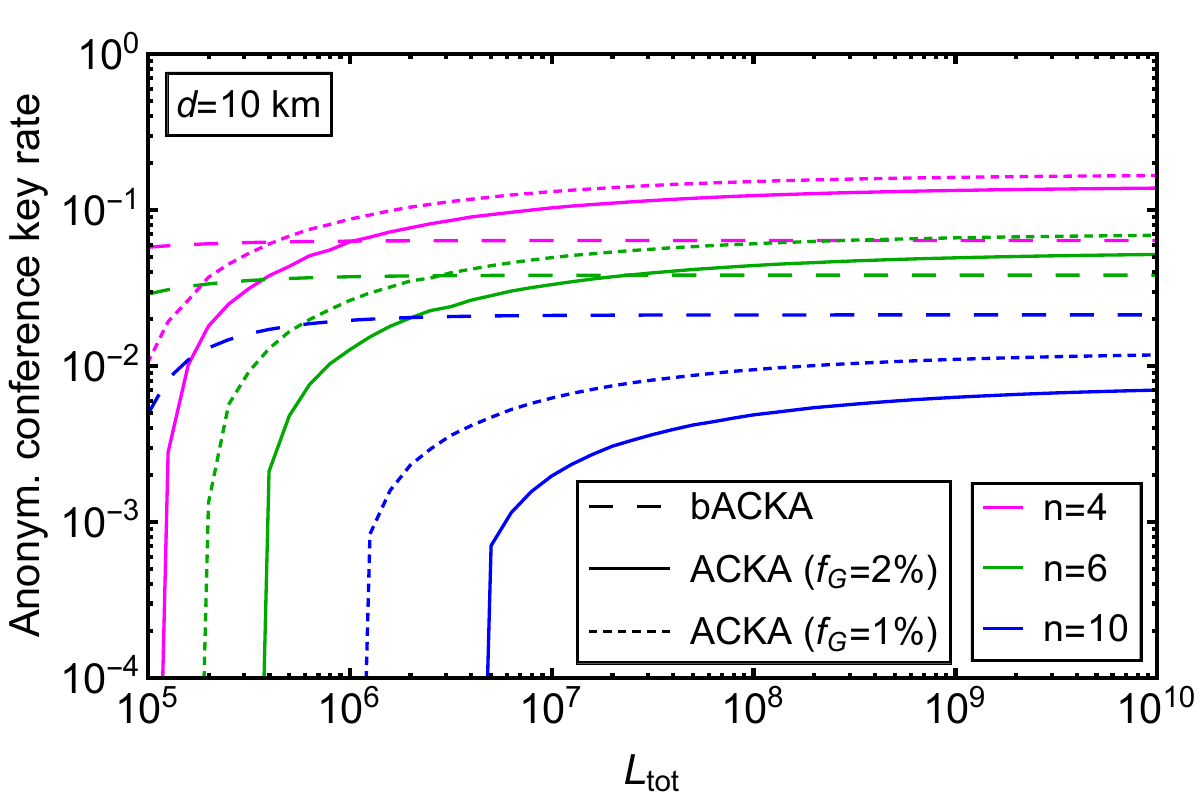}
		\end{minipage}\hfill\vline\hfill
		\begin{minipage}{0.495\textwidth}
			\centering
			\textbf{fully-ACKA vs bifully-ACKA}\par\medskip
			\includegraphics[width=1\linewidth,keepaspectratio]{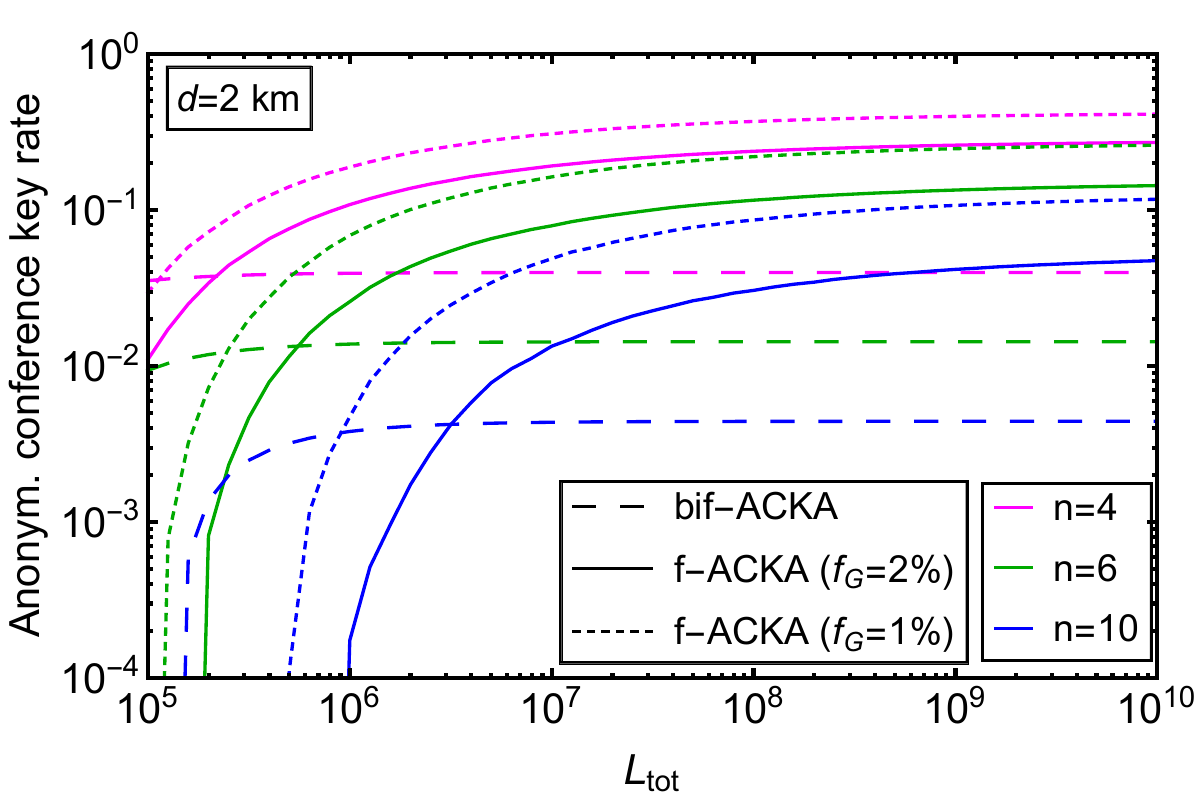}
			\includegraphics[width=1\linewidth,keepaspectratio]{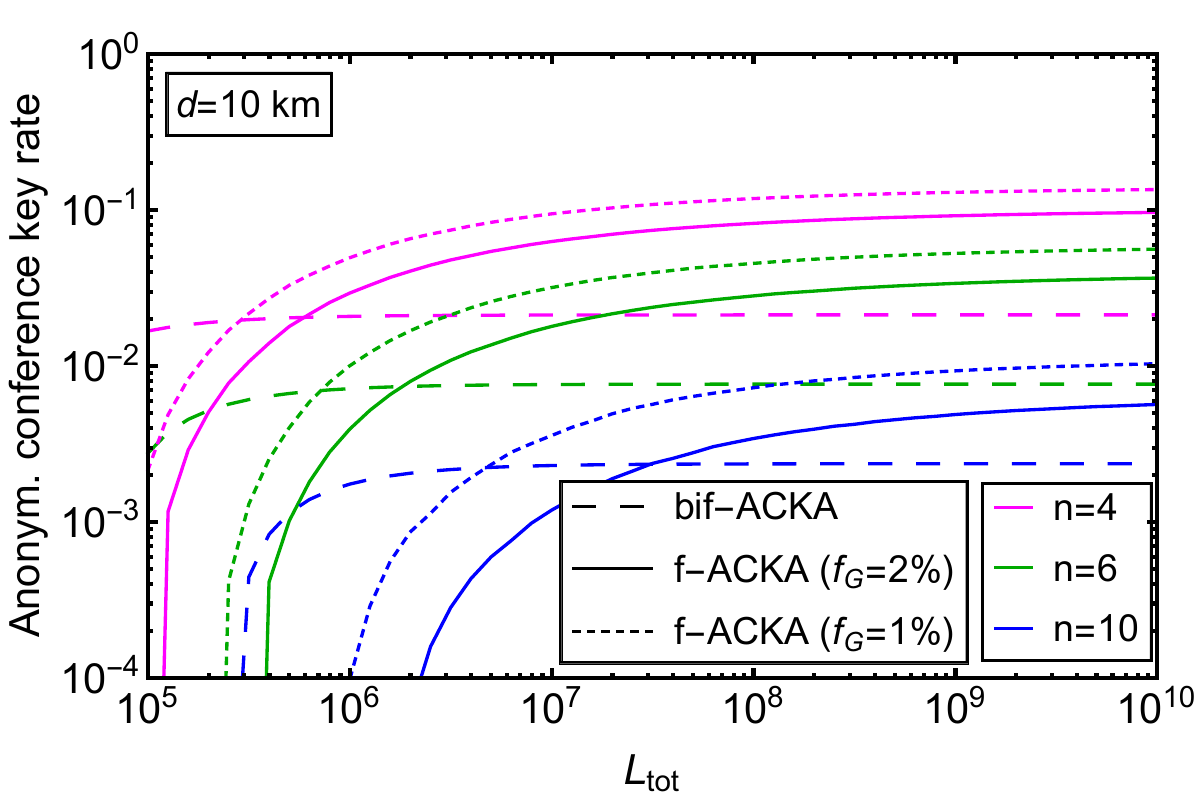}
		\end{minipage}
		\caption{Secret conference key rates of our anonymous protocols as a function of the total number of network uses ($L_{\rm tot}$), for different numbers of parties in the network ($n$). The security parameter is fixed to $\varepsilon_{\mathrm{tot}}=10^{-8}$. The protocols based on GHZ states (ACKA and fully-ACKA) outperform the ones based on Bell pairs (bACKA and bifully-ACKA) already for a relatively low number of network uses. In our network model a quantum server prepares either GHZ states or Bell pairs by repeatedly applying a faulty CNOT gate with failure probability $f_G$. The qubits are encoded on single photons and distributed to the equally-distanced parties via ultra-low-loss fiber. The gate failure probability is either fixed to $f_G=0.02$ or $f_G=0.01$ (the difference in the rates of bACKA and bifully-ACKA is negligible, hence we only reported the $f_G=0.02$ case) and the distance party-server is fixed to $d=2$ km (top) and $d=10$ (bottom).} \label{fig:finite-key}
	\end{figure}\vspace{-0.1cm}

	\twocolumngrid
	
	We compare the four protocols --ACKA vs bACKA and fully-ACKA vs bifully-ACKA-- in terms of their conference key rate, that is, the number of secret conference bits shared by Alice and the intended Bobs per network use. We define one network use to be the preparation and attempted distribution of photons from the quantum server to the parties, regardless of whether the photons are lost or not. We assume that each channel between the server and a party can transmit at most one photon per network use. However, multiple photons can be transmitted in parallel from the server to the parties in a single network use, namely entangled in a GHZ state shared between all parties or in multiple Bell pairs shared between disjoint pairs of parties (see Fig.~\ref{fig:network}).
	
	The conference key rate of the ACKA (fully-ACKA) protocol reads: $r=\ell_{\rm net}/L_{\rm tot}$ ($r=\ell/L_{\rm tot}$), where $\ell_{\rm net}$ ($\ell$) is given in \eqref{skeylength-acka} (resp. \eqref{skeylength}) and $L_{\rm tot}$ is the total number of network uses. Recall that the ACKA and fully-ACKA protocols employ GHZ states (for key generation) and Bell pairs (for the bipartite private channels), both contributing to the total number of network uses $L_{\rm tot}$. However, the contribution of Bell pairs is marginal since, in the majority of the rounds, the parties are generating conference key bits by measuring a GHZ state. Conversely, all the network uses in bACKA and bifully-ACKA are devoted to the distribution of Bell pairs, which are also used to establish the conference key.
	
	In Fig.~\ref{fig:finite-key} we plot the conference key rates of the four protocols in the finite-key regime. The key rates are numerically optimized over the protocols' parameters for each value of $L_{\rm tot}$, having fixed the security parameter (Definition~\ref{def:security-informal}) to $\varepsilon_{\mathrm{tot}}= 10^{-8}$.
	
	In the optimizations, we set the CNOT gate failure probability to $f_G=0.02$ and to $f_G=0.01$ \cite{gate-fidelity}, while the transmittance is given by $\eta=10^{-\gamma d/10}$, where $d$ is the length of the channel party-server and $\gamma=0.17$~dB/km (each channel is assumed to be an ultra-low-loss fiber). We set the distance to $d=2$~km and $d=10$~km to simulate common metropolitan communication scenarios. Further details on the conference key rates and their optimization are given in the Supplemental Material \cite{supplmat}.
	
	From Fig.~\ref{fig:finite-key} we deduce that the ACKA and fully-ACKA protocols, based on GHZ states, can yield higher conference key rates than the protocols exclusively based on Bell pairs (bACKA and fully-ACKA), especially in the high-$L_{\rm tot}$ regime, where finite-key effects can be neglected. In order to understand the intrinsic reason behind this, we compute the asymptotic conference key rates of our anonymous protocols. Indeed, the asymptotic key rates are devoid of statistical corrections and are independent of the sub-protocols requiring a fixed number of network uses regardless of the key length (e.g. Veto or ID). Hence, they reveal the bare scaling of the protocols' performance with respect to the number of parties and the quality of the source states.
	
	The asymptotic conference key rates of the ACKA and bACKA protocol are given by:
	\begin{align}
		r^{\infty} &= \lim_{L_{\rm tot} \to \infty} r=\eta^n \left[1-h(Q_X) -h(Q_Z)\right] \label{asymp-rate-acka} \\
		r^{\infty}_b &= \lim_{L_{b\rm tot} \to \infty} r_b=\frac{ \left\lfloor \frac{n}{2} \right\rfloor \eta^2\left[1-h(Q_{Xb}) -h(Q_{Zb})\right]}{n(n-1)} \label{asymp-rate-backa} 
	\end{align}
	respectively, while the asymptotic conference key rates of the fully-ACKA and bifully-ACKA protocol read:
	\begin{align}
		r^{\infty}_f &= \lim_{L_{f\rm tot} \to \infty} r_f=\frac{\eta^n \left[1-h(Q_X) -h(Q_Z)\right]}{1+\frac{n(n-1)\eta^{n-2}h(Q_Z)}{\left\lfloor \frac{n}{2} \right\rfloor \left[1-h(Q_{Xb}) -h(Q_{Zb})\right]}} \label{asymp-rate-f-acka} \\
		r^{\infty}_{bf} &= \lim_{L_{bf\rm tot} \to \infty} r_{bf}=\frac{ \left\lfloor \frac{n}{2} \right\rfloor \eta^2\left[1-h(Q_{Xb}) -h(Q_{Zb})\right]}{n(n-1)^2}. \label{asymp-rate-bif-acka} 
	\end{align}
	For completeness, we also report the asymptotic conference key rate of a standard CKA protocol (namely, the multipartite BB84 protocol introduced in \cite{Grasselli_2018}), which distils a secret conference key for the $n$ parties in the network by distributing single photons entangled in GHZ states:
	\begin{equation}
		r^{\infty}_{\rm cka} = \eta^n \left[1-h(Q_X) -h(Q_Z)\right] \label{asymp-rate-cka}.
	\end{equation}
	Interestingly, we note that the above rate is identical to the asymptotic rate of the ACKA protocol \eqref{asymp-rate-acka}, even though the latter protocol is more involved as it guarantees anonymity for the participants.
	
	Even in the case of standard CKA, we can devise an alternative protocol which only uses Bell pairs to establish the conference key. Note that such a protocol would require at least two network uses for every shared conference key bit, regardless of the number of parties. For instance, in the case of four parties --Alice and three Bobs-- the first network use distributes Bell states to Alice-Bob$_1$ and  Bob$_2$-Bob$_3$, while the second network use distributes a Bell pair to Bob$_1$-Bob$_2$. By employing a one-time pad the $n$ parties can establish a shared conference key from their pairwise secret keys. If we then consider that the  Bell pairs distributed by the source can be noisy or lost, the asymptotic key rate of a CKA protocol exclusively based on Bell pairs is independent of $n$ and reads:
	\begin{equation}
		r^{\infty}_{b\rm cka} = \frac{1}{2}\eta^2 \left[1-h(Q_{Xb}) -h(Q_{Zb})\right] \label{asymp-rate-bcka}.
	\end{equation}
	
	\begin{figure}[htb]
		\centering
		\includegraphics[width=1\linewidth,keepaspectratio]{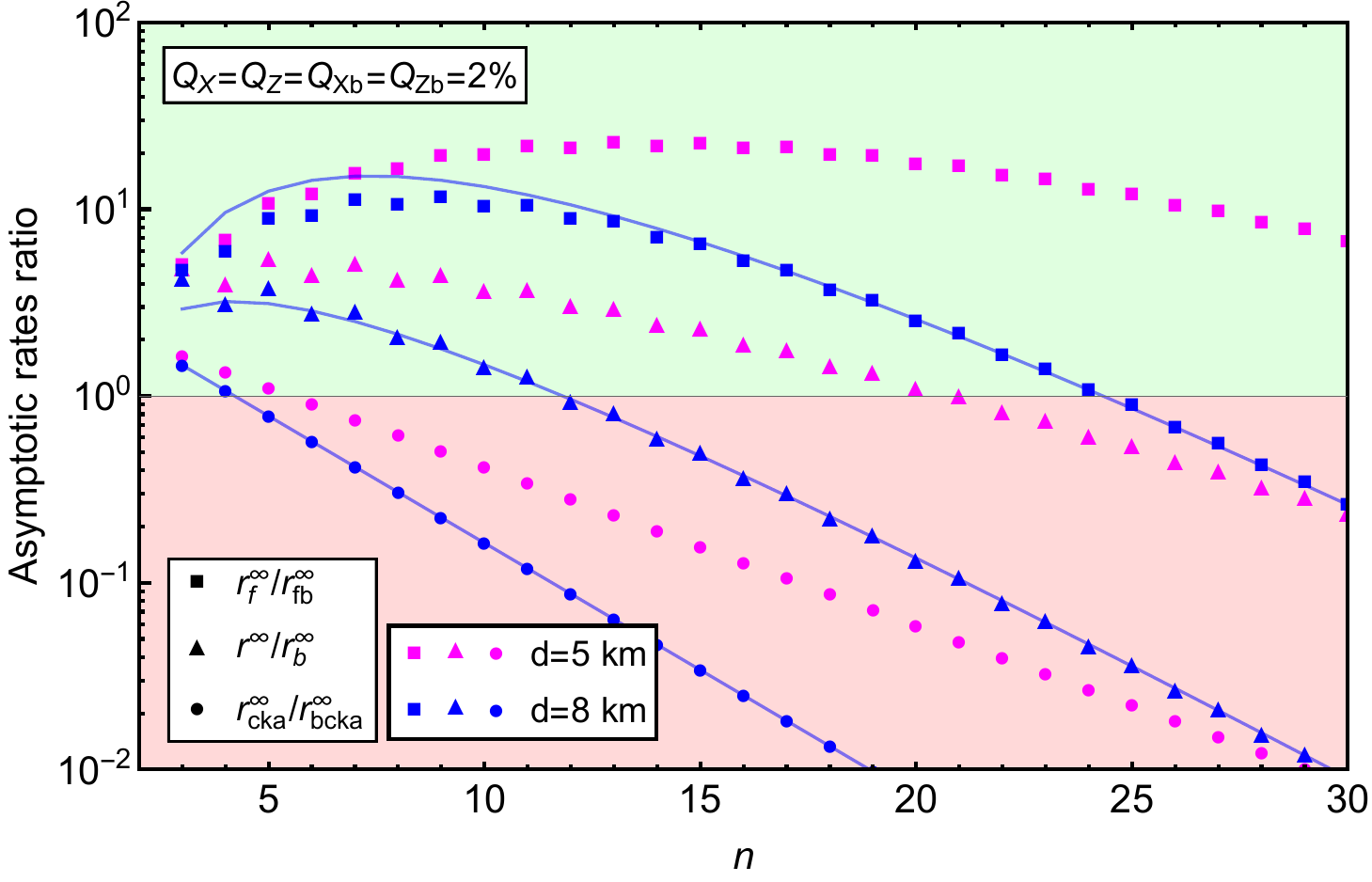}
		\caption{Ratios between the asymptotic ($L_{\rm tot}\to\infty$) secret conference key rates of the fully-ACKA (squares), ACKA (triangles) and CKA (circles) protocols with GHZ states and the rates of the corresponding protocols with Bell pairs, as a function of the number of parties in the network. The ratios of ACKA and fully-ACKA are well into the green region (i.e., greater than one), indicating that the use of GHZ states is advantageous compared to only using Bell pairs, when anonymity requirements are added to a standard CKA scheme. The error rates of the GHZ ($Q_Z,Q_X$) and Bell states ($Q_{Zb},Q_{Xb}$) are fixed to $2\%$, the distance party-server to $d=5$~km and $d=8$~km. The blue lines are given by Eq.~\eqref{scaling} with $d=8$~km.}
		\label{fig:asymptotic-rates}
	\end{figure}

	In Fig.~\ref{fig:asymptotic-rates} we plot the ratios $r^{\infty}/r^{\infty}_b$, $r^{\infty}_f/r^{\infty}_{bf}$ and $r^{\infty}_{\rm cka}/r^{\infty}_{b\rm cka}$ as a function of the number of parties $n$, for a distance party-server of $d=5$~km and $d=8$~km. We fixed the error rates of the GHZ state and Bell state to the same value ($2\%$), rather than modelling the state preparation with CNOT gates, to highlight implementation-independent behaviors. The ratios for the task of ACKA and fully-ACKA are well above 1 and can comfortably exceed one order of magnitude, indicating the advantage of using GHZ states over Bell pairs to anonymously establish conference keys. Conversely, the use of GHZ states barely brings any benefit for standard CKA, at least in our network model where the server can simultaneously distribute several Bell pairs (this is not necessarily true when considering different network models \cite{Epping}).
	\allowdisplaybreaks
	
	\section{Discussion} \label{sec:discussion}
	
	\noindent We introduced a security definition that reflects all the desired properties of an anonymous conference key agreement protocol and encompasses different levels of anonymity. Even though our security definition is inspired by the composable security paradigm, it remains an open point whether the three conditions given in Definition~\ref{def:security-informal} imply composability for an ACKA (fully-ACKA) protocol.
	
	We designed efficient and noise-robust protocols exploiting the multipartite entanglement of GHZ states, proved their security according to our security definition, and benchmarked their performance with counterpart protocols exclusively relying on the bipartite entanglement of Bell pairs.
	
	Our security proofs rely on the assumption that the $n$ parties either hold a short-lived quantum memory or have no quantum memory at all (bounded storage model \cite{bounded_storage}). The eavesdropper, instead, holds a perfect quantum memory and can perform coherent attacks as in standard QKD. While the bounded storage assumption is crucial for the security of ACKA and fully-ACKA, it is irrelevant for the bACKA and bifully-ACKA protocols, which would still remain secure even if this assumption is dropped. Potentially, there might exist protocols based on multipartite entanglement that are secure without the bounded storage assumption for the $n$ parties, and at the same time retain the valuable properties of efficiency, noise-robustness (untrusted source) and anonymity featured by our ACKA and fully-ACKA protocols. However, we think that there could be a fundamental reason forbidding the existence of such protocols. A hint in this direction comes from the impossibility of quantum bit commitment \cite{Broadbent2016}. Indeed, our ACKA and fully-ACKA protocols essentially require each non-participant to commit to a bit for each distributed state, where the bit is their $X$ measurement outcome, and to reveal the bit when the state distribution is over. Thus, quantum bit commitment can be viewed as a resource in our protocols, which, however, can only be implemented when other assumptions (e.g. bounded storage) are made.
	
	From the plots of Fig.~\ref{fig:finite-key}, we observe that the protocols based on GHZ states (ACKA and fully-ACKA) outperform the protocols exclusively based on Bell pairs (bACKA and bifully-ACKA) for an experimentally feasible \cite{CKAexperiment} number of network uses, starting from values as low as $L_{\mathrm{tot}}=10^5$. However, increasing the distance between the parties and the source of entanglement is more detrimental for the protocols based on GHZ states (bottom plots), due to the higher probability of losing at least a photon in a GHZ state compared to a Bell state. This effect can be partially mitigated if the preparation quality of the entangled states is improved (dotted lines in Fig.~\ref{fig:finite-key}). Indeed, reducing the gate failure probability from $2\%$ to $1\%$ significantly impacts the rates of the protocols based on GHZ states, while leaving the rates of the Bell-state protocols almost unchanged --hence in Fig.~\ref{fig:finite-key} we only reported the $f_G=2\%$ case for the protocols based on Bell pairs.
	
	Moreover, we observe that the protocols based on GHZ states require a higher number of network uses to yield a non-zero key rate, compared to the protocols based on Bell states. This is partially explained by the fact that we did not include finite-key effects in the rate at which the bipartite private channels distribute secret bits, while we performed a full finite-key analysis for the conference key rate with GHZ states. Since the bACKA and bifully-ACKA protocols rely on the bipartite private channels much more prominently than ACKA and fully-ACKA, the former protocols are  advantaged compared to the latter for low numbers of network uses. Even if we performed a full finite-key analysis for the bACKA and bifully-ACKA rates, they would still outperform the rates of ACKA and fully-ACKA in the low-$L_{\mathrm{tot}}$ regime (see Supplemental Material \cite{supplmat} for a detailed discussion).
	
	Finite-key effects aside, Fig.~\ref{fig:asymptotic-rates} clearly displays the benefit of employing multipartite entanglement, in the form of GHZ states, over Bell pairs, for the cryptographic tasks of ACKA and fully-ACKA. Interestingly, the superiority of GHZ-based protocols does not increase monotonically with $n$ and instead displays an optimal value of $n$ for which GHZ states are most beneficial. This is due to the interplay between two effects.
	
	On the one hand, establishing one conference key bit while maintaining \textit{anonymity} requires only 1 GHZ state in both ACKA and fully-ACKA, while it requires $n(n-1)$ Bell pairs in the case of bACKA and $n(n-1)^2$ Bell pairs in bifully-ACKA (due to the iteration of Parity protocols). This is only partially mitigated by the parallel distribution of Bell pairs from the source in each network use, which compensates by a factor $\lfloor n/2\rfloor^{-1}$. This effect dominates at low $n$ and causes the ratios of ACKA and fully-ACKA to increase with $n$. On the other hand, when the number of parties --and hence the number of simultaneously transmitted photons in a GHZ state-- increases, the rate of the GHZ-based protocols exponentially decreases due to photon loss with a factor $\eta^n$, while the rates of bACKA and bifully-ACKA present a constant factor $\eta^2$. Overall, the ratios exponentially decrease at high $n$.
	
	The interplay between a polynomial increase due to the efficiency of GHZ states and an exponential suppression due to photon loss becomes clear when computing the ratios, having set the error rates to zero and ignoring the floor functions:
	\begin{align}
		\frac{r^{\infty}_{\rm cka}}{r^{\infty}_{b\rm cka}} &\sim 2 \eta^{n-2} \nonumber\\
		\frac{r^{\infty}}{r^{\infty}_b} &\sim 2(n-1) \eta^{n-2} \nonumber \\
		\frac{r^{\infty}_f}{r^{\infty}_{bf}} &\sim 2(n-1)^2 \eta^{n-2} \label{scaling}.
	\end{align}
	The above functions are plotted as solid lines in Fig.~\ref{fig:asymptotic-rates} for $d=8$~km and reproduce the scaling of the plot points well.
	
	From \eqref{scaling} we conclude that, starting from a standard CKA scenario, if we require the participants to be anonymous with respect to the other parties and to Eve (ACKA scenario), the conference key rate of a GHZ-based protocol gains a factor of $n-1$ over the rate of a protocol based on Bell pairs. If we additionally require anonymity among the participants (fully-ACKA scenario), the key rate of a GHZ-based protocol gains a factor of \mbox{$(n-1)^2$} compared to just using Bell pairs. This holds despite allowing the source to simultaneously distribute multiple Bell pairs in one network use. Overall, this suggests that adding anonymity requirements significantly increases the advantage of multipartite entanglement over bipartite entanglement.
	
	
	Furthermore, we point out that the exponential suppression in \eqref{scaling} due to photon loss is heavily dependent on the length in kilometers $d$ of each party-server channel. This implies that the maximum number of parties $n$ for which the GHZ-based protocols provide an advantage (the ratios are above one) increases rapidly as $d$ decreases, as suggested by Fig.~\ref{fig:asymptotic-rates}. Indeed, for applications where the distance between parties is of few kilometers, the use of GHZ states is still advantageous in networks of more than $100$ parties (see discussion in Supplemental Material \cite{supplmat}). Additionally, the suppression due to photon loss could be avoided altogether by resorting to a different class of multipartite entanglement, namely W states \cite{Wstate}. Indeed, it has been shown that CKA can also be achieved when the parties share a W state post-selected after single-photon interference \cite{WstateProtocol}. This means that only one out of $n$ photons needs to be successfully transmitted, yielding a conference key rate that scales with $\eta$ (instead of $\eta^n$ when using GHZ states).
	
	The performance advantage provided by GHZ states in the fully-ACKA scenario, however, comes at the expense of a slightly weaker anonymity claim with respect to honest-but-curious receivers. Indeed, the fully-ACKA protocol is anonymous according to Definition~\ref{def:sweak-anonym} while the bifully-ACKA protocol satisfies Definition~\ref{def:security-informal}. This is the result of a trade-off with the robustness to noise and efficiency featured by the fully-ACKA protocol. Indeed, the fact that our protocol efficiently verifies the source state only in a small fraction of rounds and is robust against noisy state preparations, allows it to succeed even when asymmetries affect the multipartite states distributed to the parties. On one hand, this makes our protocol very efficient, noise-robust and practical, as opposed to previous anonymous protocols based on multipartite entanglement \cite{Wehner2005,lipinska_anonymous_2018,acka_old,ACKA,Brassard2007,unnikrishnan_anonymity_2019,yang_examining_2020}. On the other hand, it prevents the protocol from satisfying the strong anonymity condition when asymmetric multipartite states are distributed by the source, as already discussed in Sec.~\ref{sec:results}~A.
	
	In summary, our work identifies the anonymous communication tasks of ACKA and fully-ACKA, provides them with a rigorous security framework, and  demonstrates that the multipartite quantum correlations of GHZ states can increase the rate at which conference keys are anonymously generated within a quantum network, compared to solely relying on bipartite entanglement. The gain in the rate increases as anonymity requirements are added to the protocol, it can comfortably exceed one order of magnitude, and scales with the square of the number of parties in the network ($\sim n^2$) in the case of a fully-ACKA protocol. This is a striking result when compared to previously-known scaling improvements due to multipartite entanglement \cite{Epping,CKAexperiment,WalkCVQSS}, which expected at most a linear gain in the rate scaling ($\sim n$). Moreover, previous results apply in the case of particularly favourable network structures (e.g. networks with bottlenecks), whereas in this work we consider a network which is symmetric and does not privilege any party.  Additionally, we obtain such a scaling advantage despite the fact that the simultaneous distribution of multiple Bell pairs and of a single GHZ state contribute equally to the count of network uses, differently from previous works. 
	
	Therefore, our results provide strong evidence for the superiority of multipartite entanglement over bipartite entanglement for multi-user cryptographic tasks and pave the way for the implementation of quantum communication protocols beyond QKD and CKA.
	
	\section*{Acknowledgments}
	\noindent We thank Lennart Bittel and Nathan Walk for fruitful discussions. F.G., G.M., D.B. and H.K. acknowledge support by the Deutsche Forschungsgemeinschaft (DFG, German Research Foundation) under Germany’s Excellence Strategy - Cluster of Excellence Matter and Light for Quantum Computing (ML4Q) EXC 2004/1 -390534769. D.B. and H.K. acknowledge support by the QuantERA project QuICHE, via the German Ministry for Education and Research (BMBF Grant No. 16KIS1119K). A.P. and J.d.J. acknowledge support from the German Research Foundation (DFG, Emmy Noether grant No. 418294583) and F.H. from the German Academic Scholarship Foundation.
	
	F.G. and G.M. contributed equally to this work.
	
	\appendix

	\section{NOTATION} \label{sec:notation}
	
	\noindent Here we describe the formalism and notation used in the appendices and the Supplemental Material \cite{supplmat}.
	
	\begin{itemize}
		\item Strings of numbers and bitstrings are denoted with the vector sign $\vec{\phantom{a}}$, while symbols that are in boldface denote the tensor product of multiple subsystems, e.g. $\ketbra{\boldsymbol{\emptyset}}{\boldsymbol{\emptyset}}_{K_1 K_2 \dots K_n}=\bigotimes_{t=1}^n\ketbra{{\emptyset}}{{\emptyset}}_{K_t}$.
		
		\item If $\Delta$ indicates an event, then $\Delta^c$ indicates the complementary event and $\Pr[\Delta]$ denotes the probability that event $\Delta$ occurs.
		
		\item We use the running indices $i$ and $\vec{j}$ to indicate the sender and the set of receivers, respectively, for different protocol instances. Both indices run in the set of all parties: $\{i,\vec{j}\}\subset\{1,\dots,n\}$. We refer to the sender and receivers ($\{i,\vec{j}\}$) as the \emph{participants} of the protocol, while the remaining parties ($\{1,\dots,n\} \setminus \{i,\vec{j}\}$) are called \emph{non-participants}. In the case of multiple candidate senders, we indicate them with $\vec{i}$. Note that the size of the vectors $\vec{i}$ and $\vec{j}$ is not fixed.
		
		\item With $\mathcal{D}\subset \{1,\ldots,n\}$ we indicate the set of dishonest parties that may collaborate with the eavesdropper Eve.
		
		\item We denote by Identity Designation (ID) the sub-protocol of any ACKA and fully-ACKA protocol that either unambiguously assigns the identities of sender and receivers to the parties in the network or aborts. In the case of ACKA, every receiver is also informed about the identity of the other participants when ID does not abort.
		
		\item Let $p(\vec{i},\vec{j})$ be the probability that, when running the ID protocol, the parties in the set $\vec{i}$ apply to become the sender and select their receivers in $\vec{j}$ (for instance, $\vec{j}=\{\vec{j}_{i_1},\vec{j}_{i_2},\dots,\vec{j}_{i_{|\vec{i}|}}\}$, where $\vec{j}_{i_k}$ are the receivers selected by party $i_k$ if they become the sender). If there is only one party applying to become the sender ($|\vec{i}|=1$), then $\vec{j}$ is the set of receivers selected by the candidate sender $i$. The probability of this instance is indicated by $p(i,\vec{j})$.
		
		\item We define the following events.
		\begin{itemize}
			\item $\Gamma$: The ID sub-protocol aborts from the point of view of every party in the network.
			\item $\Phi$: The ID sub-protocol does not abort from the point of view of any party in the network and correctly designates the identity of the participants. That is, if party $i$ is the only one applying to become the sender with intended receivers $\vec{j}$, then the protocol designates party $i$ as the sender and $\vec{j}$ as the receivers. In the case of ACKA, the receivers are correctly informed about the identities of the other participants.
			\item $\Omega_{\mathcal{P}}$: The ACKA (fully-ACKA) protocol ends without aborting from the point of view of every participant.
			\item $\Gamma_{\mathcal{P}}$: The ACKA (fully-ACKA) protocol aborts from the point of view of every participant.
		\end{itemize}
		
		\item Every party $t \in \{1,\dots,n\}$ holds three personal classical registers: $P_t$, $K_t$ and $C_t$.
		
		\begin{itemize}
			\item The register $P_t$ stores information about the identity of party $t$ and eventual information on the identity of the other participants, as assigned by the ID sub-protocol. More specifically, $P_t=r$ if $t$ is a receiver of a fully-ACKA protocol, $P_t=r,i,\vec{j}$ if $t$ is a receiver of an ACKA protocol, $P_t=s,\vec{j}$ if $t$ is the sender with intended receivers in $\vec{j}$, and $P_t=\perp$ if $t$ is a non-participant. We emphasize that $r$, $s$ and $\perp$ are just symbols used to discriminate the identity of the parties. Finally, $P_t=\emptyset$ if the protocol aborts during ID.
			
			\item The register $K_{t}$ stores either the conference key if $t$ is a participant, or any information that the party might use to compute a guess of the conference key once the protocol is over if $t$ is not a participant (a dishonest or honest-but-curious non-participant might want to learn the conference key). If the protocol aborts for party $t$, we set $K_t=\emptyset$.
			
			\item The register $C_t$ stores all the classical side information held by party $t$ at any point in the protocol, which includes their private inputs and outputs to classical protocols and the public outputs.
		\end{itemize}

		\item We indicate with $K$ the set of registers $K_t$ of every party, i.e. $K=K_1 K_2 \dots K_n$, and similarly for $C$ and $P$. When we insert subscripts, it means that we restrict to the registers of the parties in the subscript. For instance, if $\mathcal{G}\subset\{1,\dots,n\}$ is a subset of parties, $K_{\mathcal{G}}$ is the set of conference key registers of the parties in $\mathcal{G}$. {Similarly, $(PKC)_{\mathcal{G}}$ indicates the content of the registers $P_t$, $K_t$ and $C_t$ for every party $t\in\mathcal{G}$.} Finally, with the superscript $^c$ in $K_{\mathcal{G}}^c$, we denote the registers of the complement of $\mathcal{G}$.
		
		\item $E$ is the quantum register of Eve.
		
		\item If the state of some registers depends on the value of a random variable $X$, we can express it as \mbox{$\rho=\sum_x \Pr[X=x] \rho_{|x}$}, where $\rho_{|x}$ is the state of the registers when $X=x$.\\
		Let $\Delta$ be an event for the variable $X$, i.e. $\Pr[\Delta]=\sum_{x\in\Delta}\Pr[X=x]$. With $\rho_{|\Delta}$ we indicate the \emph{normalized} state of the registers conditioned on the event $\Delta$, $\rho_{|\Delta}:= (1/\Pr[\Delta])\sum_{x\in\Delta}\Pr[X=x] \rho_{|x}$. With $\rho_{\wedge\Delta}$ we indicate the \emph{sub-normalized} state conditioned on $\Delta$ whose trace corresponds to the probability of event $\Delta$ occurring: $\rho_{\wedge\Delta}:= \sum_{x\in\Delta}\Pr[X=x] \rho_{|x}$.

		\item With $\rho_{PKCE|i,\vec{j}}$ ($\rho_{PKCE|\vec{i},\vec{j}}$) we indicate the state of registers $P$, $K$, $C$ and $E$ conditioned on the event where party(ies) $i$ ($\vec{i}$) applied to be the sender with the corresponding intended receivers in $\vec{j}$.

		\item If Alice is the sender and Bob$_1$, Bob$_2$, \dots, Bob$_m$ are the $m$ receivers, we replace the indices $i$ and $\vec{j}$ pointing to their registers with $A$ and $\vec{B}$, respectively. Thus, for instance, $K_A$ is Alice's key register containing her conference key $k_A$ and $K_{B_l}$ is Bob$_l$'s key register containing his conference key  $\vec{k}_{B_l}$ (for $l \in \{1,\dots,m\}$).
		
		\item Let $\abs{\vec{x}}$ be the number of entries of the string $\vec{x}$. Let $\omega_r(\vec{x})$ be the relative Hamming weight of the bitstring $\vec{x}$, i.e. $\omega_r(\vec{x}):=\abs{\{k:x_k=1\}}/\abs{\vec{x}}$.
		
		\item The trace distance between two states $\rho$ and $\sigma$ is given by $\norm{\rho -\sigma}=\frac{1}{2}\Tr[\sqrt{(\rho -
			\sigma)^2}]$ and is proportional to the trace norm of the operator $\rho-\sigma$.
		
		\item $\rho^{f}_{PKCE}$ is the output state, or final state, of an ACKA (fully-ACKA) protocol. We also simply indicate it as $\rho^f$.
	\end{itemize}

	\section{SECURITY DEFINITIONS} \label{sec:security-defs}
	
	\noindent As discussed in Sec.~\ref{sec:results}, we identify three properties that an ideal ACKA (fully-ACKA) protocol is expected to satisfy, namely: integrity, CKA-security and anonymity. Here we introduce a formal security definition which quantifies, for every property, how close the output state of the real protocol is from a state with the required property. Note that the output state $\rho^f$ of a generic ACKA (fully-ACKA) protocol can always be decomposed as the following:
	\begin{align}
		\rho^f&=\sum_{i,\vec{j}} p(i,\vec{j}) \rho^f_{PKCE|i,\vec{j}}  + \sum_{\vec{i},\vec{j}} p(\vec{i},\vec{j}) \rho^f_{PKCE|\vec{i},\vec{j}}  \label{rhof},
	\end{align}
	where the first term contains the output states of the protocol when only one party applied to become the sender (party $i$), while the second term groups the output states when multiple candidate senders applied (parties $\vec{i}$).
	The output state of an integrous protocol is of the form:
	\begin{align}\label{idealstateIN}
		\sigma^{\rm IN} &=\sum_{i,\vec{j}} p(i,\vec{j})  \left(\Pr[\Phi|i,\vec{j}]\,\xi^{(i,\vec{j})}_P \otimes \rho^{f}_{KCE|i,\vec{j},\Phi} \right.\nonumber\\
		&\hspace{0.5cm}\left.+ \Pr[\Phi^c|i,\vec{j}]\,\proj{\boldsymbol{\emptyset}}_P \otimes \rho^{f}_{KCE|i,\vec{j},\Phi^c}\right)  \nonumber\\
		&+ \sum_{\vec{i},\vec{j}} p(\vec{i},\vec{j})  \proj{\boldsymbol{\emptyset}}_P \otimes \rho^{f}_{KCE|\vec{i},\vec{j}},
	\end{align}
	where $\xi^{(i,\vec{j})}_P$ is the ideal state of registers $P$, provided that party $i$ applied to become sender with receivers in $\vec{j}$ and the identities were correctly assigned. For an ACKA protocol the state $\xi^{(i,\vec{j})}_P$ reads:
	\begin{align}
		\xi^{(i,\vec{j})}_P &:= \ketbra{s,\vec{j}}{s,\vec{j}}_{P_i} \otimes \ketbra{r,i,\vec{j}}{r,i,\vec{j}}_{P_{\vec{j}}} \otimes \ketbra{\pmb{\bot}}{\pmb{\bot}}_{P^c_{i,\vec{j}}} , \label{xi-acka}
	\end{align}
	while for a fully-ACKA protocol it reads:
	\begin{align}
		\xi^{(i,\vec{j})}_P &:= \ketbra{s,\vec{j}}{s,\vec{j}}_{P_i} \otimes \ketbra{r}{r}_{P_{\vec{j}}} \otimes \ketbra{\pmb{\bot}}{\pmb{\bot}}_{P^c_{i,\vec{j}}} \label{xi-facka}.
	\end{align}
	The distinction between the two states in \eqref{xi-acka} and \eqref{xi-facka} is always clear from the context of the protocol being addressed (either an ACKA or a fully-ACKA protocol).
	
	We remark that the integrous state \eqref{idealstateIN} is only constrained by the states of its $P$ registers. As such, it is obtained from the output state of the real protocol \eqref{rhof} by replacing the $P$ register of every party with the abort symbol $\emptyset$, whenever event $\Phi^c$ occurs or when there are multiple candidate senders or no sender ($|\vec{i}|\neq 1$).
	
	The output state of a CKA-secure protocol, conditioned on having honest participants ($\{i,\vec{j}\}\cap\mathcal{D}=\emptyset$) and on $\Phi$, is of the form:
	\begin{align}
		&\sigma_{KC_{i,\vec{j}}^cE|i,\vec{j},\Phi}^{{\rm CKA}}=  \Pr[\Omega_{\mathcal{P}}|i,\vec{j},\Phi]\tau_{K_{i,\vec{j}}}\otimes \rho^{f}_{(KC)_{i,\vec{j}}^c E|i,\vec{j},\Phi,\Omega_{\mathcal{P}}} \nonumber\\
		&+\Pr[\Omega^c_{\mathcal{P}}|i,\vec{j},\Phi]\proj{\boldsymbol{\emptyset}}_{{K}_{i,\vec{j}}}\otimes \rho^{f}_{(KC)_{i,\vec{j}}^c E|i,\vec{j},\Phi,\Omega^c_{\mathcal{P}}}  \label{idealstateCKA},
	\end{align}
	where the key registers $K_{i,\vec{j}}$ of the participants, for a key of $\ell$ bits, are perfectly correlated and random: 
	\begin{equation}
		\tau_{{K}_{i,\vec{j}}}:=\frac{1}{|\mathcal{K}|}\sum_{\vec{k}\in\mathcal{K}}\ketbra{\vec{k}}{\vec{k}}_{K_i} \otimes \ketbra{\vec{\pmb{k}}}{\vec{\pmb{k}}}_{K_{\vec{j}}}\quad,\quad\mathcal{K}=\{0,1\}^\ell. \label{tau}
	\end{equation}
	Note that, similarly to \eqref{idealstateIN}, the output state of a CKA-secure protocol \eqref{idealstateCKA} is obtained by replacing the $K_{i,\vec{j}}$ registers in the state of the real protocol with the ideal outputs.
	
	Finally, an ideal ACKA (fully-ACKA) protocol must be anonymous, i.e., the identity of each participant must be kept secret from the non-participants and Eve --and from the other receivers in the case of fully-ACKA. In this case we cannot simply replace the real output registers by their ideal counterparts, as in \eqref{idealstateIN} and \eqref{idealstateCKA}, since anonymity is a property of the global quantum state accounting for different instances of sender and receivers. For example, if a protocol is anonymous with respect to a non-participant $t$, its output state satisfies: $\rho^f_{P_t K_t C_t|i,\vec{j}}=\rho^f_{P_t K_t C_t|i',\vec{j}'}$ for $i\neq i'$ and $\vec{j} \neq \vec{j}'$, i.e. it is independent of the choice of sender and receivers. We remark that dishonest participants may broadcast their identity, or the identity of all the participants in the case of ACKA. Hence, in these scenarios we cannot require the output state to be independent of the identities that could be revealed.
	
	We now introduce the formal definitions of anonymity for an ACKA and a fully-ACKA protocol. We denote the output state of an anonymous ACKA (fully-ACKA) protocol by $\sigma^{\mathcal{D}}$. This emphasizes the fact that the anonymity requirements on $\sigma^{\mathcal{D}}$ crucially depend on the set of dishonest parties $\mathcal{D}$. For instance, no anonymity requirement is imposed on the output state of an ACKA protocol for the instances in which some participant is dishonest.

	\begin{Def}[ACKA anonymity]\label{def:anonym-acka}
		Let $\mathcal{D}$ be the set of dishonest parties taking part in an ACKA protocol, with output state $\sigma^{\mathcal{D}}$ of the form \eqref{rhof}. Then the ACKA protocol is anonymous if for any subset of parties $\mathcal{G}\subseteq \{1, \ldots, n\}$ it holds that:
		\begin{align}
			\sigma^{\mathcal{D}}_{P_{\mathcal{G}} K_{\mathcal{G}} C_{\mathcal{G}}E|i,\vec{j}}&=\sigma^{\mathcal{D}}_{P_{\mathcal{G}} K_{\mathcal{G}} C_{\mathcal{G}}E|i',\vec{j'}} \quad\forall\, i,i',\vec{j},\vec{j}'\notin  \mathcal{G} \cup {\mathcal{D}} \label{state:anony-dishonest} \\
			\sigma^{\mathcal{D}}_{P_{\mathcal{G}} K_{\mathcal{G}} C_{\mathcal{G}}E|\vec{i},\vec{j}}&=\sigma^{\mathcal{D}}_{P_{\mathcal{G}} K_{\mathcal{G}} C_{\mathcal{G}}E|\vec{i'},\vec{j'}} \quad\forall\, \vec{i},\vec{i'},\vec{j},\vec{j}'\notin  \mathcal{G}\cup {\mathcal{D}} \label{state:anony-dishonest-multiple-senders},
		\end{align}
		where $\sigma^{\mathcal{D}}_{P_{\mathcal{G}}K_{\mathcal{G}}C_{\mathcal{G}}E|i,\vec{j}}=\Tr_{P_{\mathcal{G}}^c K_{\mathcal{G}}^cC_{\mathcal{G}}^c}[\sigma^{\mathcal{D}}_{PKCE|i,\vec{j}}]$. 
	\end{Def}
	The anonymity conditions \eqref{state:anony-dishonest} and \eqref{state:anony-dishonest-multiple-senders} establish that the final state of any subset of non-participants and Eve is independent of the identity of the remaining parties. In other words, the reduced output states $\sigma^{\mathcal{D}}_{(PKC)^c_{i,\vec{j}}E|i,\vec{j}}$ and $\sigma^{\mathcal{D}}_{(PKC)^c_{\vec{i},\vec{j}}E|\vec{i},\vec{j}}$ do not contain any information about the identities of the participants. Note that if a participant would be contained in ${\mathcal{G}}$, the register $P_{\mathcal{G}}$ would also carry information about the identities of all the other participants (recall that an ACKA protocol reveals the participants' identities to each participant) and therefore there would be no condition to be satisfied. Moreover, note that \eqref{state:anony-dishonest} and \eqref{state:anony-dishonest-multiple-senders} do not impose any condition if the sender(s) or receivers are dishonest, since nothing prevents them from broadcasting the identities of all the participants. 
	
	\begin{widetext}
		\begin{Def}[Fully-ACKA anonymity]\label{def:anonym-f-acka}
			Let $\mathcal{D}$ be the set of dishonest parties taking part in a fully-ACKA protocol, with output state $\sigma^{\mathcal{D}}$ of the form \eqref{rhof}. Then the fully-ACKA protocol is anonymous if for any subset of parties $\mathcal{G}\subseteq \{1, \ldots, n\}$ it holds that:
			
			\begin{align}
				\sigma^{\mathcal{D}}_{P_{\mathcal{G}} K_{\mathcal{G}} C_{\mathcal{G}} E|i,\vec{j}}&=\sigma^{\mathcal{D}}_{P_{\mathcal{G}} K_{\mathcal{G}} C_{\mathcal{G}}E|i',\vec{j'}} \quad\forall\,i,i',\vec{j},\vec{j}':\,i,i'\notin  \mathcal{G}\cup \mathcal{D}\,,\,\vec{j}\cap {\mathcal{G}} =\vec{j'}\cap {\mathcal{G}}\,,\,\vec{j}\cap  \mathcal{D}=\vec{j'}\cap \mathcal{D}  \label{state:f-anony-dishonest}\\
				\sigma^{\mathcal{D}}_{P_{\mathcal{G}} K_{\mathcal{G}} C_{\mathcal{G}}E|\vec{i},\vec{j}}&=\sigma^{\mathcal{D}}_{P_{\mathcal{G}} K_{\mathcal{G}} C_{\mathcal{G}}E|\vec{i'},\vec{j'}} \quad\forall\,\vec{i},\vec{i}',\vec{j},\vec{j}':\,\vec{i},\vec{i}'\notin  \mathcal{G}\cup {\mathcal{D}}\,,\,\vec{j}\cap {\mathcal{G}} =\vec{j'}\cap {\mathcal{G}}\,,\,\vec{j}\cap  \mathcal{D}=\vec{j'}\cap \mathcal{D}. \label{state:f-anony-dishonest-multiple-senders}
			\end{align}
		\end{Def}
		The anonymity conditions \eqref{state:f-anony-dishonest} and \eqref{state:f-anony-dishonest-multiple-senders} extend the requirements of \eqref{state:anony-dishonest} and \eqref{state:anony-dishonest-multiple-senders} due to the fact that in a fully-ACKA protocol the receiver is unaware of the identity of the other participants. Indeed, conditions \eqref{state:f-anony-dishonest} and \eqref{state:f-anony-dishonest-multiple-senders} establish that in all the instances of the protocol in which the parties in subsets ${\mathcal{G}}$ and ${\mathcal{D}}$ have a fixed role (except for the role of sender), their reduced state is independent of the identities of the other parties.
		
		Definition~\ref{def:suff-cond-security}, which is a formal re-statement of Definition~\ref{def:security-informal}, defines the security of an ACKA (fully-ACKA) protocol through the trace distances of the output state of the protocol from an integrous state \eqref{idealstateIN}, a CKA-secure state \eqref{idealstateCKA}, and an anonymous state satisfying Definition~\ref{def:anonym-acka} (Definition~\ref{def:anonym-f-acka}), respectively.

		\begin{Def}[Security (rigorous)] \label{def:suff-cond-security}
			An ACKA (fully-ACKA) protocol with dishonest parties in $\mathcal{D}$ is $\varepsilon$-secure, with $\varepsilon=\varepsilon_{\rm IN}+\varepsilon_{\rm CKA}+\varepsilon_{\rm AN}$, if it satisfies the following three conditions.
			\begin{itemize}
				\item $\varepsilon_{\rm IN}$-integrity:
				\begin{align}
					\max_{i,\vec{j}} \Pr[\Gamma^c \cap \Phi^c|i,\vec{j}]\leq \varepsilon_{\rm IN} \quad \wedge\quad
					\max_{\vec{i},\vec{j}} \Pr[\Gamma^c|\vec{i},\vec{j}] \leq \varepsilon_{\rm IN},
					\label{integrity} 
				\end{align}
				\item $\varepsilon_{\rm CKA}$-CKA-security:
				\begin{align}\label{CKA-sec}
					\max_{\{i,\vec{j}\} \cap\mathcal{D} =\emptyset} &\Pr[\Omega_{\mathcal{P}}|{i,\vec{j}},\Phi]\,\norm{\rho^f_{KC_{i,\vec{j}}^cE|i,\vec{j},\Phi,\Omega_{\mathcal{P}}}- \tau_{{K}_{i,\vec{j}}}\otimes \rho^f_{(KC)_{i,\vec{j}}^c E|i,\vec{j},\Phi,\Omega_{\mathcal{P}}}} \nonumber\\
					&+ \Pr[\Omega^c_{\mathcal{P}} \cap \Gamma^c_{\mathcal{P}}|{i,\vec{j}},\Phi]\leq \varepsilon_{\rm CKA},
				\end{align}
				\item $\varepsilon_{\rm AN}$-anonymity:
				\begin{align}
					&\max_{i,\vec{j}} \norm{\rho^{f}_{(PKC)^c_{i,\vec{j}} E|i,\vec{j}}-\sigma^{\mathcal{D}}_{(PKC)^c_{i,\vec{j}} E|i,\vec{j}}} \leq \varepsilon_{\rm AN} \,\wedge\, \max_{\vec{i},\vec{j}}  \norm{\rho^{f}_{(PKC)^c_{\vec{i},\vec{j}}E|\vec{i},\vec{j}}-\sigma^{\mathcal{D}}_{(PKC)^c_{\vec{i},\vec{j}} E|\vec{i},\vec{j}}}\,\leq \varepsilon_{\rm AN} \nonumber\\
					&\hspace{13cm}\mbox{(ACKA)}\label{anonymity-acka} \\
					&\max_{i,\vec{j}} \norm{\rho^{f}_{(PKC)_i^c E|i,\vec{j}}-\sigma^{\mathcal{D}}_{(PKC)_i^c E|i,\vec{j}}} \leq \varepsilon_{\rm AN} \quad\wedge\quad \max_{\vec{i},\vec{j}}  \norm{\rho^{f}_{(PKC)_{\vec{i}}^c E|\vec{i},\vec{j}}-\sigma^{\mathcal{D}}_{(PKC)_{\vec{i}}^cE|\vec{i},\vec{j}}}\,\leq \varepsilon_{\rm AN} \nonumber\\
					&\hspace{12cm}\mbox{(fully-ACKA)}\label{anonymity-f-acka}
				\end{align}
			\end{itemize}
			where $\rho^f$ is the output state of the ACKA (fully-ACKA) protocol, $\sigma^{\mathcal{D}}$ is the output state of any ACKA (fully-ACKA) protocol which satisfies Definition~\ref{def:anonym-acka} (Definition~\ref{def:anonym-f-acka}), and $\tau_{K_{i,\vec{j}}}$ is given in \eqref{tau}.
		\end{Def}
	\end{widetext}

	We remark that to simplify the integrity condition, we replaced the trace distance from the integrous state \eqref{idealstateIN} with a sufficient condition on the probabilities of unwanted events.
	
	The CKA-security condition \eqref{CKA-sec} establishes that the final state of the protocol, conditioned on honest parties $i$ and $\vec{j}$ correctly designated as sender and receivers, is close to a state in which an ideal key \eqref{tau} is distributed to the participants or where the protocol aborts for every participant. Indeed, in the anonymous conference key agreement scenario, it could happen that the protocol aborts for some participants but not for others (event $\Omega^c_{\mathcal{P}}\cap \Gamma^c_{\mathcal{P}}$). This would spoil the CKA-security of the protocol, since not all the participants would end up with the same conference key (recall that if the protocol aborts for party $t$, then $K_t=\emptyset$). We remark that in the standard QKD and CKA scenarios there is no need to account for such instances, since the identities of the participants are public and they can communicate over the authenticated classical channel, thus agreeing on when the protocol aborts.
	
	Finally, the anonymity conditions \eqref{anonymity-acka} and \eqref{anonymity-f-acka} account for deviations of the real protocol from an anonymous protocol, as defined by Definitions~\ref{def:anonym-acka} and \ref{def:anonym-f-acka}, respectively. Nevertheless, since the property of anonymity is intertwined with integrity and CKA-security, any issue of the real protocol with regards to the latter may at the same time increase its deviation from a perfectly anonymous protocol.
	
	We observe that the conditions \eqref{integrity}-\eqref{anonymity-f-acka} are independent of the distribution $\{p(i,\vec{j}), p(\vec{i},\vec{j})\}$, which describes the probability that certain parties apply to become senders with specified receivers. This is a desirable feature since the distribution may not be accessible to a party who wants to prove the protocol's security.\vspace{1em}
	
	As discussed in Sec.~\ref{sec:results}, our fully-ACKA protocol (Protocol~\ref{prot:sf-ACKA}) does not satisfy the anonymity condition \eqref{anonymity-f-acka} of Definition~\ref{def:suff-cond-security}, but rather a weaker anonymity condition provided in Definition~\ref{def:sweak-anonym} and formally stated here.
	
	\begin{Def}[Weak-anonymity for fully-ACKA (rigorous)] \label{def:weak-anonym}
		Let $\mathcal{D}$ be the set of dishonest parties taking part in a fully-ACKA protocol, with output state $\rho^{f}$. The fully-ACKA protocol is $\varepsilon_{\rm w AN}$-weak-anonymous, with $\varepsilon_{\rm wAN}=\varepsilon_{\rm npAN} + \varepsilon_{\rm pAN}$, if the following two conditions are satisfied:
		\begin{enumerate}
			\item (Anonymity with respect to non-participants and Eve) The output state of the protocol satisfies the anonymity condition \eqref{anonymity-acka} for ACKA protocols:
			\begin{align}
				&\max_{i,\vec{j}} \norm{\rho^{f}_{(PKC)^c_{i,\vec{j}} E|i,\vec{j}}-\sigma^{\mathcal{D}}_{(PKC)^c_{i,\vec{j}} E|i,\vec{j}}} \leq \varepsilon_{\rm npAN} \quad \wedge \nonumber\\
				&\max_{\vec{i},\vec{j}}  \norm{\rho^{f}_{(PKC)^c_{\vec{i},\vec{j}} E|\vec{i},\vec{j}}-\sigma^{\mathcal{D}}_{(PKC)^c_{\vec{i},\vec{j}} E}}\,\leq \varepsilon_{\rm npAN},\label{anonymity-weak}
			\end{align}
			where $\sigma^{\mathcal{D}}$ satisfies Definition~\ref{def:anonym-acka}.
			\item (Anonymity with respect to honest-but-curious receivers) When the protocol's specifications are invariant under permutations of parties, any subset of honest-but-curious receivers $\mathcal{R}\subseteq\vec{j}$ cannot guess the identity of other participants with a higher probability than the trivial guess, except for a small deviation $\varepsilon_{\rm pAN}$. In formulas, the probability of correctly guessing the set of participants is bounded as follows:
			\begin{align}
				p_{guess} &\leq \max_{i,\vec{j}\supseteq \mathcal{R}} p_{\mathcal{R}}(i,\vec{j}) + \varepsilon_{\rm pAN} \label{weak-pguess1}\\
				p_{guess} &\leq \max_{\vec{i},\vec{j}\supseteq \mathcal{R}} p_{\mathcal{R}}(\vec{i},\vec{j}) + \varepsilon_{\rm pAN}, \label{weak-pguess2}
			\end{align}
			where $p_{\mathcal{R}}(i,\vec{j})$ ($p_{\mathcal{R}}(\vec{i},\vec{j})$) is defined as: $p_{\mathcal{R}}(i,\vec{j}):= p(i,\vec{j})/(\sum_{i,\vec{j}\supseteq\mathcal{R}} p(i,\vec{j}))$ ($p_{\mathcal{R}}(\vec{i},\vec{j}):= p(\vec{i},\vec{j})/(\sum_{\vec{i},\vec{j}\supseteq\mathcal{R}} p(\vec{i},\vec{j}))$).
		\end{enumerate}
	\end{Def}

	\section{SUB-PROTOCOLS} \label{sec:sub-protocols}
	
	\noindent Before presenting the ACKA and fully-ACKA protocols, we introduce the sub-protocols on which they built upon. Such protocols involve many different bitstrings, whose length in many cases is not defined to be an integer number, unless a ceiling or floor function is applied. In order not to increase the complexity of the notation, we omit the ceiling and floor functions. Note that, from the point of view of the plots, this omission is irrelevant since few bits of difference do not sensibly modify the key rates (consider that all plots start from $L_{\rm tot}=10^5)$.
	
	To start with, we make use of the classical Parity, Veto and Collision Detection protocols from \cite{Broadbent2007}. The first two protocols compute the parity and the logical OR of their inputs, respectively, while Collision Detection detects the presence of multiple parties applying to become the sender. Importantly, in our version of the Parity protocol we do not require simultaneous broadcast, contrary to Ref.~\cite{Broadbent2007}.\\
	
	\begin{protocol}\label{prot:parity}
		\caption{Parity \cite{Broadbent2007}}
		\noindent Let $x_t \in \{0,1\}$ be the input of party $t$ and $y_t$ the output of the protocol for party $t$. Then $y_t=x_1 \oplus x_2 \oplus \dots \oplus x_n$ for every $t$.\\
		Every party $t\in \{1,\ldots, n\}$ does the following.
		\begin{enumerate}
			\item Select uniformly at random an $n$-bit string $\vec{r}_t=r_t^1 r_t^2 \dots r_t^n$ such that $x_t=\oplus_{j=1}^n r^j_t$.
			\item Send $r^j_t$ to party $j$ and keep $r^t_t$.
			\item Compute $z_t=\oplus_{j=1}^n r^t_j$, i.e. the parity of the bits received (including $r^t_t$).
			\item Broadcast $z_t$.
			\item Compute $y_t=\oplus_{k=1}^n z_k$ to obtain the parity of the inputs $\{x_t\}_t$.
		\end{enumerate}
	\end{protocol}\vspace{0.5cm}
	\newpage

	\begin{protocol}\label{prot:Veto}
		\caption{Veto \cite{Broadbent2007}}
		\noindent Let $x_t \in \{0,1\}$ be the input of party $t$ and $y_t$ the output of the protocol for party $t$. Then $y_t=x_1\lor x_2 \lor \dots \lor x_n$ for every $t\in \{1,\ldots, n\}$.
		\begin{enumerate}
			\item Initialize $y_t=0$ $\forall \,t$.
			\item For every party $t$, repeat the following ${r_V}$ times:
			\begin{enumerate}
				\item Each party $j$ sets the value of $q_j$ according to the following     
				\begin{align}
					\left\lbrace\begin{array}{ll}
						q_j =0 &\mbox{if } x_j=0   \\
						q_j \in_R \{0,1\}  &\mbox{if } x_j=1
					\end{array}\right.
				\end{align}
				where $q_j \in_R \{0,1\}$ denotes that $q_j$ is picked uniformly at random in $\{0,1\}$.
				\item The parties execute the Parity protocol (Protocol~\ref{prot:parity}) with inputs $q_1, q_2 ,\ldots, q_n$, such that party $t$ is the last to broadcast. If the outcome of the Parity protocol is 1 or any party refuses to broadcast, then set $y_t=1$ $\forall\, t \in \{1,\ldots, n\}$.
			\end{enumerate}
		\end{enumerate}
	\end{protocol}\vspace{0.5cm}
	
	Note that, due to the probabilistic action in step 2.1, Protocol~\ref{prot:Veto} allows a given party $t$ to learn whether any other party has input $1$ even when the input of party $t$ is $x_t=1$, which would not happen in an ideal implementation of the veto function \cite{Broadbent2007}. This crucial fact allows a candidate sender to detect the presence of other candidate senders in the first Veto performed in Collision Detection.\vspace{0.5cm}
	
	\begin{protocol}\label{prot:Collision}
		\caption{Collision Detection \cite{Broadbent2007}}
		\noindent This protocol is used to detect the presence of multiple candidate senders, or no senders at all. Let $x_t \in \{0,1,2\}$ be the input of party $t$ and $y_t$ the output of the protocol for party $t$. Then the output of every party $t\in \{1,\ldots, n\}$ is interpreted as follows:
		\begin{align}
			\left\lbrace\begin{array}{ll}
				y_t =0 &\mbox{no sender applies and the protocol aborts}    \\
				y_t =1 &\mbox{one sender applied and the protocol proceeds}\\
				y_t =2 &\mbox{a collision is detected (multiple senders)}\\
				& \mbox{and the protocol aborts}
			\end{array}\right.
		\end{align}
		The protocol steps are given in the following.
		\begin{enumerate}
			\item Veto A:
			\begin{enumerate}
				\item Each party $j$ sets $a_j=\min \{x_j,1\}$. 
				\item All participants perform the Veto protocol (Protocol~\ref{prot:Veto}) with input $a_1,a_2, \ldots, a_n$.
			\end{enumerate}
			\item Veto B (skip if Veto A outputs 0):
			\begin{enumerate}
				\item Each party $j$ sets
				\begin{align}
					\left\lbrace\begin{array}{ll}
						b_j =1 &\mbox{if } x_j=2, \mbox{ or if } x_j=1 \mbox{ and party $j$ detected}\\
						& \mbox{that another party had input 1 in Veto A} \\
						b_j =0 &\mbox{otherwise}
					\end{array}\right.
				\end{align}
				\item All participants perform the Veto protocol (Protocol~\ref{prot:Veto}) with input $b_1,b_2, \ldots, b_n$.
			\end{enumerate}
			\item Every party $t\in \{1,\ldots, n\}$ sets their output to:
			\begin{align}
				\left\lbrace\begin{array}{ll}
					y_t =0 &\mbox{if the output of Veto A is 0}     \\
					y_t =1 &\mbox{if the outputs of Veto A and B sum to 1} \\
					y_t =2 &\mbox{if the outputs of Veto A and B sum to 2}
				\end{array}\right.
			\end{align}
		\end{enumerate}
	\end{protocol}\vspace{0.5cm}
	
	In the Collision Detection protocol (Protocol~\ref{prot:Collision}), the input $x_t=2$ can describe the action of a dishonest party that wants to force the protocol to detect a collision and output $y_t=2$.
	\vspace{1em}
	
	In the ACKA protocol, Alice must disclose her identity and the identities of the other Bobs to every Bob. This is achieved by transmitting a bitstring containing the positions of the sender and the chosen receivers in the network. In particular, let $\vec{d}_t$ be an $(n-1+\log_2 n)$-bit long string that Alice transmits to every other party $t$. If Alice is transmitting $d_t$ to a Bob, then the first bit of $d_t$ is 1, the following $\log_2 n$ bits contain the position of Alice with respect to some predefined ordering of the $n$ parties, and the remaining $n-2$ bits indicate the identities of the other parties (bit~$1$ for receiver and bit~$0$ for a non-participant), except for Alice and the Bob receiving $d_t$. If instead Alice is transmitting $d_t$ to a non-participant, then the first bit is 0 and the remaining bits are random.
	
	In order to ensure that the transmission of $\vec{d}_t$ is error-free, Alice encodes the bitstring with a public encoding algorithm $(F,G)$ composed of an encoding function $F$ and a decoding function $G$, denoted Algebraic Manipulation Detection (AMD) code \cite{AMD,Broadbent2007}. An AMD code has the important property of detecting, with high probability, any tampering with the encrypted message, as stated in the following lemma.
	\begin{Lmm}[AMD code \cite{AMD,Broadbent2007}] \label{lmm:AMD}
		Let $\vec{x}$ be a bitstring. There exists an AMD code $(F,G)$, whose encoded string $F(\vec{x})$ has length:
		\begin{align}
			|F(\vec{x})|=|\vec{x}|+2\left(\log_2 |\vec{x}| + \log_2 \frac{1}{\varepsilon_{\mathrm{enc}}}\right),
		\end{align}
		such that $G(F(\vec{x}))=\vec{x}$ for every $\vec{x}$ and such that $\Pr[G(F(\vec{x})\oplus\vec{b})\neq \top]\leq \varepsilon_{\mathrm{enc}}$ for every bitstring $\vec{b}\neq \vec{0}$ chosen without prior knowledge of the encoded bitstring $F(\vec{x})$, while a complete knowledge of $\vec{x}$ is allowed \cite{AMD} (note that, due to the probabilistic nature of an AMD code, $F(\vec{x})$ can still be unknown even when the input $\vec{x}$ is known).
		
	\end{Lmm}
	For the use in ACKA, the encoded string $F(\vec{d}_t)$ has length:
	\begin{align}
		|F(\vec{d}_t)|&=n-1+\log_2 n \nonumber \\
		&+2\left(\log_2 (n-1+{\log_2 n}) + \log_2 \frac{1}{\varepsilon_{\mathrm{enc}}}\right) \label{F(d)-length},
	\end{align}
	and it is then used by Alice as her input in a sequence of Parity protocols in order to anonymously transmit it to party $t$. This is the core of the identity designation sub-protocol used by the ACKA protocol, named ACKA-ID.\\
	
	\begin{protocol}\label{prot:ID}
		\caption{Identity Designation for ACKA (\textbf{ACKA-ID})}
		\noindent Before the protocol starts, the $n$ parties set $v_t=0\; \forall t$.
		\begin{enumerate}[label=\arabic*., ref=\arabic*]
			\item\label{ID:step1} The $n$ parties perform the Collision Detection protocol (Protocol~\ref{prot:Collision}), where party $t$ inputs 1 if they want to be the sender or 0 otherwise. If the Collision Detection outputs 0 or 2 then the ACKA protocol aborts for all the $n$ parties. Else, they proceed with the next steps.
			\item\label{ID:step2} If the protocol does not abort in step~\ref{ID:step1}, then there is a single sender, whom we identify as Alice, except for a small probability.
			\item\label{ID:step3} For every party $t \in \{1,\dots,n\}$, do the following.
			\begin{enumerate}
				\item The $n$ parties perform the Parity protocol for $|F(\vec{d}_t)|$ times with the following inputs. If $t \neq A$, Alice uses $F(\vec{d}_t)$, party $t$ uses a random bitstring $\vec{r}_t$ and the other parties input $\vec{0}$. If $t=A$, then Alice uses $F(\vec{0}) \oplus \vec{r}$, where $\vec{r}$ is a random bitstring, and the other parties input $\vec{0}$. Let $\vec{o}$ be the output of the Parity protocols.
				\item Party $t$ computes $G(\vec{r}_t\oplus \vec{o})$. If $G$ returns $\top$, party $t$ sets $v_t=1$, otherwise party $t$ recovers their identity as assigned by Alice (and eventually the identity of the other participants if they are a receiver). If $t=A$, then Alice computes $G(\vec{r}\oplus \vec{o})$. If $G$ returns $\top$, Alice sets $v_t=1$.
			\end{enumerate}
			\item\label{ID:step4} The $n$ parties perform the Veto protocol (Protocol~\ref{prot:Veto}) with inputs $v_t$. If the Veto outputs 1, the ACKA protocol aborts for all the $n$ parties.
		\end{enumerate}
	\end{protocol}\vspace{0.5cm}
	
	The total number of bipartite channel uses required by the ACKA-ID protocol is: $n^2(n-1)(3 r_V \,+ |F(\vec{d}_t)|)$, where  $|F(\vec{d}_t)|$ is given by \eqref{F(d)-length}.
	
	Recall that, differently from ACKA, in a fully-ACKA protocol Alice only needs to communicate to each party whether they are a Bob or a non-participant. Thus, we present the fully-ACKA-ID protocol, where we denote $d_t$ the bit that Alice sends to every other party $t$ to communicate their identity. In particular, $d_t=1$ ($d_t=0$) means that party $t$ is a receiver (non-participant). In order to detect bit-flips in the transmission of $d_t$, Alice encodes $d_t$ with an AMD code \cite{Broadbent2007,AMD} $(F,G)$ which satisfies the statement of Lemma~\ref{lmm:AMD} and is such that:
	\begin{align}
		|F(d_t)| = 1+2\log_2 \frac{1}{\varepsilon_{\rm enc}} \label{F(dt)}.
	\end{align}
	\newpage 
	\begin{protocol}\label{prot:f-ID}
		\caption{Identity Designation for fully-ACKA (\textbf{fully-ACKA-ID})}
		\noindent Before the protocol starts, the $n$ parties set $v_t=0\; \forall t$.
		\begin{enumerate}[label=\arabic*., ref=\arabic*]
			\item\label{f-ID:step1} The $n$ parties perform the Collision Detection protocol (Protocol~\ref{prot:Collision}), where party $t$ inputs 1 if they want to be the sender or 0 otherwise. If the Collision Detection outputs 0 or 2 then the fully-ACKA protocol aborts for all the $n$ parties. Else, they proceed with the next steps.
			\item\label{f-ID:step2} If the protocol does not abort in step~\ref{ID:step1}, then there is a single sender, whom we identify as Alice, except for a small probability.
			\item\label{f-ID:step3} For every party $t \in \{1,\dots,n\}$, do the following.
			\begin{enumerate}
				\item The $n$ parties perform the Parity protocol for $|F(d_t)|$ times with the following inputs. If $t \neq A$, Alice uses $F(d_t)$, party $t$ uses a random bitstring $\vec{r}_t$ and the other parties input $\vec{0}$. If $t=A$, then Alice uses $F(0) \oplus \vec{r}$, where $\vec{r}$ is a random bitstring, and the other parties input $\vec{0}$. Let $\vec{o}$ be the output of the Parity protocols.
				\item Party $t$ computes $G(\vec{r}_t\oplus \vec{o})$. If $G$ returns $\top$, party $t$ sets $v_t=1$, otherwise party $t$ recovers their identity as assigned by Alice. If $t=A$, then Alice computes $G(\vec{r}\oplus \vec{o})$ and if $G$ returns $\top$, Alice sets $v_t=1$.
			\end{enumerate}
			\item\label{f-ID:step4} The $n$ parties perform the Veto protocol (Protocol~\ref{prot:Veto}) with inputs $v_t$. If the Veto outputs 1, the fully-ACKA protocol aborts for all the $n$ parties.
		\end{enumerate}
	\end{protocol}\vspace{0.5cm}
	
	The total number of bipartite channel uses required by the fully-ACKA-ID protocol is: $n^2(n-1)(3 r_V \,+ |F(d_t)|)$, where $|F(d_t)|$ is given by \eqref{F(dt)}.\vspace{1em}

	
	As discussed in Sec.~\ref{sec:results}, the fully-ACKA scenario prevents Alice from distributing the testing key through a pre-established conference key shared by all participants. For this, we introduce another sub-protocol called Testing Key Distribution (TKD) protocol. With this protocol, Alice anonymously provides each Bob$_l$ (for $l\in\{1,\dots,m\}$) with a key $\vec{k}_l$ that is given by the concatenation of two independent bitstrings: $\vec{k}_l=(\vec{k}_{T},\vec{r}_l)$, where $\vec{k}_T$ ($|\vec{k}_T|=Lh(p)$) is the testing key and $\vec{r}_l$ is additional randomness used in the error correction phase of fully-ACKA. The string $\vec{r}_l$ is obtained by encoding the concatenated random string $(b_l,\vec{r}_\emptyset)$ with an AMD code $(F,G)$: $\vec{r}_l=F(b_l,\vec{r}_\emptyset)$. In principle, Bob$_l$ can recover the concatenated string by computing $G(\vec{r}_l)=(b_l,\vec{r}_\emptyset)$ and use it in error correction. In particular, the random bit $b_l$ and the random string $\vec{r}_\emptyset$, with $|\vec{r}_\emptyset|=1+2\log_2 1/ \varepsilon_{\mathrm{enc}}$, are employed in two separate steps of error correction. By adding the lengths of the two bitstrings composing $\vec{k}_l$, we obtain the total length of $\vec{k}_l$:
	\begin{equation}
		|\vec{k}_l|= Lh(p) + 4\left(1+\log_2 \frac{1}{\varepsilon_{\mathrm{enc}}}\right)+2\log_2 \left(1+\log_2 \frac{1}{\varepsilon_{\mathrm{enc}}}\right) \label{testing-key-length}.
	\end{equation}
	
	In order not to reveal the number $m$ of Bobs chosen by Alice, the distribution of the key $\vec{k}_l$ is repeated for $n-1$ times. In each iteration, Alice first notifies one party, say party $s$, to be the recipient of the key $\vec{k}_s=(\vec{k}_T,\vec{r}_s)$ and then runs a sequence of Parity protocols to transmit the bits of the key, with $s\neq A$. In case the recipient is not a Bob, Alice sums modulo two the bits of the key with random bits before transmitting them. Alice's notification procedure is inspired by the Notification protocol in \cite{Broadbent2007} and it successfully notifies party $s$ with probability at least $1-2^{-r_N}$, thanks to the iteration of $r_N$ rounds.\\
	
	\begin{protocol}\label{prot:TKD}
		\caption{Testing Key Distribution (\textbf{TKD})}
		\noindent Let $y_t \in\{0,1\}$ be the output of the Notification sub-protocol for party $t$: If $y_t=1$, party $t$ has been notified as the recipient.
		Before the TKD protocol starts, we set the verification bits of Alice and Bob$_l$ to $v_A=v_{B_l}=0$ for every $l\in\{1,\dots,m\}$.
		The following sequence of steps is repeated $n-1$ times.
		\begin{enumerate}[label=\arabic*., ref=\arabic*]
			\item Alice randomly picks a party $s \in\{1,\dots,n\}$ that has not been notified in previous iterations.
			\item\label{TKD:step2} \textbf{Notification:} The $n$ parties repeat the following steps for every party $t \in\{1,\dots,n\}$.
			\begin{enumerate}
				\item Initialize $y_t=0$.
				\item Repeat $r_N$ times:
				\begin{enumerate}
					\item Every party $j \neq A$ sets $p_j=0$. Alice sets
					\begin{align}
						\left\lbrace\begin{array}{ll}
							p_A =0 &\mbox{if } t\neq s   \\
							p_A \in_R \{0,1\}  &\mbox{if } t=s.
						\end{array}\right.
					\end{align}
					\item The $n$ parties perform Parity (Protocol~\ref{prot:parity}) with inputs $p_1,p_2,\dots,p_n$ but party $t$ does not broadcast. In this way party $t$ is the only who can compute the outcome of the Parity protocol. If the Parity outcome is 1, party $t$ sets $y_t=1$.
				\end{enumerate}
			\end{enumerate}
			\item The party $s$ is notified ($y_s=1$) with high probability.
			\item\label{TKD:step4}\textbf{Distribution:} The $n$ parties perform the Parity protocol for $|\vec{k}_l|$ times with the following inputs. If party $s$ corresponds to Bob$_l$ (for $l \in\{1,\dots,m\}$), Alice inputs $\vec{k}_l$, Bob$_l$ inputs a random bitstring $\vec{r}$, and the other parties input $\vec{0}$. Otherwise, if $s$ is a non-participant, Alice inputs $\vec{k}_s \oplus\vec{r}$, where $\vec{r}$ is a random string, and the other parties input $\vec{0}$. Let $\vec{o}$ be the output of the Parity protocols.
			\item If $s$ corresponds to Bob$_l$, he computes $\vec{o}\oplus \vec{r}$ and recovers the testing key $\vec{k}_T$. Moreover, Bob$_l$ applies the decoding function $G$ on the last $|\vec{r}_l|$ bits of $\vec{o} \oplus \vec{r}$. If $G$ returns $\top$, then Bob$_l$ sets $v_{B_l}=1$. Otherwise, Bob$_l$ recovers the bit $b_l$ and the string $\vec{r}_\emptyset$ to be used in error correction.\\
			If $s$ is a non-participant, Alice computes $\vec{o}\oplus \vec{r}$ and applies the decoding function $G$ on the last $|\vec{r}_s|$ bits of $\vec{o} \oplus \vec{r}$. If $G$ returns $\top$, then Alice sets $v_A=1$.
		\end{enumerate}
		If Bob$_l$ (for $l \in\{1,\dots,m\}$) has not been notified in none of the $n-1$ Notification rounds, he sets $v_{B_l}=1$.
	\end{protocol}\vspace{0.5cm}
	
	The total number of bipartite channel uses required for the TKD protocol is: $n^2(n-1)^2r_N \,+ \, n(n-1)^2 |\vec{k}_l|$, where $|\vec{k}_l|$ is given in \eqref{testing-key-length}.
	
	The verification bits $v_A$ and $v_{B_l}$ are used to abort the fully-ACKA protocol in step~\ref{f-ACKA:step7} (c.f. Protocol~\ref{prot:sf-ACKA}) if a dishonest party attempts to modify the random bit $b_l$ or the random string $\vec{r}_\emptyset$ destined to Bob$_l$. The fact that Alice and the Bobs verify this fact in the exact same way, prevents non-participants and Eve from learning the identity of the participant who causes fully-ACKA to abort.
	
	Additionally, the bits $v_{B_l}$ are used to abort the fully-ACKA protocol in step~\ref{f-ACKA:step7} if some Bob has not been notified and has not received $\vec{k}_l$. Note that, in order not to reveal the number of Bobs from the outcome of step~\ref{f-ACKA:step7}, we will require the non-participants to behave like a Bob if they have not been notified in TKD.\vspace{1em}
	
	Finally, we illustrate the one-way error correction protocols adopted in ACKA (Protocol~\ref{prot:err-corr-acka}) and fully-ACKA (Protocol~\ref{prot:err-corr}), where Alice provides the Bobs with a syndrome that allows them to correct their faulty raw keys and match Alice's. Together with the syndrome, Alice distributes a hash of her raw key so that each Bob can verify the success of his error correction procedure.\\
	
	\begin{protocol}\label{prot:err-corr-acka}
		\caption{Error Correction for ACKA (\textbf{ACKA-EC})}
		\begin{enumerate}[label=\arabic*., ref=\arabic*]
			\item Alice computes the syndrome $\vec{y}$, with $|\vec{y}|= L(1-p)h(Q_Z)$, from her raw key, i.e. from the string of measurement outcomes corresponding to key generation rounds in $\vec{k}'_T$.
			\item All the parties broadcast a random string of $|\vec{y}|$ bits, while Alice broadcasts $\vec{y}\oplus\vec{k}_2$, where $\vec{k}_2$ is extracted from a previously-established conference key.
			\item From the knowledge of $\vec{k}_2$, each Bob recovers $\vec{y}$ from Alice's broadcast and uses it to correct his raw key.
			\item In order to verify if the error correction is successful, the public randomness picks a two-universal hash function mapping keys of $L(1-p)$ bits to keys of $b_h:= \log_2 \frac{n-1}{\varepsilon_{\mathrm{EC}}}$ bits. Alice and the Bobs compute the hashes $\vec{h}_A$ and $\vec{h}_{B_l}$ (for $l\in\{1,\dots,m\}$) by applying the hash function on their (error-corrected) raw keys.
			\item All the parties broadcast a $b_h$-bit random string, except for Alice who broadcasts $\vec{h}_A\oplus \vec{k}_3$, where $\vec{k}_3$ is extracted from a previously-established conference key.
			\item Each Bob recovers $\vec{h}_A$ and compares it with his hash $\vec{h}_{B_l}$. If $\vec{h}_A\neq \vec{h}_{B_l}$, then the error correction procedure failed for Bob$_{l}$ and he sets $v_{B_l}=1$, otherwise he sets $v_{B_l}=0$.
			\item\label{EC:step7} All the parties broadcast a bit. Alice (Bob$_{l}$, for $l\in\{1,\dots,m\}$) broadcasts $v_A \oplus b_A$ ($v_{B_l} \oplus b_l$), where $b_A$ ($b_l$) is extracted from a previously-established conference key. The predefined ordering of the $n$ parties could be used to assign the bits $b_A$ and $b_l$ of a previous key to Alice and Bob$_l$, respectively.
			\item From the knowledge of the bits $b_A$ and $b_l$, Alice and the Bobs learn if the verification bit $v_A$ or $v_{B_l}$ of any participant is equal to $1$, in which case they consider the protocol aborted and set their conference keys to $\vec{k}_A=\vec{k}_{B_l}=\emptyset$ for $l\in\{1,\dots,m\}$.
		\end{enumerate}
	\end{protocol}\vspace{0.5cm}
	
	\begin{protocol}\label{prot:err-corr}
		\caption{Error Correction for fully-ACKA (\textbf{fully-ACKA-EC})}
		\begin{enumerate}[label=\arabic*., ref=\arabic*]
			\item The public source of randomness picks a two-universal hash function. Alice uses it to compute the syndrome $\vec{y}$, an $L(1-p)h(Q_Z)$-bit long string, from her raw key, i.e. from the string of measurement outcomes corresponding to key generation rounds in $\vec{k}'_T$.
			\item The $n$ parties repeatedly perform the Parity protocol for $L(1-p)h(Q_Z)$ times. Alice uses as an input the bits of $\vec{y}$, while the other parties input $\vec{0}$. The output string of the Parity protocols is $\vec{o}_1$.
			\item Each Bob uses $\vec{o}_1$ as the syndrome to correct his raw key \cite{RennerThesis,ECwithHashFunc}.
			\item In order to verify if the error correction is successful, the public randomness picks a two-universal hash function mapping keys of $L(1-p)$ bits to keys of $b_h:= \log_2 \frac{n-1}{\varepsilon_{\mathrm{EC}}}$ bits. Alice and the Bobs compute the hashes $\vec{h}_A$ and $\vec{h}_{B_l}$ (for $l\in\{1,\dots,m\}$) by applying the hash function on their (error-corrected) raw keys.
			\item The $n$ parties perform the Parity protocol $b_h$ times, using input $\vec{0}$ except for Alice who uses the bits of $\vec{h}_A$. The output string of the Parity protocols is $\vec{o}_2$.
			\item\label{f-EC:step6} Each Bob compares $\vec{o}_2$ with his hash $\vec{h}_{B_l}$. If $\vec{o}_2\neq \vec{h}_{B_l}$ then Bob$_{l}$ sets $v_{B_l}=1$, otherwise he sets $v_{B_l}=0$. Then, all the parties broadcast a random bit, except for Bob$_{l}$ who broadcasts $b_l \oplus v_{B_l}$ (for $l\in\{1,\dots,m\}$).
			\item From the knowledge of the bits $b_l$, Alice retrieves the verification bits $v_{B_l}$ from the broadcast. 
			\item The $n$ parties execute the Parity protocol $|\vec{r}_{\emptyset}|$ times, where every party except for Alice inputs $\vec{0}$. If $v_{B_l}=0$ for every $l$ and if $\vec{o}_2=\vec{h}_A$, Alice inputs $\vec{r}_{\emptyset}\oplus F(0)$, otherwise she inputs $\vec{r}_{\emptyset}\oplus F(1)$. Let $\vec{o}_3$ be the output of the Parity protocols.
			\item\label{f-EC:step9} Alice considers the fully-ACKA protocol aborted if $\vec{o}_3$ differs from her input or if she inputs $\vec{r}_{\emptyset}\oplus F(1)$, while every Bob computes $G(\vec{r}_{\emptyset}\oplus\vec{o}_3)$. If $G(\vec{r}_{\emptyset}\oplus\vec{o}_3)\in\{\top,1\}$, the Bobs consider the fully-ACKA protocol aborted.
		\end{enumerate}
	\end{protocol}\vspace{0.5cm}
	
	Note that if a dishonest party flipped some of the bits of $\vec{o}_1$ or  $\vec{o}_2$ or $\vec{o}_3$ by inputting $1$ in the Parity protocols, then the fully-ACKA protocol aborts with high probability for Alice and all the Bobs.

	\section{PROTOCOLS WITH GHZ STATES} \label{sec:GHZ-protocols}
	
	\noindent Here we provide a more detailed description of Protocol~\ref{prot:sACKA} (ACKA) and Protocol~\ref{prot:sf-ACKA} (fully-ACKA), which anonymously extract a conference key thanks to the multipartite entanglement of GHZ states. The protocols' parameters are summarized in Table~\ref{tab:parameters}. We prove the protocols' security in the Supplemental Material \cite{supplmat}.\\
	
	\newcounter{copyalgorithm}
	\setcounter{copyalgorithm}{\value{algorithm}}
	\setcounter{algorithm}{0}
	
	\begin{protocol}
		\caption{Anonymous Conference Key Agreement  (\textbf{ACKA})}
		\begin{enumerate}[label=\arabic*., ref=\arabic*]
			\item\label{ACKA:step1} The parties perform the ACKA-ID protocol (Protocol~\ref{prot:ID}). If the ACKA-ID protocol does not abort, Alice is guaranteed to be the only sender and the Bobs learn the identities of Alice and of each other, except for a small probability.
			\item Alice and the Bobs recover a shared conference key previously established.
			\item\label{ACKA:step3} Alice generates a random bitstring of length $L$ where 1 corresponds to a test round and 0 to a key generation round. Given that $p$ is the probability that a round is identified as a test round, she compresses the string to a testing key $\vec{k}_T$ of length $Lh(p)$ (typically $h(p)< 8\%$). All the parties broadcast a random string of $Lh(p)$ bits, except for Alice who broadcasts $\vec{k}_T \oplus\vec{k}_1$, where $\vec{k}_1$ is extracted from a previously established conference key. Thanks to the knowledge of $\vec{k}_1$, each Bob recovers the testing key $\vec{k}_T$ from Alice's broadcast.
			\item\label{ACKA:step4} Repeat the following for $L$ rounds.
			\begin{enumerate}
				\item An $n$-party GHZ state is distributed to the $n$ parties.
				\item Alice and the Bobs measure their qubits according to the testing key $\vec{k}_T$. They measure in the $Z$ basis if the round is a key generation round, or in the $X$ basis if the round is a test round. All the other parties measure $X$.
			\end{enumerate}
			\item\label{ACKA:step5} Once all the qubits have been measured (bounded storage assumption), the testing key $\vec{k}_T$ is publicly revealed. This is done anonymously by iterating the Parity protocol (Protocol~\ref{prot:parity}) $Lh(p)$ times. In each instance of the Parity protocol, Alice inputs a bit of $\vec{k}_T$, while the other parties input $0$. The output of the Parity protocols is $\vec{k}'_T$.
			\item\label{ACKA:step6} For every round in step~\ref{ACKA:step3} that is labelled as a test round by $\vec{k}'_T$, the $n$ parties perform the Parity protocol with the following inputs. Let $X_t$ (for $t\in\{1,\dots,n\}$) be the outcome of party $t$ if they measured $X$ in that round, otherwise $X_t \in_R \{0,1\}$. Every party except for Alice ($t \neq A$) inputs $X_t$, while Alice inputs $T_A \in_R \{0,1\}$. Let $\vec{o}_T$ be the output of the Parity protocols for all the test rounds in $\vec{k}'_T$. Alice computes $Q^{\mathrm{obs}}_X=\omega_r (\vec{X}_A \oplus \vec{T}_A \oplus \vec{o}_T)$. 
			\item\label{ACKA:step7} Verification of secrecy: Alice compares $Q^{\mathrm{obs}}_X$ with the predefined value $Q_X$. If $Q_X^{\mathrm{obs}}+ \gamma(Q_X^{\mathrm{obs}}) >Q_X + \gamma(Q_X)$ Alice sets $v_A=1$, otherwise she sets $v_A=0$.
			\item\label{ACKA:step8} ACKA-EC (Protocol \ref{prot:err-corr-acka}):
			Alice broadcasts $ L(1-p)h(Q_Z)$ bits of error correction (EC) information, in order for the Bobs to correct their raw keys and match Alice's. Additionally, she broadcasts a hash of $\log_2 \frac{n-1}{\varepsilon_{\mathrm{EC}}}$ bits so that each Bob can verify the success of the EC procedure. Alice's broadcasts are encrypted and only the Bobs can decrypt them. If the EC or the verification of secrecy failed, the participants abort the protocol, but this information is encrypted and only available to them.
			\item\label{ACKA:step9} PA: The public randomness outputs a two-universal hash function that maps keys of length $L (1-p)$ to keys of length $\ell$, where $\ell$ is given by:
			\begin{equation}
				\ell =  L (1-p) \left[1-h\left(Q_X + \gamma(Q_X) \right)\right] -2\log_2 \frac{1}{2\varepsilon_{\mathrm{PA}}}  \label{keylength-acka}.
			\end{equation}
			Alice and each Bob$_l$ apply the two-universal hash function on their error-corrected keys and obtain the secret conference keys $\vec{k}_A$ and $\vec{k}_{B_l}$. However, if the protocol aborted in the previous step, they do nothing.
		\end{enumerate}
	\end{protocol}\vspace{0.5cm}
	We remark that, in order to fairly compare the performance of Protocol~\ref{prot:sACKA}, for the plots we consider the net number of generated bits, i.e. the number of conference key bits produced by one execution of the protocol, minus the bits consumed from previously-established conference keys. The latter amounts to $|\vec{k}_1| + |\vec{k}_2| + |\vec{k}_3| +n$ bits, which yield a net conference key length given by \eqref{skeylength-acka}.

	\onecolumngrid

	\begin{table}[hb]
		\caption{Parameters used in Protocols~\ref{prot:sACKA} and \ref{prot:sf-ACKA}.}
		\centering
		\begin{tabular}{|c|l|}
			\hline
			$L$ & total number of GHZ states distributed and detected (including noisy states) \\
			\hline
			$p$ & probability of a test round (typically $p \leq 0.01$)  \\
			\hline
			$Q_Z$ &  estimation of the largest error rate between the $Z$ outcomes of Alice \\
			& and of any Bob \\
			\hline
			$Q_X$ & threshold value of the test error rate, picked in the interval $ [0,1/2)$ and \\
			&  defined as the frequency of non-passed test rounds. A test round is passed \\
			& if $\oplus_{t=1}^n X_t=0$, where $X_t\in \{0,1\}$ is party $t$'s outcome in the $X$ basis  \\
			&  mapped to a binary value \\
			\hline
			$v_A$, $v_t$ and $v_{B_l}$ & verification bits of Alice, a generic party $t\in\{1,\dots,n\}$ and Bob$_l$  \\
			& (for $l=1,\dots,m$), respectively  \\
			\hline
			$r_V$ & number of iterations in the Veto protocol (Protocol~\ref{prot:Veto}) \\
			\hline
			$r_N$ & number of iterations in the TKD protocol (Protocol~\ref{prot:TKD}) \\ 
			\hline
			$\varepsilon_{\mathrm{EC}}$ & failure probability of the error correction sub-protocol  \\
			\hline
			$\varepsilon_{\mathrm{enc}}$ & failure probability of the AMD code $(F,G)$ \cite{Broadbent2007,AMD}  \\
			\hline
			$\gamma(Q_X)$ & statistical fluctuation, defined as the positive root of the equation \\
			&  \cite{tail_bound,WstateProtocol}: $\ln \binom{L(1-p)\gamma+L Q_X}{Lp Q_X} + \ln \binom{L(1-Q_X)-L(1-p) \gamma}{Lp(1-Q_X)} = \ln \binom{L }{Lp} + 2\ln \varepsilon_x $ \\
			\hline
			$\varepsilon^2_x$ &  upper bound on the probability that the test error rate affecting the \\
			& key-generation rounds is larger than the observed test error rate ($Q_X^{\mathrm{obs}}$) \\
			& corrected by the statistical fluctuation $\gamma$ \\
			\hline
			$\varepsilon_{\mathrm{PA}}$ & probability related to the success of privacy amplification \\
			\hline
		\end{tabular}
		\label{tab:parameters}
	\end{table}

	\twocolumngrid

	Since one of the major novelties of our ACKA protocol is the efficient parameter estimation (or source verification) in step~\ref{ACKA:step6} of Protocol~\ref{prot:sACKA}, we would like to spend few words on its functioning.
	
	First of all, we emphasize that the parameter $Q^{\rm obs}_X$ estimates the phase error rate of the state distributed by the untrusted source. If in a test round a GHZ state is distributed, the parity of the outcomes obtained by measuring every qubit in the $X$ basis is zero: $\oplus_{t=1}^n X_t=0$. Thus these instances do not contribute to the error rate $Q^{\rm obs}_X$. As a matter of fact, one can simplify the error rate expression provided in step~\ref{ACKA:step6} by noting that: $\vec{o}_T=\vec{T}_A\oplus (\oplus_{t \neq A} \vec{X}_t)$, which substituted in the error rate yields: $Q^{\rm obs}_X=\omega_r (\oplus_{t=1}^n \vec{X}_t)$.
	
	The reason for which we require Alice to input a random bit $T_A$ in place of her outcome $X_A$ in the Parity protocol is that, in this way, we prevent any dishonest party from artificially decreasing the error rate ($Q^{\rm obs}_X=\omega_r (\oplus_{t=1}^n \vec{X}_t)$) based on the $X_t$ outcomes of all the other parties. Indeed, if Alice would input $X_A$, a dishonest party who is the last to broadcast in the Parity protocol can arbitrarily set the output of Parity --and thus $\oplus_{t=1}^n X_t$-- to zero.
	
	Finally, we employ the Parity protocol to compute the parity  $\vec{o}_T=\vec{T}_A\oplus (\oplus_{t \neq A} \vec{X}_t)$ instead of using a regular broadcast in order to preserve the participants' anonymity.
	
	To see why their identities would be under threat, let us suppose that we replace the Parity protocol with a regular broadcast. Then, based on the state that Eve --who controls the source-- distributed in a given test round, she could make predictions on the $X$ outcomes of the parties and compare them with the broadcast bits. Since Alice always broadcasts a random bit instead of her $X$ outcome as the other parties do, Eve could distinguish her broadcast and learn her identity. Similarly, if a Bob did not measure in the $X$ basis for some test round in $\vec{k}'_T$ (due to a mismatch between $\vec{k}'_T$ and his testing key), he must broadcast a random bit and Eve could learn his identity.\\

	\begin{protocol}
		\caption{Fully Anonymous Conference Key Agreement (\textbf{fully-ACKA})}
		\begin{enumerate}[label=\arabic*., ref=\arabic*]
			\item\label{f-ACKA:step1} The parties perform the fully-ACKA-ID protocol (Protocol~\ref{prot:f-ID}). If the fully-ACKA-ID protocol does not abort, Alice is guaranteed to be the only sender and the Bobs are notified to be receivers, except for a small probability.
			\item\label{f-ACKA:step2} Alice generates a random bitstring of length $L$ where 1 corresponds to a test round and 0 to a key generation round. Given that $p$ is the probability that a round is identified as a test round, she compresses the string to a testing key $\vec{k}_T$ of length $L h(p)$. Additionally, Alice generates the strings $\vec{r}_l=F(b_l,\vec{r}_\emptyset)$ (for $l\in\{1,\dots,m\}$) where $(F,G)$ is an AMD code and $(b_l,\vec{r}_\emptyset)$ is a concatenated random string.
			\item\label{f-ACKA:step3} The parties perform the TKD protocol (Protocol~\ref{prot:TKD}) in order to distribute the key $\vec{k}_l=(\vec{k}_T,\vec{r}_l)$, which includes the testing key $\vec{k}_T$, to every Bob.
			\item\label{f-ACKA:step4} Repeat the following for $L$ rounds.
			\begin{enumerate}
				\item An $n$-party GHZ state is distributed to the $n$ parties.
				\item Alice and the Bobs measure their qubits according to the testing key $\vec{k}_T$. They measure in the $Z$ basis if the round is a key generation round, or in the $X$ basis if the round is a test round. All the other parties measure $X$.
			\end{enumerate}
			\item\label{f-ACKA:step5} Once all the qubits have been measured (bounded storage assumption), the testing key $\vec{k}_T$ is publicly revealed. This is done anonymously by iterating the Parity protocol (Protocol~\ref{prot:parity}) $Lh(p)$ times. In each instance of the Parity protocol, Alice inputs a bit of $\vec{k}_T$, while the other parties input $0$. The output of the Parity protocols is $\vec{k}'_T$.
			\item\label{f-ACKA:step6} For every round in step~\ref{f-ACKA:step4} that is labelled as a test round by $\vec{k}'_T$, the $n$ parties perform the Parity protocol with the following inputs. Let $X_t$ (for $t\in\{1,\dots,n\}$) be the outcome of party $t$ if they measured $X$ in that round, otherwise $X_t \in_R \{0,1\}$. Every party except for Alice ($t \neq A$) inputs $X_t$, while Alice inputs $T_A \in_R \{0,1\}$. Let $\vec{o}_T$ be the output of the Parity protocols for all the test rounds in $\vec{k}'_T$. Alice computes $Q^{\mathrm{obs}}_X=\omega_r (\vec{X}_A \oplus \vec{T}_A \oplus \vec{o}_T)$.
			
			\item\label{f-ACKA:step7} Verification of secrecy: Alice compares $Q^{\mathrm{obs}}_X$ with the predefined value $Q_X$ and sets $v_s=1$ if $Q_X^{\mathrm{obs}}+ \gamma(Q_X^{\mathrm{obs}}) >Q_X + \gamma(Q_X)$ and $v_s=0$ otherwise. All the parties perform Veto (Protocol~\ref{prot:Veto}), where Alice inputs $v_s \vee v_A$, Bob$_l$ inputs $v_{B_l}$ and the non-participants input 1 if they have not been notified in TKD, otherwise they input $0$. If Veto outputs $1$, the protocol aborts for every party.
			\item\label{f-ACKA:step8} fully-ACKA-EC (Protocol \ref{prot:err-corr}): Alice anonymously broadcasts $L(1-p)h(Q_Z)$ bits of EC information, in order for the Bobs to correct their raw keys and match Alice's. Additionally, she anonymously broadcasts a hash of $\log_2 \frac{n-1}{\varepsilon_{\mathrm{EC}}}$ bits so that each Bob can verify the success of the EC procedure. If the EC fails for at least one Bob, the protocol aborts but this information is only available to Alice and the Bobs.
			\item\label{f-ACKA:step9} PA: The public randomness outputs a two-universal hash function that maps keys of length $L (1-p)$ to keys of length $\ell$, where $\ell$ is given by:
			\begin{align}
				\ell =  &L (1-p) \left[1-h\left(Q_X + \gamma(Q_X) \right) -h(Q_Z)\right] \nonumber\\
				&-\log_2 \frac{2(n-1)}{\varepsilon_{\mathrm{EC}}}-2\log_2 \frac{1}{2\varepsilon_{\mathrm{PA}}}  \label{keylength-f-acka}.
			\end{align}
			Alice and each Bob$_l$ apply the two-universal hash function on their error-corrected keys and obtain the secret conference keys $\vec{k}_A$ and $\vec{k}_{B_l}$. However, if the protocol aborted in the previous step, they do nothing.
		\end{enumerate}
	\end{protocol}
	
	\setcounter{algorithm}{\value{copyalgorithm}}
	
	\section{PROTOCOLS WITHOUT MULTIPARTITE ENTANGLEMENT} \label{sec:Bell-protocols}
	
	\noindent In order to evaluate the benefits of using GHZ states to perform ACKA and fully-ACKA, we develop alternative protocols which only rely on bipartite private channels, implemented with Bell pairs. For this reason, we denote these protocols as bACKA and bifully-ACKA, respectively. We remark that bACKA and bifully-ACKA are not mere extensions of the Anonymous Message Transmission protocol \cite{Broadbent2007} to multiple parties, as they are thoroughly optimized to achieve the tasks under consideration.\\
	
	\begin{protocol}\label{prot:bACKA}
		\caption{ACKA without multiparty entanglement (\textbf{bACKA})}
		\begin{enumerate}[label=\arabic*., ref=\arabic*]
			\item The parties perform the ACKA-ID protocol (Protocol~\ref{prot:ID}). If the ACKA-ID protocol does not abort, Alice is guaranteed to be the only sender and the Bobs learn the identities of Alice and of each other, except for a small probability.
			\item Alice generates uniformly at random an $L_b$-bit conference key, $\vec{k}_A$.
			\item\label{bACKA:step3} Every party --except for Alice-- sends a random string of $L_b$ bits to every other party through the bipartite private channels. Alice sends the conference key $\vec{k}_A$ to the Bobs and a random string to the other parties. Bob$_l$ identifies the string $\vec{k}_A$ received from Alice as his conference key: $\vec{k}_{B_l}=\vec{k}_A$, for $l\in\{1,\dots,m\}$.
		\end{enumerate}
	\end{protocol}\vspace{0.5cm}

	\begin{protocol}\label{prot:bif-ACKA}
		\caption{fully-ACKA without multiparty entanglement (\textbf{bifully-ACKA})}
		\begin{enumerate}[label=\arabic*., ref=\arabic*]
			\item\label{bf-ACKA:step1} The parties perform the fully-ACKA-ID protocol (Protocol~\ref{prot:f-ID}). If the fully-ACKA-ID protocol does not abort, Alice is guaranteed to be the only sender and the Bobs are notified to be receivers, except for a small probability.
			\item Alice generates uniformly at random an $L_b$-bit conference key, $\vec{k}_A$, and encodes it as $F(\vec{k}_A)$ with an AMD code $(F,G)$, such that $|F(\vec{k}_A)|=L_b+2(\log_2 L_b + \log_2 1/\varepsilon_{\mathrm{enc}})$. She also sets her verification bit to $v_A=0$.
			\item\label{bf-ACKA:step3} Repeat for $n-1$ times.
			\begin{enumerate}
				\item Alice randomly picks a party $s\in \{1,\dots,n\}$ that has not been notified in previous iterations. 
				\item\label{bf-ACKA:step3.2} The $n$ parties perform the Notification protocol (step~2 in Protocol~\ref{prot:TKD}), such that party $s$ is notified with high probability.
				\item\label{bf-ACKA:step3.3} The $n$ parties execute $|F(\vec{k}_A)|$ rounds of the Parity protocol with the following inputs. If the notified party $s$ corresponds to Bob$_l$ (for $l \in\{1,\dots,m\}$), Alice inputs $F(\vec{k}_A)$, Bob$_l$ inputs a random bitstring $\vec{r}$, and the other parties input $\vec{0}$. Otherwise, if $s$ is a non-participant, Alice inputs $F(\vec{k}_A) \oplus\vec{r}$, where $\vec{r}$ is a random string, and the other parties input $\vec{0}$. Let $\vec{o}$ be the output of the Parity protocols.
				\item If $s$ corresponds to Bob$_l$, he retrieves the conference key by computing $\vec{k}_{B_l}= G(\vec{r}\oplus\vec{o})$. If $\vec{k}_{B_l}=\top$, then he sets $v_{B_l}=1$, otherwise $v_{B_l}=0$. If $s$ is a non-participant, Alice computes $G(\vec{o}\oplus \vec{r})$. If the computation returns $\top$, then Alice sets $v_A=1$.
			\end{enumerate}
			\item\label{bf-ACKA:step4} Bob$_l$ (for $l\in\{1,\dots,m\}$) sets $v_{B_l}=1$ if he has not been notified in step~\ref{bf-ACKA:step3}. The $n$ parties perform Veto (Protocol~\ref{prot:Veto}) where Bob$_l$ inputs $v_{B_l}$ (for $l\in\{1,\dots,m\}$), Alice inputs $v_A$ and the other parties input $1$ if they have not been notified in step~\ref{bf-ACKA:step3}, otherwise they input $0$. If Veto outputs $1$, the protocol aborts for every party and the participants set $\vec{k}_A=\vec{k}_{B_l}=\emptyset \,\, \forall\, l$.
		\end{enumerate}
	\end{protocol}\vspace{0.5cm}
	
	We emphasize that the bifully-ACKA protocol (Protocol~\ref{prot:bif-ACKA}) satisfies the anonymity condition for fully-ACKA protocols \eqref{anonymity-f-acka} given in Definition~\ref{def:suff-cond-security}, opposed to the fully-ACKA protocol based on GHZ states (Protocol~\ref{prot:sf-ACKA}) which satisfies a weaker anonymity property (Definition~\ref{def:weak-anonym}). More specifically, the bifully-ACKA protocol is $\varepsilon_{\rm AN}$-anonymous with respect to (dishonest) receivers in the sense of conditions \eqref{state:f-anony-dishonest} and \eqref{state:f-anony-dishonest-multiple-senders}, i.e. the reduced state of the receivers is close to a state independent of the identity of the other participants. Conversely, Protocol~\ref{prot:sf-ACKA} satisfies a weaker anonymity condition with respect to the receivers, formalized in terms of their guessing probability by conditions \eqref{weak-pguess1} and \eqref{weak-pguess2}.
	
	Moreover, note that the security of bACKA and bifully-ACKA can be proved without the assumption about the $n$ parties  holding a short-lived quantum memory (bounded storage assumption), while this assumption is crucial for the security of the ACKA and fully-ACKA protocols (Protocols~1 and 2), as discussed in Sec.~\ref{sec:discussion}.
	
	\begin{Thm}[Security of bACKA and bifully-ACKA] \label{th:backa-security-proof}
		The bACKA protocol (Protocol~\ref{prot:bACKA}), exclusively based on bipartite private channels,  yields a secret conference key of length $L_b$ and is $\varepsilon_{\mathrm{tot}}$-secure according to Definition~\ref{def:suff-cond-security}, with $\varepsilon_{\mathrm{tot}}=2^{-r_V}+(n-1)\,\varepsilon_{\rm enc}$.
		
		The bifully-ACKA protocol (Protocol~\ref{prot:bif-ACKA}), exclusively based on bipartite private channels, yields a secret conference key of length $L_b$ and is $\varepsilon_{\mathrm{tot}}$-secure according to Definition~\ref{def:suff-cond-security}, with $\varepsilon_{\mathrm{tot}}= 3\cdot2^{-r_V}+ (n-1)(2^{-(r_N-1)}+3\varepsilon_{\rm enc})$.
	\end{Thm}

	%
	%
	%
	\nocite{TL17}
	\nocite{QLHL}
	\nocite{entropicUncert}
	\nocite{tight-finitekey-QKD}
	\nocite{Tomamichel-book}

	\bibliography{bibliography}

\ifarXiv


\section*{Supplementary material}

## II. SUFFICIENT CONDITIONS FOR CKA-SECURITY

In standard QKD and CKA, the security of the established key is often deduced by proving that the protocol is both secret and correct. Here, we introduce analogous definitions of *correctness* and *secrecy* for a generic ACKA (fully-ACKA) protocol with $m$ receivers contained in $\vec{j} = (j_1, j_2, \ldots, j_m)$. We then prove that correctness and secrecy are sufficient to imply CKA-security.

**Definition 1** (Correctness). An ACKA (fully-ACKA) protocol that outputs secret keys $\vec{k}_i$ for sender $i$ and $\vec{k}_{j_l}$ for receiver $j_l$ ($l = 1, \ldots, m$) is said to be $\varepsilon_{\mathrm{cor}}$-correct if

$$\max_{\{i,\vec{j}\} \cap \mathcal{D} = \emptyset} \Pr[\Omega_{\mathcal{P}} | i, \vec{j}, \Phi] \Pr[\cup_{l=1}^m \vec{k}_i \neq \vec{k}_{j_l} | i, \vec{j}, \Phi, \Omega_{\mathcal{P}}] + \Pr[\Omega_{\mathcal{P}}^c \cap \Gamma_{\mathcal{P}}^c | i, \vec{j}, \Phi] \leq \varepsilon_{\mathrm{cor}}. \tag{S1}$$

**Definition 2** (Secrecy). An ACKA (fully-ACKA) protocol is said to be $\varepsilon_{\mathrm{sec}}$-secret if the following inequality holds:

$$\max_{\{i,\vec{j}\} \cap \mathcal{D} = \emptyset} \Pr[\Omega_{\mathcal{P}} | i, \vec{j}, \Phi] \left\| \rho^f_{K_i(KC)^c_{i,\vec{j}}E | i, \vec{j}, \Phi, \Omega_{\mathcal{P}}} - \tau_{K_i} \otimes \rho^f_{(KC)^c_{i,\vec{j}}E | i, \vec{j}, \Phi, \Omega_{\mathcal{P}}} \right\|_{\mathrm{tr}} \leq \varepsilon_{\mathrm{sec}}, \tag{S2}$$

where $\tau_{K_i} = \frac{1}{|\mathcal{K}|} \sum_{\vec{k}_i \in \mathcal{K}} |\vec{k}_i\rangle\langle\vec{k}_i|_{K_i}$ is the maximally mixed state over all the possible realizations of the sender's key ($\mathcal{K} = \{0,1\}^\ell$).

**Lemma 1** (CKA-security). An ACKA (fully-ACKA) protocol that is both $\varepsilon_{\mathrm{cor}}$-correct and $\varepsilon_{\mathrm{sec}}$-secret is also $\varepsilon_{\mathrm{CKA}}$-CKA-secure according to Eq. (B12), with $\varepsilon_{\mathrm{CKA}} = \varepsilon_{\mathrm{cor}} + \varepsilon_{\mathrm{sec}}$.

*Proof.* The proof follows the corresponding one valid for QKD protocols [35]. We start by noting that the final state of the protocol, conditioned on the event $\Phi \cap \Omega_{\mathcal{P}}$, is classical on subsystems $K_{i,\vec{j}}$ and can be expressed as follows:

$$\rho^f_{KC^c_{i,\vec{j}}E | i, \vec{j}, \Phi, \Omega_{\mathcal{P}}} = \sum_{\vec{k}_i, \vec{k}_{j_l} \in \mathcal{K}} \Pr[\vec{k}_i, \vec{k}_{j_1}, \ldots, \vec{k}_{j_m} | i, \vec{j}, \Phi, \Omega_{\mathcal{P}}] |\vec{k}_i\rangle\langle\vec{k}_i|_{K_i} \otimes_{l=1}^m |\vec{k}_{j_l}\rangle\langle\vec{k}_{j_l}|_{K_{j_l}} \otimes \rho^f_{(KC)^c_{i,\vec{j}}E | i, \vec{j}, \Phi, \Omega_{\mathcal{P}}}, \tag{S3}$$

for some distribution $\Pr[\vec{k}_i, \vec{k}_{j_1}, \ldots, \vec{k}_{j_m} | i, \vec{j}, \Phi, \Omega_{\mathcal{P}}]$ of keys. Let us now define the state:

$$\gamma_{KC^c_{i,\vec{j}}E} = \sum_{\vec{k}_i, \vec{k}_{j_l} \in \mathcal{K}} \Pr[\vec{k}_i, \vec{k}_{j_1}, \ldots, \vec{k}_{j_m} | i, \vec{j}, \Phi, \Omega_{\mathcal{P}}] |\vec{\mathbf{k}_i}\rangle\langle\vec{\mathbf{k}_i}|_{K_{i,\vec{j}}} \otimes \rho^f_{(KC)^c_{i,\vec{j}}E | i, \vec{j}, \Phi, \Omega_{\mathcal{P}}}, \tag{S4}$$

which is obtained from (S3) by just replacing the keys of the receivers with the sender's key. Note that the reduced states of (S3) and (S4) are identical and equal to one of the states that appears in (S2):

$$\gamma_{K_i(KC)^c_{i,\vec{j}}E} = \mathrm{Tr}_{K_{\vec{j}}}[\gamma_{KC^c_{i,\vec{j}}E}] = \mathrm{Tr}_{K_{\vec{j}}}\left[\rho^f_{KC^c_{i,\vec{j}}E | i, \vec{j}, \Phi, \Omega_{\mathcal{P}}}\right] = \rho^f_{K_i(KC)^c_{i,\vec{j}}E | i, \vec{j}, \Phi, \Omega_{\mathcal{P}}}. \tag{S5}$$

Let us now use the triangle inequality to upper bound the first term in the CKA-security condition Eq. (B12) as follows:

$$\begin{aligned}
\Pr[\Omega_{\mathcal{P}} | i, \vec{j}, \Phi] \left\| \rho^f_{KC^c_{i,\vec{j}}E | i, \vec{j}, \Phi, \Omega_{\mathcal{P}}} - \tau_{K_{i,\vec{j}}} \otimes \rho^f_{(KC)^c_{i,\vec{j}}E | i, \vec{j}, \Phi, \Omega_{\mathcal{P}}} \right\|_{\mathrm{tr}} &\leq \Pr[\Omega_{\mathcal{P}} | i, \vec{j}, \Phi] \left\| \rho^f_{KC^c_{i,\vec{j}}E | i, \vec{j}, \Phi, \Omega_{\mathcal{P}}} - \gamma_{KC^c_{i,\vec{j}}E} \right\|_{\mathrm{tr}} \\
&\quad + \Pr[\Omega_{\mathcal{P}} | i, \vec{j}, \Phi] \left\| \gamma_{KC^c_{i,\vec{j}}E} - \tau_{K_{i,\vec{j}}} \otimes \rho^f_{(KC)^c_{i,\vec{j}}E | i, \vec{j}, \Phi, \Omega_{\mathcal{P}}} \right\|_{\mathrm{tr}},
\end{aligned} \tag{S6}$$

and let us upper bound the two terms on the rhs individually.

For the second term, we notice that the subsystems $K_{\vec{j}}$ are perfect copies of the subsystem $K_i$ in both states (S4) and Eq. (B6). Therefore, the trace distance of the states is not increased when the subsystems $K_{\vec{j}}$ are accounted for [35]. In other words, it holds that:

$$\left\| \gamma_{KC^c_{i,\vec{j}}E} - \tau_{K_{i,\vec{j}}} \otimes \rho^f_{(KC)^c_{i,\vec{j}}E | i, \vec{j}, \Phi, \Omega_{\mathcal{P}}} \right\|_{\mathrm{tr}} = \left\| \gamma_{K_i(KC)^c_{i,\vec{j}}E} - \tau_{K_i} \otimes \rho^f_{(KC)^c_{i,\vec{j}}E | i, \vec{j}, \Phi, \Omega_{\mathcal{P}}} \right\|_{\mathrm{tr}}. \tag{S7}$$

By combining the last expression with (S5) and by using the secrecy condition (S2), we can upper bound the second term in (S6) as follows:

$$\Pr[\Omega_{\mathcal{P}}|i,\vec{j},\Phi]\,\left\|\gamma_{KC^c_{i,j}E}-\tau_{K_{i,j}}\otimes\rho^f_{(KC)^c_{i,j}E|i,\vec{j},\Phi,\Omega_{\mathcal{P}}}\right\|_{\mathrm{tr}}\leq\varepsilon_{\mathrm{sec}},\tag{S8}$$

which is valid for every choice of sender $i$ and receivers $\vec{j}$.

The first term in (S6) can be bounded by using the convexity of the trace distance and the fact that $\|A\otimes B_1 - A\otimes B_2\|=\|B_1-B_2\|_{\mathrm{tr}}$. We obtain:

$$\begin{aligned}
\left\|\rho^f_{KC^c_{i,j}E|i,\vec{j},\Phi,\Omega_{\mathcal{P}}}-\gamma_{KC^c_{i,j}E}\right\|_{\mathrm{tr}} &\leq \sum_{\vec{k}_i,\vec{k}_{j_l}\in\mathcal{K}}\Pr[\vec{k}_i,\vec{k}_{j_1},\ldots,\vec{k}_{j_m}|i,\vec{j},\Phi,\Omega_{\mathcal{P}}]\,\left\|\otimes_{l=1}^m|\vec{k}_{j_l}\rangle\langle\vec{k}_{j_l}|_{K_{j_l}}-|\vec{\boldsymbol{k}_i}\rangle\langle\vec{\boldsymbol{k}_i}|_{K_{\vec{j}}}\right\|_{\mathrm{tr}} \\
&= \sum_{\vec{k}_i,\vec{k}_{j_l}\in\mathcal{K}}\Pr[\vec{k}_i,\vec{k}_{j_1},\ldots,\vec{k}_{j_m}|i,\vec{j},\Phi,\Omega_{\mathcal{P}}](1-\delta_{\vec{k}_{j_1},\vec{k}_i}\delta_{\vec{k}_{j_2},\vec{k}_i}\cdots\delta_{\vec{k}_{j_m},\vec{k}_i}) \\
&= \Pr[\cup_{l=1}^m\vec{k}_i\neq\vec{k}_{j_l}|i,\vec{j},\Phi,\Omega_{\mathcal{P}}],
\end{aligned}\tag{S9}$$

which is again valid for every $i$ and $\vec{j}$. By employing (S8) and (S9) in (S6), we can upper bound the lhs of the CKA-security condition Eq. (B12) as follows:

$$\begin{aligned}
&\max_{\{i,\vec{j}\}\cap\mathcal{D}=\emptyset}\Pr[\Omega_{\mathcal{P}}|i,\vec{j},\Phi]\,\left\|\rho^f_{KC^c_{i,j}E|i,\vec{j},\Phi,\Omega_{\mathcal{P}}}-\tau_{K_{i,j}}\otimes\rho^f_{(KC)^c_{i,j}E|i,\vec{j},\Phi,\Omega_{\mathcal{P}}}\right\|_{\mathrm{tr}}+\Pr[\Omega^c_{\mathcal{P}}\cap\Gamma^c_{\mathcal{P}}|i,\vec{j},\Phi] \\
&\leq\max_{\{i,\vec{j}\}\cap\mathcal{D}=\emptyset}\Pr[\Omega_{\mathcal{P}}|i,\vec{j},\Phi]\,\Pr[\cup_{l=1}^m\vec{k}_i\neq\vec{k}_{j_l}|i,\vec{j},\Phi,\Omega_{\mathcal{P}}]+\Pr[\Omega^c_{\mathcal{P}}\cap\Gamma^c_{\mathcal{P}}|i,\vec{j},\Phi]+\varepsilon_{\mathrm{sec}} \\
&\leq\varepsilon_{\mathrm{cor}}+\varepsilon_{\mathrm{sec}},
\end{aligned}\tag{S10}$$

where we used (S1) in the last inequality. This concludes the proof. $\qquad\square$

## III. SECURITY PROOF OF ACKA

In this section we prove the security of the ACKA protocol (Protocol 1).

### A. Integrity

Here we show that the integrity condition Eq. (B11) is satisfied by Protocol 1. Since the assignment and notification of the identities is done in step 1 through the ACKA-ID protocol (Protocol 6), the following lemma is actually a proof of the integrity of the ACKA-ID protocol.

**Lemma 2.** The ACKA protocol (Protocol 1) is $\varepsilon_{\mathrm{IN}}$-integrous with $\varepsilon_{\mathrm{IN}}=2^{-r_V}+(n-1)\,\varepsilon_{\mathrm{enc}}$.

*Proof.* In order to prove the integrity of the ACKA-ID protocol we employ in ACKA, we need to bound the probabilities $\Pr[\Gamma^c\cap\Phi^c|i,\vec{j}]$ and $\Pr[\Gamma^c|\vec{i},\vec{j}]$. The former is the probability that the ACKA-ID protocol does not assign correctly the identities or aborts for a strict subset of parties, when only one sender applied. The latter is the probability that the protocol does not abort for every party when multiple senders apply.

We first analyze what happens when a single party $i$ applies to become the sender. In the case of our ACKA-ID protocol (Protocol 6), the Collision Detection protocol either aborts for every party or designates party $i$ as the sender. Therefore, a wrong identity assignment can only happen in steps 3 and 4 of the ACKA-ID protocol. In particular, it happens if for any party $t\neq A$ (except for Alice) the decoding function $G$ returns a valid output different from the one sent by Alice, or if it returns $\top$ but the Veto protocol in step 4 fails and returns 0. Let us define the events:

- $e_t$: The decoding function $G$ returns a wrong output (i.e. $\neq\vec{d_t},\top$) for party $t\neq A$.

- $e_V$: The Veto protocol in step 4 fails to return 1 when at least one party inputs 1.

By using the union bound, we obtain:

$$\Pr[\Gamma^c \cap \Phi^c | i, \vec{j}] = \Pr[e_V \cup_{t \neq A} e_t | i, \vec{j}] \leq \Pr[e_V] + \sum_{t \neq A} \Pr[e_t], \tag{S11}$$

and from the properties of the AMD code we employed, we have that $\Pr[e_t] \leq \varepsilon_{\text{enc}}$. The Veto in step 4 of Protocol 6 can only fail if, in every round of the Veto protocol in which the party who inputs 1 is the last to broadcast, the input of the other parties is such that each Parity protocol outputs zero. This happens with probability $2^{-r_V}$, where $r_V$ is a parameter of the Veto protocol. Thus we have that: $\Pr[e_V] \leq 2^{-r_V}$. Then, we bound (S11) as follows:

$$\Pr[\Gamma^c \cap \Phi^c | i, \vec{j}] \leq 2^{-r_V} + (n-1)\,\varepsilon_{\text{enc}} \quad \forall\, i, \vec{j}. \tag{S12}$$

When more than one party (or no party) applies to become the sender, the ACKA-ID protocol may abort in step 2 or in step 4. Only in the case where the ACKA-ID protocol does not abort in neither of these steps, we get an integrity issue. However, for simplicity, we bound $\Pr[\Gamma^c | \vec{i}, \vec{j}]$ by the probability that the protocol does not abort in step 2 only. This happens when Veto B in the Collision Detection protocol outputs 0 when at least one party inputs 1, which occurs with probability $2^{-r_V}$. Thus we obtain:

$$\Pr[\Gamma^c | \vec{i}, \vec{j}] \leq 2^{-r_V} \quad \forall\, \vec{i}, \vec{j}. \tag{S13}$$

By combining (S12) and (S13), we can upper bound the lhs of the integrity condition Eq. (B11) and show the claim:

$$\max_{i, \vec{j}; \vec{i}, \vec{j}} \left\{ \Pr[\Gamma^c \cap \Phi^c | i, \vec{j}], \Pr[\Gamma^c | \vec{i}, \vec{j}] \right\} \leq 2^{-r_V} + (n-1)\,\varepsilon_{\text{enc}}. \tag{S14}$$

This concludes the proof. $\qquad\qquad\square$

## B. CKA-security

In order to prove the CKA-security Eq. (B12) of Protocol 1, we make use of Lemma 1 that allows one to prove CKA-security by independently showing that the ACKA protocol is both correct and secret according to Definitions 1 and 2.

**Lemma 3.** The ACKA protocol (Protocol 1) is $\varepsilon_{\text{cor}}$-correct, with $\varepsilon_{\text{cor}} = \varepsilon_{\text{EC}}$.

*Proof.* The proof integrates the analogous one valid for standard CKA protocols [19], which guarantees correctness except for probability $\varepsilon_{\text{EC}}$. However, the correctness definition of an ACKA protocol (Definition 1) imposes us to also account for the events where the protocol aborts only for a strict subset of participants, given that the ACKA-ID protocol (Protocol 6) was successful ($\Pr[\Omega_{\mathcal{P}}^c \cap \Gamma_{\mathcal{P}}^c | i, \vec{j}, \Phi]$).

We first bound the probability $\Pr[\cup_{l=1}^m \vec{k}_A \neq \vec{k}_{B_l} \cap \Omega_{\mathcal{P}} | A, \vec{B}, \Phi]$, where we relabelled the sender $i$ and the receivers $\vec{j}$ as Alice ($i = A$) and Bobs ($\vec{j} = \vec{B}$). Recall from error correction that when the hash of some Bob differs from Alice's hash, $\vec{h}_A \neq \vec{h}_{B_l}$, the ACKA protocol aborts. Thus, when the protocol does not abort (event $\Omega_{\mathcal{P}}$), it holds that $\vec{h}_A = \vec{h}_{B_l}$ for $l \in \{1, \ldots, m\}$. Thus we can recast the probability as:

$$\Pr[\cup_{l=1}^m \vec{k}_A \neq \vec{k}_{B_l} \cap \Omega_{\mathcal{P}} | A, \vec{B}, \Phi] = \Pr[\cup_{l=1}^m (\vec{k}_A \neq \vec{k}_{B_l} \cap \vec{h}_A = \vec{h}_{B_l}) \cap \Omega_{\mathcal{P}} | A, \vec{B}, \Phi]$$
$$\leq \Pr[\cup_{l=1}^m (\vec{k}_A \neq \vec{k}_{B_l} \cap \vec{h}_A = \vec{h}_{B_l}) | A, \vec{B}, \Phi]. \tag{S15}$$

By the properties of two-universal hash functions (see the correctness proof in [19]), we can bound the rhs of the last expression and obtain:

$$\Pr[\Omega_{\mathcal{P}} | A, \vec{B}, \Phi]\Pr[\cup_{l=1}^m \vec{k}_A \neq \vec{k}_{B_l} | A, \vec{B}, \Phi, \Omega_{\mathcal{P}}] = \Pr[\cup_{l=1}^m \vec{k}_A \neq \vec{k}_{B_l} \cap \Omega_{\mathcal{P}} | A, \vec{B}, \Phi] \leq \varepsilon_{\text{EC}}. \tag{S16}$$

Now we observe that the only way in which the ACKA protocol can abort for some Bob –but not for the other participants– is when Bob uses a different set of random bits $b_A, b_l$ to encode and decode the broadcast in step 7 of EC. This, however, can only happen if Bob thinks he is performing the protocol with a wrong set of participants, which happens if ACKA-ID fails. Since the probability $\Pr[\Omega_{\mathcal{P}}^c \cap \Gamma_{\mathcal{P}}^c | A, \vec{B}, \Phi]$ is conditioned on the success of ACKA-ID (event $\Phi$), the event just described never occurs in ACKA. Thus we have $\Pr[\Omega_{\mathcal{P}}^c \cap \Gamma_{\mathcal{P}}^c | A, \vec{B}, \Phi] = 0$.

By repeating the above arguments for any set of sender and receivers, we can upper bound the lhs of (S1) as follows:

$$\max_{\{i, \vec{j}\} \cap \mathcal{D} = \emptyset} \Pr[\Omega_{\mathcal{P}} | i, \vec{j}, \Phi]\Pr[\cup_{l=1}^m \vec{k}_i \neq \vec{k}_{j_l} | i, \vec{j}, \Phi, \Omega_{\mathcal{P}}] + \Pr[\Omega_{\mathcal{P}}^c \cap \Gamma_{\mathcal{P}}^c | i, \vec{j}, \Phi] \leq \varepsilon_{\text{EC}}, \tag{S17}$$

which concludes the proof. $\qquad\qquad\square$

We now prove that Protocol 1 is secret according to Definition 2. In doing so we must bound the test error rate[1] affecting the key-generation rounds, $Q_X^{\text{KG}}$, with the corresponding observed error rate, $Q_X^{\text{obs}}$.

Denote $\vec{X}_t^{\text{obs}}$ the random variable representing the string of $X$ outcomes of party $t$, collected in the test rounds. Denote $\vec{X}_t^{\text{KG}}$ the string of $X$ outcomes of party $t$ had that party measured in the $X$ basis for every key-generation round. Note that if some party did not measure $X$ in a test round or in a key-generation round (e.g. due to a faulty testing key or if the party is dishonest), we still indicate as $X_t$ the bit that the party would input in the Parity protocol of step 6 in Protocol 1 –except for Alice, who always inputs a random bit $T_A$ but computes the parameter $Q_X^{\text{obs}}$ with her outcomes $X_A$. Then we have:

$$Q_X^{\text{obs}} = \omega_r \left( \bigoplus_{t=1}^{n} \vec{X}_t^{\text{obs}} \right) \tag{S18}$$

$$Q_X^{\text{KG}} = \omega_r \left( \bigoplus_{t=1}^{n} \vec{X}_t^{\text{KG}} \right) \equiv \omega_r \left( \vec{X}_A^{\text{KG}} \oplus \vec{X}_{\oplus A^c}^{\text{KG}} \right), \tag{S19}$$

where we defined $\vec{X}_{\oplus A^c}^{\text{KG}}$ as the string obtained by adding modulo two all the strings of $X$ outcomes in key generation rounds except for Alice's:

$$\vec{X}_{\oplus A^c}^{\text{KG}} := \bigoplus_{t \neq A} \vec{X}_t^{\text{KG}}. \tag{S20}$$

Since the test rounds are drawn randomly, we can estimate the value of $Q_X^{\text{KG}}$ through $Q_X^{\text{obs}}$ with the theory of random sampling without replacement. In particular, we employ a recent tail inequality to bound the probability that $Q_X^{\text{KG}}$ exceeds $Q_X^{\text{obs}}$ corrected with an appropriate statistical fluctuation:

$$\Pr \left[ Q_X^{\text{KG}} > Q_X^{\text{obs}} + \gamma(Q_X^{\text{obs}}) \right] \leq \varepsilon_x^2 \tag{S21}$$

where $\gamma(Q_X^{\text{obs}})$ is the statistical fluctuation, defined as the positive root of the following equation [10,46]:

$$\ln \binom{L(1-p)\gamma + LQ_X^{\text{obs}}}{LpQ_X^{\text{obs}}} + \ln \binom{L(1-Q_X^{\text{obs}}) - L(1-p)\gamma}{Lp(1-Q_X^{\text{obs}})} = \ln \binom{L}{Lp} + 2\ln \varepsilon_x . \tag{S22}$$

We remark that the inequality (S21) is valid for any probability distribution governing the random variables $\vec{X}_t^{\text{obs}}$ and $\vec{X}_t^{\text{KG}}$, hence also for the distribution characterizing Protocol 1.

In order to prove the secrecy of the ACKA protocol, we make use of the following lemma, which is an adaptation of Proposition 8 in [47].

**Lemma 4.** Let $\Omega_{S1}$ and $\Omega_{S2}$ indicate the events $Q_X^{\text{obs}} + \gamma(Q_X^{\text{obs}}) \leq Q_X + \gamma(Q_X)$ and $Q_X^{\text{KG}} \leq Q_X^{\text{obs}} + \gamma(Q_X^{\text{obs}})$, respectively. Let $\rho_{\vec{X}_A^{\text{KG}} \vec{X}_{\oplus A^c}^{\text{KG}} Q_X^{\text{obs}} \wedge \Omega_{S1}}$ be the sub-normalized state of the random variables $\vec{X}_A^{\text{KG}}$, $\vec{X}_{\oplus A^c}^{\text{KG}}$ and $Q_X^{\text{obs}}$ in the fictitious scenario where the participants measured in the $X$ basis even in the key generation rounds. Then it holds:

$$H_{\max}^{\varepsilon_x} (\vec{X}_A^{\text{KG}} | \vec{X}_{\oplus A^c}^{\text{KG}})_{\wedge \Omega_{S1}} \leq L(1-p)\, h(Q_X + \gamma(Q_X)), \tag{S23}$$

where we emphasized that the smooth[2] max-entropy is computed on the sub-normalized state $\rho_{\vec{X}_A^{\text{KG}} \vec{X}_{\oplus A^c}^{\text{KG}} \wedge \Omega_{S1}}$ and where $L(1-p)$ is the average length of the random strings $\vec{X}_A^{\text{KG}}$, $\vec{X}_{\oplus A^c}^{\text{KG}}$.

*Proof.* We start by observing that only event $\Omega_{S1}$ is actually verified during the execution of the protocol (the verification of secrecy done by Alice), while the event $\Omega_{S2}$ cannot be checked since the participants measure in the $Z$ basis in the key generation rounds. Nevertheless, both events are necessary to ensure the secrecy of Alice's raw key, since their combination provides an upper bound on the "phase error rate" $Q_X^{\text{KG}}$ affecting Alice's raw key: $Q_X^{\text{KG}} \leq Q_X + \gamma(Q_X)$. Luckily, we can bound the probability that such a bound is not verified while the verification of secrecy is successful (event $\Omega_{\text{Sfail}}$) through the following chain of inequalities:

$$\Pr \left[ \Omega_{\text{Sfail}} \right] = \Pr \left[ \Omega_{S1} \cap Q_X^{\text{KG}} > Q_X + \gamma(Q_X) \right] \leq \Pr \left[ \Omega_{S1} \cap \Omega_{S2}^c \right] \leq \Pr \left[ \Omega_{S2}^c \right] \leq \varepsilon_x^2, \tag{S24}$$

---

[1] The equivalent of the phase error rate in the standard BB84 protocol.

[2] We remark that the notion of distance we use in the definition of smooth min- and max-entropy is the purified distance [48].

where in the last inequality we used (S21). We stress the fact that the probabilities in the last expression are computed from the same probability distribution defining the state $\rho_{\vec{X}_A^{\mathrm{KG}} \vec{X}_{\oplus A^c}^{\mathrm{KG}} Q_X^{\mathrm{obs}} \wedge \Omega_{\mathrm{S1}}}$, which governs the random variables $\vec{X}_A^{\mathrm{KG}}$, $\vec{X}_{\oplus A^c}^{\mathrm{KG}}$ and $Q_X^{\mathrm{obs}}$. In particular, by labelling $\vec{x}$, $\vec{x}'$ and $\vec{q}$ the realizations of the variables $\vec{X}_A^{\mathrm{KG}}$, $\vec{X}_{\oplus A^c}^{\mathrm{KG}}$ and $Q_X^{\mathrm{obs}}$, respectively, we can write the sub-normalized state of the protocol as follows:

$$\rho_{\vec{X}_A^{\mathrm{KG}} \vec{X}_{\oplus A^c}^{\mathrm{KG}} Q_X^{\mathrm{obs}} \wedge \Omega_{\mathrm{S1}}} = \sum_{\substack{\vec{x}, \vec{x}', \vec{q}: \\ \vec{q} + \gamma(\vec{q}) \leq Q_X + \gamma(Q_X)}} \Pr[\vec{x}, \vec{x}', \vec{q}] \, |\vec{x}\rangle\langle\vec{x}| \otimes |\vec{x}'\rangle\langle\vec{x}'| \otimes |\vec{q}\rangle\langle\vec{q}|. \tag{S25}$$

The state on which the smooth max-entropy is computed is then obtained by tracing out the subsystem $Q_X^{\mathrm{obs}}$ in (S25):

$$\rho_{\vec{X}_A^{\mathrm{KG}} \vec{X}_{\oplus A^c}^{\mathrm{KG}} \wedge \Omega_{\mathrm{S1}}} = \sum_{\vec{x}, \vec{x}'} \Pr[\vec{x}, \vec{x}', \Omega_{\mathrm{S1}}] \, |\vec{x}\rangle\langle\vec{x}| \otimes |\vec{x}'\rangle\langle\vec{x}'|. \tag{S26}$$

The inequality in (S24) tells us that the probability of the unwanted event $\Omega_{\mathrm{Sfail}}$ in the above state is bounded. In the following, we find a sub-normalized state $\zeta$ such that the event $\Omega_{\mathrm{Sfail}}$ never occurs while being $\varepsilon_x$-close to $\rho_{\vec{X}_A^{\mathrm{KG}} \vec{X}_{\oplus A^c}^{\mathrm{KG}} \wedge \Omega_{\mathrm{S1}}}$ in terms of purified distance [48]: $P(\zeta, \rho) \leq \varepsilon_x$. In this way, thanks to the smoothing of the max-entropy, we can upper bound the lhs of (S23) with the max-entropy of the state $\zeta$:

$$H_{\max}^{\varepsilon_x}(\vec{X}_A^{\mathrm{KG}} | \vec{X}_{\oplus A^c}^{\mathrm{KG}})_{\wedge \Omega_{\mathrm{S1}}} \leq H_{\max}(\vec{X}_A^{\mathrm{KG}} | \vec{X}_{\oplus A^c}^{\mathrm{KG}})_{\zeta}. \tag{S27}$$

One can then further bound the max-entropy of $\zeta$ by following the proof of Proposition 8 in [47] as follows:

$$H_{\max}(\vec{X}_A^{\mathrm{KG}} | \vec{X}_{\oplus A^c}^{\mathrm{KG}})_{\zeta} \leq L(1-p) \, h(Q_X + \gamma(Q_X)). \tag{S28}$$

The combination of (S27) and (S28) yields the claim (S23). The crucial fact that allows us to obtain the bound in (S28) is that the fraction of errors between the strings $\vec{X}_A^{\mathrm{KG}}$ and $\vec{X}_{\oplus A^c}^{\mathrm{KG}}$, which by definition is $Q_X^{\mathrm{KG}}$ (S19), is bounded by the maximum number of errors $Q_X + \gamma(Q_X)$ for *every* realization of the strings in $\zeta$, while this is not true for the original state (S26).

The desired state $\zeta$ is constructed from the same distribution appearing in (S26) and reads:

$$\zeta := \sin^2\theta \sum_{\substack{\vec{x}, \vec{x}': \\ \omega_r(\vec{x} \oplus \vec{x}') \leq Q_X + \gamma(Q_X)}} \frac{\Pr[\vec{x}, \vec{x}', \Omega_{\mathrm{S1}}]}{\Pr\left[\Omega_{\mathrm{S1}} \cap Q_X^{\mathrm{KG}} \leq Q_X + \gamma(Q_X)\right]} \, |\vec{x}\rangle\langle\vec{x}| \otimes |\vec{x}'\rangle\langle\vec{x}'|, \tag{S29}$$

where the pre-factor $\sin^2\theta$ corresponds to the trace of $\zeta$ and is fixed by the requirement that $P(\zeta, \rho) \leq \varepsilon_x$, where $\rho$ is a shorthand for $\rho_{\vec{X}_A^{\mathrm{KG}} \vec{X}_{\oplus A^c}^{\mathrm{KG}} \wedge \Omega_{\mathrm{S1}}}$. The purified distance is defined through the generalized fidelity [48]:

$$\bar{F}(\zeta, \rho) := \mathrm{Tr}\left[\sqrt{\sqrt{\rho}\zeta\sqrt{\rho}}\right] + \sqrt{1 - \mathrm{Tr}[\zeta]}\sqrt{1 - \mathrm{Tr}[\rho]}, \tag{S30}$$

as follows:

$$P(\zeta, \rho) := \sqrt{1 - \bar{F}^2(\zeta, \rho)}. \tag{S31}$$

We fix the pre-factor $\sin^2\theta$ by maximizing the generalized fidelity between the states (S29) and (S26) over $\theta$:

$$\begin{aligned} \bar{F}(\zeta, \rho) &= \sin\theta \sqrt{\Pr\left[\Omega_{\mathrm{S1}} \cap Q_X^{\mathrm{KG}} \leq Q_X + \gamma(Q_X)\right]} + \cos\theta \sqrt{1 - \Pr[\Omega_{\mathrm{S1}}]} \\ &\leq \max_\theta \sin\theta \sqrt{\Pr\left[\Omega_{\mathrm{S1}} \cap Q_X^{\mathrm{KG}} \leq Q_X + \gamma(Q_X)\right]} + \cos\theta \sqrt{1 - \Pr[\Omega_{\mathrm{S1}}]} \\ &= \sqrt{\Pr\left[\Omega_{\mathrm{S1}} \cap Q_X^{\mathrm{KG}} \leq Q_X + \gamma(Q_X)\right] + 1 - \Pr[\Omega_{\mathrm{S1}}]} \\ &= \sqrt{1 - \Pr[\Omega_{\mathrm{Sfail}}]}. \end{aligned} \tag{S32}$$

The maximized generalized fidelity in the last expression, combined with (S24), yields $P(\zeta, \rho) \leq \varepsilon_x$ as required. This concludes the proof. □

We are now ready to prove the secrecy of Protocol 1.

**Lemma 5.** The ACKA protocol (Protocol 1) is $\varepsilon_{\mathrm{sec}}$-secret, with $\varepsilon_{\mathrm{sec}} = 2\varepsilon_x + \varepsilon_{\mathrm{PA}}$.

*Proof.* For consistency with the labelling of the parties' roles in the protocol's description, we identify the sender $i$ as Alice ($i \mapsto A$) and the $m$ receivers $\vec{j}$ as Bobs ($\vec{j} \mapsto \vec{B}$).

Note that the secrecy condition (S2) only considers the events where the participants are honest, the ACKA-ID protocol is successful (event $\Phi$), and Protocol 1 does not abort for any participant (event $\Omega_{\mathcal{P}}$), thus we restrict the analysis to these events. Conditioning on the event $\Phi$ implies that the bits $b_A$ and $b_l$ used by Alice and Bob$_l$ (for $l \in \{1, \ldots, m\}$) to communicate to each other in the last two steps of error correction are correctly known to all participants. This enables the participants to agree on whether the protocol aborts or not in the last step of EC. Therefore, the event $\Omega_{\mathcal{P}}$ corresponds to the fact that the verification of secrecy in step 7 ($v_A = 0$) and the error correction of every Bob ($v_{B_l} = 0$ for every $l$) are successful. We indicate these two events as $\Omega_{\mathrm{S1}}$ and $\Omega_{\mathrm{EC}}$, respectively, and emphasize that, conditioned on $\Phi$, we have $\Omega_{\mathcal{P}} = \Omega_{\mathrm{S1}} \cap \Omega_{\mathrm{EC}}$.

Similarly to CKA, the secrecy of our ACKA protocol hinges on the well-established Quantum Leftover Hash Lemma [44,47], which provides an upper bound on the lhs of (S2) for every $\varepsilon \geq 0$, as follows:

$$
\begin{aligned}
\Pr[\Omega_{\mathcal{P}} | A, \vec{B}, \Phi] \, & \left\| \rho^f_{K_A (KC)^c_{A,\vec{B}} E | A, \vec{B}, \Phi, \Omega_{\mathcal{P}}} - \tau_{K_A} \otimes \rho^f_{(KC)^c_{A,\vec{B}} E | A, \vec{B}, \Phi, \Omega_{\mathcal{P}}} \right\|_{\mathrm{tr}} \\
& = \left\| \rho^f_{K_A (KC)^c_{A,\vec{B}} E \wedge \Omega_{\mathcal{P}} | A, \vec{B}, \Phi} - \tau_{K_A} \otimes \rho^f_{(KC)^c_{A,\vec{B}} E \wedge \Omega_{\mathcal{P}} | A, \vec{B}, \Phi} \right\|_{\mathrm{tr}} \\
& \leq 2\varepsilon + \frac{1}{2} \sqrt{2^{\ell - H^{\varepsilon}_{\min}(\vec{Z}^{\mathrm{KG}}_A | (KC'C^{\mathrm{EC}})^c_{A,\vec{B}} E)_{\wedge \Omega_{\mathcal{P}} | A, \vec{B}, \Phi}}},
\end{aligned}
\tag{S33}
$$

where in the first equality we used the definition of sub-normalized states, $\ell$ is given in Eq. (D1) and where we emphasized that the smooth min-entropy is computed on the sub-normalized state $\rho_{\vec{Z}^{\mathrm{KG}}_A (KC'C^{\mathrm{EC}})^c_{A,\vec{B}} E \wedge \Omega_{\mathcal{P}} | A, \vec{B}, \Phi}$. The Quantum Leftover Hash Lemma allows us to reduce the trace distance of the final state of the ACKA protocol (hence after PA) to the computation of the smooth min-entropy of the same state evaluated before PA. In particular, we aim at deriving a lower bound on the smooth min-entropy.

We start by recalling that the classical register $C^c_{A,\vec{B}}$ of the non-participants comprises: the classical inputs and outputs of each non-participant from step 1 until step 7 ($C'^c_{A,\vec{B}}$), the inputs and outputs in step 8 (when error correction takes place, $(C^{\mathrm{EC}})^c_{A,\vec{B}}$), and the public random hash function $F^{\mathrm{PA}}$ used by the participants in PA. Thus: $C^c_{A,\vec{B}} = (C'C^{\mathrm{EC}})^c_{A,\vec{B}} F^{\mathrm{PA}}$. We observe that the state of the register $(C^{\mathrm{EC}})^c_{A,\vec{B}}$ is in tensor product with the rest of the quantum state on which the smooth min-entropy is computed:

$$
\rho_{\vec{Z}^{\mathrm{KG}}_A (KC'C^{\mathrm{EC}})^c_{A,\vec{B}} E \wedge \Omega_{\mathcal{P}} | A, \vec{B}, \Phi} = \rho_{(C^{\mathrm{EC}})^c_{A,\vec{B}}} \otimes \rho_{\vec{Z}^{\mathrm{KG}}_A (KC')^c_{A,\vec{B}} E \wedge \Omega_{\mathcal{P}} | A, \vec{B}, \Phi}.
\tag{S34}
$$

Indeed, the public communication in error correction is completely random from the point of view of the non-participants and thus uncorrelated from everything else. In particular, their inputs and outputs are either fixed to a predefined value or random. This is because the information about the syndrome $\vec{y}$ and the hash $\vec{h}_A$ are encoded by Alice before being transmitted. Then, by virtue of (S34), we have that:

$$
H^{\varepsilon}_{\min}(\vec{Z}^{\mathrm{KG}}_A | (KC'C^{\mathrm{EC}})^c_{A,\vec{B}} E)_{\wedge \Omega_{\mathcal{P}} | A, \vec{B}, \Phi} = H^{\varepsilon}_{\min}(\vec{Z}^{\mathrm{KG}}_A | (KC')^c_{A,\vec{B}} E)_{\wedge \Omega_{\mathcal{P}} | A, \vec{B}, \Phi}.
\tag{S35}
$$

We remark that having removed the conditioning on $(C^{\mathrm{EC}})^c_{A\vec{B}}$ allows us to compute the min-entropy in the rhs of (S35) on the state $\rho_{\vec{Z}^{\mathrm{KG}}_A \vec{Z}^{\mathrm{KG}}_{\vec{B}} (KC')^c_{A,\vec{B}} E \wedge \Omega_{\mathcal{P}} | A, \vec{B}, \Phi}$, which describes the protocol's classical registers and Eve's quantum system at step 7 of ACKA.

We now bound the rhs of (S35) with the uncertainty relation for smooth entropies [49]. In order to apply the uncertainty relation, there are three remarks to be made.

Firstly, we need to consider the state of the protocol in a fictitious scenario, where the parties have already measured their quantum systems in all the test rounds but have not yet measured their quantum systems in the key generation rounds. The considered state would be: $\rho_{A^{\mathrm{KG}} B^{\mathrm{KG}} (KC')^c_{A,\vec{B}} E \wedge \Omega_{\mathcal{P}} | A, \vec{B}, \Phi}$, where $A^{\mathrm{KG}}$ and $B^{\mathrm{KG}}$ are Alice's and Bobs' quantum systems before the key generation measurements which transform them into $\vec{Z}^{\mathrm{KG}}_A$ and $\vec{Z}^{\mathrm{KG}}_{\vec{B}}$ and where $\Omega_{\mathcal{P}} = \Omega_{\mathrm{S1}} \cap \Omega_{\mathrm{EC}}$. Clearly, this quantum state is not well-defined. Indeed, the event $\Omega_{\mathrm{EC}}$ is related to the classical registers $\vec{Z}^{\mathrm{KG}}_A$ and $\vec{Z}^{\mathrm{KG}}_{\vec{B}}$ and becomes meaningless when such registers are replaced by the quantum registers $A^{\mathrm{KG}}$ and

$B^{\mathrm{KG}}$. This fact, often overlooked, is properly accounted for in the QKD proof presented in [47]. To solve the issue, we employ Lemma 10 from [47] and remove the conditioning on $\Omega_{\mathrm{EC}}$ from the min-entropy in (S35):

$$H_{\min}^{\varepsilon}(\vec{Z}_A^{\mathrm{KG}}|(KC')_{A,\vec{B}}^c E)_{\wedge\Omega_{\mathcal{P}}|A,\vec{B},\Phi} = H_{\min}^{\varepsilon}(\vec{Z}_A^{\mathrm{KG}}|(KC')_{A,\vec{B}}^c E)_{\wedge\Omega_{\mathrm{S1}}\cap\Omega_{\mathrm{EC}}|A,\vec{B},\Phi}$$
$$\geq H_{\min}^{\varepsilon}(\vec{Z}_A^{\mathrm{KG}}|(KC')_{A,\vec{B}}^c E)_{\wedge\Omega_{\mathrm{S1}}|A,\vec{B},\Phi}. \tag{S36}$$

We can now consider the well-defined state $\rho_{A^{\mathrm{KG}}B^{\mathrm{KG}}(KC')_{A,\vec{B}}^c E\wedge\Omega_{\mathrm{S1}}|A,\vec{B},\Phi}$ of the fictitious scenario described above for the uncertainty relation.

The second remark is that although the fictitious scenario does not occur in the real experiment, we can still consider it as far as the experimental setup theoretically allows the participants to delay their key generation measurements until after step 7 while maintaining the statistics of their outcomes unchanged[3].

The third remark regards the classical registers $(C')_{A,\vec{B}}^c$. We point out that $(C')_{A,\vec{B}}^c$ includes the classical registers $(\vec{X}^{\mathrm{obs}})_{A,\vec{B}}^c$ and $(\vec{X}^{\mathrm{KG}})_{A,\vec{B}}^c$. The latter register represents the bits that the non-participants would announce in step 6 of Protocol 1 if the key generation rounds had been labelled as test rounds. Note that, since the parties are assumed to have no/limited quantum memory, by the time the testing key is publicly revealed in step 5, the non-participants have already measured all their quantum systems and computed the bits they would announce for every measurement round. Then, upon learning the testing key, the non-participants only announce $(\vec{X}^{\mathrm{obs}})_{A,\vec{B}}^c$. We now assume without loss of generality that the ACKA protocol produces two copies of the classical registers $(\vec{X}^{\mathrm{KG}})_{A,\vec{B}}^c$, one of which is stored in $(C')_{A,\vec{B}}^c$, as already mentioned. We are allowed to assume this since classical information can always be copied for free. Then, the quantum state of the ACKA protocol that we consider for the application of the uncertainty relation is $\rho_{A^{\mathrm{KG}}B^{\mathrm{KG}}(KC'\vec{X}^{\mathrm{KG}})_{A,\vec{B}}^c E\wedge\Omega_{\mathrm{S1}}|A,\vec{B},\Phi}$.

We are now ready to apply the uncertainty relation [49], where we group the registers $(\vec{X}^{\mathrm{KG}})_{A,\vec{B}}^c$ with Bobs' quantum systems $B^{\mathrm{KG}}$, while $(KC')_{A,\vec{B}}^c$ are grouped with $E$. We obtain the following lower bound:

$$H_{\min}^{\varepsilon}(\vec{Z}_A^{\mathrm{KG}}|(KC')_{A,\vec{B}}^c E)_{\wedge\Omega_{\mathrm{S1}}|A,\vec{B},\Phi} \geq L(1-p) - H_{\max}^{\varepsilon}(\vec{X}_A^{\mathrm{KG}}|B^{\mathrm{KG}}(\vec{X}^{\mathrm{KG}})_{A,\vec{B}}^c)_{\wedge\Omega_{\mathrm{S1}}|A,\vec{B},\Phi}, \tag{S37}$$

where $\vec{X}_A^{\mathrm{KG}}$ is the string of $X$ outcomes Alice would obtain had she measured in the $X$ basis the quantum systems devoted to key generation. By employing the data-processing inequality [51], we can further bound the smooth min-entropy as follows:

$$H_{\min}^{\varepsilon}(\vec{Z}_A^{\mathrm{KG}}|(KC')_{A,\vec{B}}^c E)_{\wedge\Omega_{\mathrm{S1}}|A,\vec{B},\Phi} \geq L(1-p) - H_{\max}^{\varepsilon}(\vec{X}_A^{\mathrm{KG}}|\vec{X}_{\vec{B}}^{\mathrm{KG}}(\vec{X}^{\mathrm{KG}})_{A,\vec{B}}^c)_{\wedge\Omega_{\mathrm{S1}}|A,\vec{B},\Phi}$$
$$\geq L(1-p) - H_{\max}^{\varepsilon}(\vec{X}_A^{\mathrm{KG}}|\vec{X}_{\oplus A^c}^{\mathrm{KG}})_{\wedge\Omega_{\mathrm{S1}}|A,\vec{B},\Phi}, \tag{S38}$$

where $\vec{X}_{\vec{B}}^{\mathrm{KG}}$ are the strings of $X$ outcomes the Bobs would obtain had they measured in the $X$ basis also in the key generation rounds and where we computed the addition modulo two between the strings $\vec{X}_{\vec{B}}^{\mathrm{KG}}$ and $(\vec{X}^{\mathrm{KG}})_{A,\vec{B}}^c$ and used (S20) to obtain the second inequality. Finally, by employing the result of Lemma 4 upon fixing $\varepsilon = \varepsilon_x$, we obtain the following lower bound on the smooth min-entropy:

$$H_{\min}^{\varepsilon_x}(\vec{Z}_A^{\mathrm{KG}}|(KC')_{A,\vec{B}}^c E)_{\wedge\Omega_{\mathrm{S1}}|A,\vec{B},\Phi} \geq L(1-p)[1 - h(Q_X + \gamma(Q_X))]. \tag{S39}$$

At the same time, from Eq. (D1) we get:

$$\ell \leq L(1-p)\left[1 - h\left(Q_X + \gamma(Q_X)\right)\right] - 2\log_2\frac{1}{2\varepsilon_{\mathrm{PA}}}. \tag{S40}$$

By employing (S39) and (S40) in (S33), we can upper bound the rhs of (S33) as follows:

$$\Pr[\Omega_{\mathcal{P}}|A,\vec{B},\Phi]\left\|\rho_{K_A(KC')_{A,\vec{B}}^c E|A,\vec{B},\Phi,\Omega_{\mathcal{P}}}^f - \tau_{K_A}\otimes\rho_{(KC')_{A,\vec{B}}^c E|A,\vec{B},\Phi,\Omega_{\mathcal{P}}}^f\right\|_{\mathrm{tr}} \leq 2\varepsilon_x + \frac{1}{2}\sqrt{2^{2\log_2(2\varepsilon_{\mathrm{PA}})}}$$
$$= 2\varepsilon_x + \varepsilon_{\mathrm{PA}}, \tag{S41}$$

where we used $\varepsilon = \varepsilon_x$. By repeating the same proof for any choice of sender $i$ and receivers $\vec{j}$, we obtain the claim for $\varepsilon_{\mathrm{sec}} = 2\varepsilon_x + \varepsilon_{\mathrm{PA}}$. This concludes the proof. $\qquad\square$

---

[3] For instance, the parties could delay their key generation measurements by storing the quantum systems in perfect quantum memories. However, in order not to change the statistics of their outcomes, we require the parties' measurement devices to be memoryless [50]. In [47] the authors simply assumed that the key generation and test measurements commute.

**Lemma 6.** The ACKA protocol (Protocol 1) is $\varepsilon_{\mathrm{CKA}}$-CKA-secure, with $\varepsilon_{\mathrm{CKA}} = \varepsilon_{\mathrm{EC}} + 2\varepsilon_x + \varepsilon_{\mathrm{PA}}$.

*Proof.* The claim follows from Lemmas 1, 3 and 5. $\qquad\square$

### C. Anonymity

Here we show that the anonymity condition Eq. (B13) is satisfied by Protocol 1.

**Lemma 7.** The ACKA protocol (Protocol 1) is $\varepsilon_{\mathrm{AN}}$-anonymous, with $\varepsilon_{\mathrm{AN}} = 0$.

*Proof.* According to the conditions in Eq. (B7) and Eq. (B8) in the anonymity definition (Definition 3), we only need to consider the instances in which the intended sender and receivers are honest, since otherwise no condition applies to the final state of Protocol 1. We now analyse the steps of the protocol and argue that, for these instances, the state of any subset of non-participants and Eve is independent of the action of the remaining parties at each step of the protocol. This implies that the output state of the ACKA protocol satisfies the conditions Eq. (B7) and Eq. (B8), hence the protocol is perfectly anonymous with $\varepsilon_{\mathrm{AN}} = 0$. We remark that –differently from QKD security– anonymity is required even in the instances in which the protocol aborts. That is, if an ACKA protocol aborts, no conference key is established but we still require that the identities of the intended participants remain unknown to the other parties.

The proof makes use of the following property stated as Proposition 1, which directly follows from the Kraus representation of CPTP maps.

**Proposition 1.** Given a system of $n$ parties and a subset $\mathcal{G} \subseteq \{1, \dots, n\}$ of the parties, the partial state of subset $\mathcal{G}$ is invariant under the action of CPTP maps on the complement set $\mathcal{G}^c$. That is, if $\mathcal{M}_{\mathcal{G}^c}$ and $\mathcal{M}'_{\mathcal{G}^c}$ are any CPTP maps that can act on the subset of parties $\mathcal{G}^c$, then

$$\rho_{\mathcal{G}} = \mathrm{Tr}_{\mathcal{G}^c}[\rho] = \mathrm{Tr}_{\mathcal{G}^c}\left[(\mathbb{1}_{\mathcal{G}} \otimes \mathcal{M}_{\mathcal{G}^c})(\rho)\right] = \mathrm{Tr}_{\mathcal{G}^c}\left[(\mathbb{1}_{\mathcal{G}} \otimes \mathcal{M}'_{\mathcal{G}^c})(\rho)\right]. \tag{S42}$$

In step 1 of Protocol 1, the parties run the ACKA-ID protocol. In the first part of the ACKA-ID protocol, the parties perform Collision Detection (Protocol 5). From the proven security of Collision Detection [20], no information about the individual inputs of each honest participant is leaked. Subsequently, the sender proceeds to notify the receivers of their role and the identity of the remaining participants. This is done through the Parity protocol with a random public output. Therefore, no information is acquired by non-participants and Eve. Finally, a Veto protocol is performed in which the parties' inputs depend on whether their communication was tampered with in the previous step of the ACKA-ID protocol, regardless of whether they are a participant or not. Therefore, the output of the Veto is independent of the identity of the parties.

In step 3, assuming that the conference key used by the participants is perfectly secure, all the parties broadcast strings that are completely random from the perspective of the non-participants and Eve.

In step 4, the action of the parties on their share of the distributed quantum state depends on whether they are a participant or not. Indeed, based on the bits of the testing key, in the key generation rounds the participants will perform a measurement in the $Z$ basis, while non-participants always measure in the $X$ basis. However, the outcomes of the key generation rounds are not announced. Therefore, the state of any subset of non-participants and Eve, by Proposition 1, is independent of whether the participants performed a measurement in the $Z$ or the $X$ basis. Thus, although in step 4 the participants act differently, no information about their role is leaked to the other parties.

The public information generated in steps 5 and 6 is again independent of the identities of the participants. Indeed, step 5 publicly outputs the testing key which is randomly chosen by Alice and the public output of step 6 is a random string, since Alice inputs random bits in the sequence of Parity protocols.

The Veto protocol in step 7 reveals information about the value of the observed parameter $Q_X^{\mathrm{obs}}$, since the protocol aborts if $Q_X^{\mathrm{obs}}$ is too large compared to the predefined value $Q_X$. However, from the fact that $Q_X^{\mathrm{obs}}$ is invariant under permutation of the parties' roles, we deduce that step 7 does not reveal any information on the parties' identities, regardless of whether the protocol aborts or not at this step.

Step 8 only consists of broadcasts encrypted with pre-established keys that only Alice and the Bobs have access to. Therefore, without the knowledge of the pre-established keys, the information publicly available to the non-participants and Eve in step 8 is completely random.

In step 9, if the protocol did not abort, the participants apply a hash to their raw keys. Since this is a local map with no public communication, by Proposition 1 it does not change the state of the non-participants and Eve.

Finally, we remark that we cannot prevent the dishonest parties from publicly disclosing their identity at any moment. Therefore, we have shown that the final state of the protocol is such that:

$$\rho^f_{(PKC)^c_{i,j}E|i,\vec{j}} = \sigma^{\mathcal{D}}_{(PKC)^c_{i,j}E|i,\vec{j}}$$

$$\rho^f_{(PKC)^c_{i,j}E|\vec{i},\vec{j}} = \sigma^{\mathcal{D}}_{(PKC)^c_{i,j}E|\vec{i},\vec{j}}$$

(S43)

where the states on the rhs satisfy the anonymity conditions Eq. (B7) and Eq. (B8). This implies that Protocol 1 is $\varepsilon_{\mathrm{AN}}$-anonymous according to condition Eq. (B13), with $\varepsilon_{\mathrm{AN}} = 0$. □

### D. Security

**Lemma 8.** The ACKA protocol (Protocol 1) is $\varepsilon_{\mathrm{tot}}$-secure according to Definition 5, with $\varepsilon_{\mathrm{tot}} = 2^{-r_V} + (n-1)\,\varepsilon_{\mathrm{enc}} + 2\varepsilon_x + \varepsilon_{\mathrm{EC}} + \varepsilon_{\mathrm{PA}}$.

*Proof.* The claim follows from combining the results of Lemmas 2, 6 and 7 with Definition 5. □

## IV. SECURITY PROOF OF bACKA

In this section we prove the security of the bACKA protocol (Protocol 11).

### A. Integrity

**Lemma 9.** The bACKA protocol (Protocol 11) is $\varepsilon_{\mathrm{IN}}$-integrous with $\varepsilon_{\mathrm{IN}} = 2^{-r_V} + (n-1)\,\varepsilon_{\mathrm{enc}}$.

*Proof.* The Identity Designation step in bACKA is performed using the same ID protocol used in ACKA (Protocol 6). Therefore the integrity proof for bACKA coincides with the proof of Lemma 2. □

### B. CKA-security

**Lemma 10.** The bACKA protocol (Protocol 11) is $\varepsilon_{\mathrm{CKA}}$-CKA-secure, with $\varepsilon_{\mathrm{CKA}} = 0$.

*Proof.* We separately consider the correctness (Definition 1) and the secrecy (Definition 2) of bACKA. For both of them, we must only consider the instances where the ACKA-ID protocol does not abort and is successful (event $\Phi$). In these instances, due to the noiseless bipartite private channels linking every pair of parties, the receivers recover the sender's conference key without errors: $\Pr[\cup_{l=1}^m \vec{k}_i \neq \vec{k}_{j_l} | i, \vec{j}, \Phi, \Omega_{\mathcal{P}}] = 0$, and without any information leak. At the same time, the strings received by the non-participants in step 3 are random and thus uncorrelated from the participants' keys. We deduce that the state of the sender's conference key is uncorrelated from the systems held by the non-participants and Eve:

$$\rho^f_{K_i(KC)^c_{i,j}E|i,\vec{j},\Phi,\Omega_{\mathcal{P}}} = \tau_{K_i} \otimes \rho^f_{(KC)^c_{i,j}E|i,\vec{j},\Phi,\Omega_{\mathcal{P}}}.$$

(S44)

Moreover, the bACKA protocol does not abort, given that event $\Phi$ occurred, hence: $\Pr[\Omega^c_{\mathcal{P}} \cap \Gamma^c_{\mathcal{P}} | i, \vec{j}, \Phi] = 0$.

We conclude that the correctness and secrecy parameters of the bACKA protocol are equal to: $\varepsilon_{\mathrm{cor}} = \varepsilon_{\mathrm{sec}} = 0$. Then the claim follows from Lemma 1. □

### C. Anonymity

**Lemma 11.** The bACKA protocol (Protocol 11) is $\varepsilon_{\mathrm{AN}}$-anonymous, with $\varepsilon_{\mathrm{AN}} = 0$.

*Proof.* Similarly to the proof of Lemma 7, the anonymity of the bACKA protocol follows from the fact that, in all the steps of Protocol 11, the outputs of the non-participants are either random strings or independent of the identity of the participants. □

#### D. Security

**Lemma 12.** The bACKA protocol (Protocol 11) is $\varepsilon_{\text{tot}}$-secure according to Definition 5, with $\varepsilon_{\text{tot}} = 2^{-r_V} + (n-1)\,\varepsilon_{\text{enc}}$.

*Proof.* The claim follows from combining the results of Lemmas 9, 10 and 11 with Definition 5. $\qquad\square$

## V. SECURITY PROOF OF fully-ACKA

In the following subsections we prove that our fully-ACKA protocol (Protocol 2) is integrous and CKA-secure according to conditions Eq. (B11) and Eq. (B12). We also prove its weak-anonymity as given by Definition 6.

### A. Integrity

Here we show that the integrity condition Eq. (B11) is satisfied by Protocol 2.

**Lemma 13.** The fully-ACKA protocol (Protocol 2) is $\varepsilon_{\text{IN}}$-integrous with $\varepsilon_{\text{IN}} = 2^{-r_V} + (n-1)\,\varepsilon_{\text{enc}}$.

*Proof.* The integrity of fully-ACKA depends on the probability that the fully-ACKA-ID protocol (Protocol 7) fails to abort or to correctly assign the participants' identities. Indeed, we need to bound the probabilities $\Pr[\Gamma^c \cap \Phi^c | \vec{i}, \vec{j}]$ and $\Pr[\Gamma^c | \vec{i}, \vec{j}]$. The former is the probability that the fully-ACKA-ID protocol does not assign correctly the identities or aborts for a strict subset of parties, when only one party applied to be the sender. The latter is the probability that the protocol does not abort for every party when multiple parties apply to be the sender. The ID protocol used in fully-ACKA (Protocol 7) only differs from that of ACKA (Protocol 6) in the size of the message transmitted in step 3, therefore the integrity of fully-ACKA follows in a similar fashion as the proof of Lemma 2. $\qquad\square$

### B. CKA-security

In order to prove the CKA-security Eq. (B12) of Protocol 2, we show that it is both correct and secret according to Definitions 1 and 2. However, before doing so, we single out the contributions of the failures of Veto in step 7, of the Notification in TKD (step 2 in Protocol 8), and of the decoding function $G$ (used in TKD and EC) to the CKA-security epsilon parameter. Indeed, if any of such instances fail, both correctness and secrecy are affected. We remark that we consider a failure for Veto if it fails to return the outcome 1 when at least one party inputs 1. We have a failure for Notification in TKD if it fails to notify the party $s$ picked by Alice in any of the $n-1$ rounds of TKD. Finally we consider a failure of the decoding function $G$ of an AMD code $(F, G)$ if, when $G$ is used in TKD and EC, it holds $G(F(\vec{x}) \oplus \vec{b}) \neq \top$, where $\vec{x}$ is the string to be encoded and $\vec{b} \neq \vec{0}$.

Regarding correctness, suppose that a dishonest party flips some bits of the bitstring $\vec{r}_l$ transmitted from Alice to Bob$_l$ in TKD without being detected, i.e. $G$ does not return $\top$ in the last step of TKD. Then there will be a misunderstanding between Alice and Bob$_l$ on whether the protocol must abort in the last steps of EC. For instance, it could happen that the protocol does not abort from Alice's viewpoint while it does from Bob$_l$'s in the last step of EC. This causes Alice's key to differ from Bob$_l$'s key (which in this case is the trivial key $K_{B_l} = \emptyset$). Even if the decoding function $G$ successfully returns $\top$ for Bob$_l$, a correctness issue arises if Veto fails to output 1 in step 7 and thus to abort the protocol. Indeed, Bob$_l$ will be unable to communicate with the other participants in the last steps of EC: e.g. the protocol might not abort even though the error correction failed for Bob$_l$, or the protocol aborts for the other participants but not for Bob$_l$. Similarly, suppose that Notification fails for Bob$_l$. This means that he does not receive the random bits transmitted by Alice in TKD needed to communicate in EC, causing correctness issues[4].

Regarding secrecy, the final conference key might be partly known to Eve if the test error rate $Q_X^{\text{obs}}$ is too large (such that $Q_X^{\text{obs}} + \gamma(Q_X^{\text{obs}}) > Q_X + \gamma(Q_X)$) but Veto failed to abort the protocol in step 7. Alternatively, if Notification

---

[4] Note that, strictly speaking, the failure of Notification must be combined with the fact that the protocol does not abort in step 7 in order to cause a correctness loophole (otherwise, if the protocol aborts, all participants share the same trivial key). The probability that the combined events occur can be bounded by: $(n-1)2^{-r_N}(2^{-r_V} + \varepsilon_{\text{enc}})$. Indeed, the protocol does not abort in step 7 when Veto fails (with probability $\leq 2^{-r_V}$) or when Bob$_l$ does not realize that he has not been notified when Alice planned to. This can happen if Bob$_l$ gets notified by a dishonest party in another TKD round destined to party $s \neq B_l$ and the decoding function $G$ applied by Bob$_l$ fails to return $\top$ (which happens with probability $\leq \varepsilon_{\text{enc}}$). To conclude, the fact that we simply consider the failure of Notification alone is due to simplicity and symmetry with the proof of anonymity, where Notification failure alone can lead to loopholes. However, performance-wise, this is not optimal as it increases the probability of the unwanted event to from $(n-1)2^{-r_N}(2^{-r_V} + \varepsilon_{\text{enc}})$ to $(n-1)2^{-r_N}$.

fails for Bob$_l$, the output of the Parity protocols in the Distribution phase of TKD publicly reveals the testing key $\vec{k}_T$, which can be used by any dishonest party to impersonate a Bob and learn the conference key[5].

Let us denote $\Psi$ the event in which: Veto in step 7, Notification in TKD, and the decoding function $G$ (every time it is used in TKD and EC) are successful –i.e., Veto returns 1 when at least one party inputs 1, every party $s$ picked by Alice is notified in TKD, and $G(F(\vec{x}) \oplus \vec{b}) = \top$ for every $\vec{b} \neq 0$. We now prove that the fully-ACKA protocol is both correct and secret conditioned on the event $\Psi$. We then deal with the general case while proving CKA-security in Lemma 16.

**Lemma 14.** The fully-ACKA protocol (Protocol 2) –conditioned on the event $\Psi$– is $\varepsilon_{\mathrm{cor}}$-correct, with $\varepsilon_{\mathrm{cor}} = \varepsilon_{\mathrm{EC}}$.

*Proof.* According to the correctness definition (Definition 1) applied to the fully-ACKA protocol conditioned on event $\Psi$, we must evaluate the probabilities $\Pr[\cup_{l=1}^m \vec{k}_A \neq \vec{k}_{B_l} \cap \Omega_{\mathcal{P}}|A, \vec{B}, \Phi, \Psi]$ and $\Pr[\Omega_{\mathcal{P}}^c \cap \Gamma_{\mathcal{P}}^c|A, \vec{B}, \Phi, \Psi]$, where we relabelled the sender $i$ and the receivers $\vec{j}$ as Alice ($i = A$) and Bobs ($\vec{j} = \vec{B}$). The former is the probability that the conference key of some Bob differs from Alice's and that the protocol does not abort for any participant. The latter is the probability that the protocol only aborts for a strict subset of participants.

Having removed the correctness loopholes described at the beginning of the subsection by conditioning on $\Psi$, the security proof is analogous to that of standard CKA protocols.

To start with, we argue that for our fully-ACKA protocol it holds $\Pr[\Omega_{\mathcal{P}}^c \cap \Gamma_{\mathcal{P}}^c|A, \vec{B}, \Phi, \Psi] = 0$. Indeed, the only possibility for the participants not agreeing on whether the protocol aborts or not is if there is some miscommunication in the last steps of EC. This happens if the Bobs are not using in error correction the random bits $(b_l, \vec{r}_\emptyset)$ transmitted by Alice, or if some dishonest party flips the outcome of the last Parity protocol in error correction and some Bob does not detect that. However, having conditioned on the event $\Psi$ guarantees that if the bits $(b_l, \vec{r}_\emptyset)$ are not correctly transmitted to every Bob in TKD, then the fully-ACKA protocol aborts for every party with the Veto in step 7 and does not reach the error correction stage. Moreover, it guarantees that any bit flip in the last Parity of error correction is detected by all participants, who then agree to abort. Therefore, the miscommunication in error correction cannot happen when conditioning on $\Psi$. Thus it holds:

$$\Pr[\Omega_{\mathcal{P}}^c \cap \Gamma_{\mathcal{P}}^c|A, \vec{B}, \Phi, \Psi] = 0. \tag{S45}$$

For the above argument, in computing $\Pr[\cup_{l=1}^m \vec{k}_A \neq \vec{k}_{B_l} \cap \Omega_{\mathcal{P}}|A, \vec{B}, \Phi, \Psi]$ we must only consider the instances where the transmission of the bits $(b_l, \vec{r}_\emptyset)$ in TKD is successful for every Bob, otherwise the protocol would abort. This avoids any misunderstanding between Alice and the Bobs on the success of the verification of error correction through the hash function. Therefore, when the protocol does not abort (event $\Omega_{\mathcal{P}}$) it can only happen that $\vec{h}_A = \vec{h}_{B_l}$ for $l \in \{1, \dots, m\}$ (otherwise the protocol would abort). Thus we can recast the probability as follows:

$$\Pr[\cup_{l=1}^m \vec{k}_A \neq \vec{k}_{B_l} \cap \Omega_{\mathcal{P}}|A, \vec{B}, \Phi, \Psi] = \Pr[\cup_{l=1}^m (\vec{k}_A \neq \vec{k}_{B_l} \cap \vec{h}_A = \vec{h}_{B_l}) \cap \Omega_{\mathcal{P}}|A, \vec{B}, \Phi, \Psi]$$
$$\leq \Pr[\cup_{l=1}^m (\vec{k}_A \neq \vec{k}_{B_l} \cap \vec{h}_A = \vec{h}_{B_l})|A, \vec{B}, \Phi, \Psi]. \tag{S46}$$

By the properties of two-universal hash functions (see correctness proof in [19]), we can bound the rhs of the last expression and obtain:

$$\Pr[\cup_{l=1}^m \vec{k}_A \neq \vec{k}_{B_l} \cap \Omega_{\mathcal{P}}|A, \vec{B}, \Phi, \Psi] \leq \varepsilon_{\mathrm{EC}}. \tag{S47}$$

The combination of (S45) with (S47) concludes the proof. $\square$

We derive the secrecy of Protocol 2 using the same nomenclature of ACKA for the random strings $\vec{X}_t^{\mathrm{obs}}$ and $\vec{X}_t^{\mathrm{KG}}$. Analogously to ACKA, the test error rate affecting the key-generation rounds, $Q_X^{\mathrm{KG}}$, and the corresponding observed error rate, $Q_X^{\mathrm{obs}}$, are defined by (S19) and (S18), respectively.

**Lemma 15.** The fully-ACKA protocol (Protocol 2) –conditioned on the event $\Psi$– is $\varepsilon_{\mathrm{sec}}$-secret, with $\varepsilon_{\mathrm{sec}} = 2\varepsilon_x + \varepsilon_{\mathrm{PA}}$.

*Proof.* For consistency with the labelling of the parties' roles in the protocol's description, we identify the sender $i$ as Alice ($i \mapsto A$) and the $m$ receivers $\vec{j}$ as Bobs ($\vec{j} \mapsto \vec{B}$).

According to the secrecy condition (S2), we must only consider the instances where the participants are honest, the fully-ACKA-ID protocol is successful (event $\Phi$), and Protocol 2 does not abort for any participant (event $\Omega_{\mathcal{P}}$). By

---

[5] As for the correctness case, the secrecy loophole actually arises only if the Notification fails and the protocol does not abort in step 7.

combining $\Omega_\mathcal{P}$ with the further conditioning on the event $\Psi$, we deduce that every Bob has been correctly notified in TKD and that the observed test error rate does not exceed its upper bound ($v_s = 0$, otherwise the protocol would have aborted). This implies that we can avoid considering the secrecy loopholes highlighted at the beginning of the subsection and the secrecy epsilon parameter reduces to the standard one of CKA protocols.

In order to proceed with the proof, we define the events $\Omega_{\mathrm{S1}}$ and $\Omega_{\mathrm{EC}}$ as the event where the protocol does not abort until step 7 is completed –which, conditioned on $\Psi$, implies $Q_X^{\mathrm{obs}} + \gamma(Q_X^{\mathrm{obs}}) \leq Q_X + \gamma(Q_X)$– and the event where the protocol does not abort for the participants in EC, respectively. Then we have: $\Omega_\mathcal{P} = \Omega_{\mathrm{S1}} \cap \Omega_{\mathrm{EC}}$.

Analogously to the secrecy proof of the ACKA protocol (Lemma 5), the secrecy of our fully-ACKA protocol relies on the Quantum Leftover Hash Lemma [44,47], which provides an upper bound on the lhs of the secrecy condition (S2) (conditioned on $\Psi$) for every $\varepsilon \geq 0$, as follows:

$$\left\| \rho^f_{K_A(KC)^c_{A,\vec{B}}E \wedge \Omega_\mathcal{P}|A,\vec{B},\Phi,\Psi} - \tau_{K_A} \otimes \rho^f_{(KC)^c_{A,\vec{B}}E \wedge \Omega_\mathcal{P}|A,\vec{B},\Phi,\Psi} \right\|_{\mathrm{tr}} \leq 2\varepsilon + \frac{1}{2}\sqrt{2^{\ell - H^\varepsilon_{\min}(\vec{Z}_A^{\mathrm{KG}}|(KC'C^{\mathrm{EC}})^c_{A,\vec{B}}E)_{\wedge \Omega_\mathcal{P}|A,\vec{B},\Phi,\Psi}}},$$
(S48)

where $\ell$ is given in Eq. (D2). Note that the classical register $C^c_{A,\vec{B}}$ comprises: the classical inputs and outputs of each non-participant from step 1 until step 7 ($C'^c_{A,\vec{B}}$), the inputs and outputs of the non-participants in error correction ($(C^{\mathrm{EC}})^c_{A,\vec{B}}$), and the public random hash function $F^{\mathrm{PA}}$ used by the participants in PA. Thus: $C^c_{A,\vec{B}} = (C'C^{\mathrm{EC}})^c_{A,\vec{B}}F^{\mathrm{PA}}$.

Now, our goal is to derive a lower bound on the smooth min-entropy in the rhs of (S48), which is computed on the state of the protocol before PA. We first employ the following chain-rule for the min-entropy [50] in order to remove the conditioning on $C^{\mathrm{EC}}$:

$$H^\varepsilon_{\min}(\vec{Z}_A^{\mathrm{KG}}|(KC'C^{\mathrm{EC}})^c_{A,\vec{B}}E)_{\wedge \Omega_\mathcal{P}|A,\vec{B},\Phi,\Psi} \geq H^\varepsilon_{\min}(\vec{Z}_A^{\mathrm{KG}}|(KC')^c_{A,\vec{B}}E)_{\wedge \Omega_\mathcal{P}|A,\vec{B},\Phi,\Psi} - \log_2|(C^{\mathrm{EC}})^c_{A,\vec{B}}|,$$
(S49)

where $\log_2|(C^{\mathrm{EC}})^c_{A,\vec{B}}|$ quantifies, in principle, all the information contained in the classical register $C$ of the non-participants in error correction (see Protocol 10). This includes the syndrome $\vec{y}$ and the hash $\vec{h}_A$ publicly broadcast by Alice, which are computed from Alice's raw key $\vec{Z}_A^{\mathrm{KG}}$ and thus correlated to it. On the other hand, all the inputs and the remaining outputs of the non-participants in error correction are either fixed or random and hence independent of $\vec{Z}_A^{\mathrm{KG}}$. Thus, their registers can be safely discarded from $(C^{\mathrm{EC}})^c_{A,\vec{B}}$ without changing the smooth min-entropy. Thus, we can restrict $\log_2|(C^{\mathrm{EC}})^c_{A,\vec{B}}|$ to only quantifying the error correction information that is correlated with $\vec{Z}_A^{\mathrm{KG}}$, namely the syndrome and the hash computed by Alice:

$$\log_2|(C^{\mathrm{EC}})^c_{A,\vec{B}}| = L(1-p)h(Q_Z) + \left\lceil \log_2 \frac{n-1}{\varepsilon_{\mathrm{EC}}} \right\rceil \leq L(1-p)h(Q_Z) + \log_2 \frac{2(n-1)}{\varepsilon_{\mathrm{EC}}}.$$
(S50)

We remark that having removed the conditioning on $(C^{\mathrm{EC}})^c_{A,\vec{B}}$ allows us to compute the min-entropy in the rhs of (S49) on the state $\rho_{\vec{Z}_A^{\mathrm{KG}}\vec{Z}_{\vec{B}}^{\mathrm{KG}}(KC')^c_{A,\vec{B}}E \wedge \Omega_\mathcal{P}|A,\vec{B},\Phi,\Psi}$, which describes the protocol's classical registers and Eve's quantum system at step 7 of fully-ACKA.

We now derive a lower bound on the min-entropy in the rhs of (S49) with the entropic uncertainty relation [49]. Following the steps in the secrecy proof of ACKA (Lemma 5), we assume that it is possible to consider the state of the protocol's registers in the fictitious scenario in which the key generation measurements are delayed, such that they are yet to be performed once all the test measurements are completed while the statistics of the outcomes are unchanged. In order to obtain a well-defined state[6], we remove the conditioning on the event $\Omega_{\mathrm{EC}}$ in (S49) by using Lemma 10 from [47]:

$$H^\varepsilon_{\min}(\vec{Z}_A^{\mathrm{KG}}|(KC')^c_{A,\vec{B}}E)_{\wedge \Omega_\mathcal{P}|A,\vec{B},\Phi,\Psi} = H^\varepsilon_{\min}(\vec{Z}_A^{\mathrm{KG}}|(KC')^c_{A,\vec{B}}E)_{\wedge \Omega_{\mathrm{S1}} \cap \Omega_{\mathrm{EC}}|A,\vec{B},\Phi,\Psi}$$
$$\geq H^\varepsilon_{\min}(\vec{Z}_A^{\mathrm{KG}}|(KC')^c_{A,\vec{B}}E)_{\wedge \Omega_{\mathrm{S1}}|A,\vec{B},\Phi,\Psi}.$$
(S51)

We can now consider the well-defined state $\rho_{A^{\mathrm{KG}}B^{\mathrm{KG}}(KC')^c_{A,\vec{B}}E \wedge \Omega_{\mathrm{S1}}|A,\vec{B},\Phi,\Psi}$, where $A^{\mathrm{KG}}, B^{\mathrm{KG}}$ are the quantum systems of Alice and the Bobs, respectively, which are measured in the key generation rounds. For the sake of applying the

---

[6] This fact is often overlooked in QKD security proofs, while [47] accounts for it, see discussion in Lemma 5's proof.

uncertainty relation, we actually consider the state $\rho_{A^{\mathrm{KG}}B^{\mathrm{KG}}(KC'\vec{X}^{\mathrm{KG}})^c_{A,\vec{B}}E \wedge \Omega_{\mathrm{S}1}|A,\vec{B},\Phi,\Psi}$, where we copied the register $(\vec{X}^{\mathrm{KG}})^c_{A,\vec{B}}$, originally contained in $(C')^c_{A,\vec{B}}$, in a new register. We are allowed to do so since classical information can be copied for free. The register $(\vec{X}^{\mathrm{KG}})^c_{A,\vec{B}}$ represents the bits that the non-participants would announce in step 6 of Protocol 2 if the key generation rounds had been labelled as test rounds.

By applying the uncertainty relation for smooth entropies [49] and the data-processing inequality, we obtain the following lower bound:

$$
\begin{aligned}
H^\varepsilon_{\min}(\vec{Z}^{\mathrm{KG}}_A|(KC')^c_{A,\vec{B}}E)_{\wedge\Omega_{\mathrm{S}1}|A,\vec{B},\Phi,\Psi} &\geq L(1-p) - H^\varepsilon_{\max}(\vec{X}^{\mathrm{KG}}_A|B^{\mathrm{KG}}(\vec{X}^{\mathrm{KG}})^c_{A,\vec{B}})_{\wedge\Omega_{\mathrm{S}1}|A,\vec{B},\Phi,\Psi} \\
&\geq L(1-p) - H^\varepsilon_{\max}(\vec{X}^{\mathrm{KG}}_A|\vec{X}^{\mathrm{KG}}_{\vec{B}}(\vec{X}^{\mathrm{KG}})^c_{A,\vec{B}})_{\wedge\Omega_{\mathrm{S}1}|A,\vec{B},\Phi,\Psi} \\
&\geq L(1-p) - H^\varepsilon_{\max}(\vec{X}^{\mathrm{KG}}_A|\vec{X}^{\mathrm{KG}}_{\oplus A^c})_{\wedge\Omega_{\mathrm{S}1}|A,\vec{B},\Phi,\Psi},
\end{aligned}
\tag{S52}
$$

where $\vec{X}^{\mathrm{KG}}_A$ ($\vec{X}^{\mathrm{KG}}_{\vec{B}}$) is the string of $X$ outcomes Alice (the Bobs) would obtain had she (they) measured in the $X$ basis the quantum systems devoted to key generation and where $\vec{X}^{\mathrm{KG}}_{\oplus A^c}$ is defined in (S20). By employing the result of Lemma 4 upon fixing $\varepsilon = \varepsilon_x$, we obtain the following lower bound on the smooth min-entropy:

$$
H^{\varepsilon_x}_{\min}(\vec{Z}^{\mathrm{KG}}_A|(KC')^c_{A,\vec{B}}E)_{\wedge\Omega_{\mathrm{S}1}|A,\vec{B},\Phi,\Psi} \geq L(1-p)[1 - h(Q_X + \gamma(Q_X))].
\tag{S53}
$$

By combining (S53) with (S51), (S50) and (S49), we obtain the following lower bound on the smooth min-entropy contained in (S48):

$$
H^{\varepsilon_x}_{\min}(\vec{Z}^{\mathrm{KG}}_A|(KC'C^{\mathrm{EC}})^c_{A,\vec{B}}E)_{\wedge\Omega_{\mathcal{P}}|A,\vec{B},\Phi,\Psi} \geq L(1-p)[1 - h(Q_X + \gamma(Q_X)) - h(Q_Z)] - \log_2 \frac{2(n-1)}{\varepsilon_{\mathrm{EC}}}.
\tag{S54}
$$

At the same time, from Eq. (D2) we get:

$$
\ell \leq L(1-p)\left[1 - h\left(Q_X + \gamma(Q_X)\right) - h(Q_Z)\right] - \log_2 \frac{2(n-1)}{\varepsilon_{\mathrm{EC}}} - 2\log_2 \frac{1}{2\varepsilon_{\mathrm{PA}}}.
\tag{S55}
$$

By employing (S54) and (S55) in (S48), we can upper bound the rhs of (S48) as follows:

$$
\left\| \rho^f_{K_A(KC)^c_{A,\vec{B}}E \wedge \Omega_{\mathcal{P}}|A,\vec{B},\Phi,\Psi} - \tau_{K_A} \otimes \rho^f_{(KC)^c_{A,\vec{B}}E \wedge \Omega_{\mathcal{P}}|A,\vec{B},\Phi,\Psi} \right\|_{\mathrm{tr}} \leq 2\varepsilon_x + \varepsilon_{\mathrm{PA}}.
\tag{S56}
$$

The last expression is an upper bound on the lhs of the secrecy condition (S2) –conditioned on $\Psi$– and is valid for any choice of sender or receiver. This concludes the proof. $\qquad\square$

**Lemma 16.** The fully-ACKA protocol (Protocol 2) is $\varepsilon_{\mathrm{CKA}}$-CKA-secure, with $\varepsilon_{\mathrm{CKA}} = \varepsilon_{\mathrm{EC}} + 2\varepsilon_x + \varepsilon_{\mathrm{PA}} + 2^{-r_V} + (n-1)2^{-r_N} + 2(n-1)\varepsilon_{\mathrm{enc}}$.

*Proof.* We first decompose the lhs of the CKA-security condition Eq. (B12) into two parts: one due to the event $\Psi$ and the other one due to $\Psi^c$. Then we bound the first part with Lemma 1 together with the results of Lemmas 14 and 15.

The lhs of Eq. (B12) can be decomposed and bounded as follows (we ignore for now the maximization over the

sender and receivers):

$$
\begin{aligned}
\text{Eq. (20)} =& \left\| \rho^f_{KCc^c_{i,\vec{j}} E \wedge \Omega_\mathcal{P} | i, \vec{j}, \Phi} - \tau_{K_{i,\vec{j}}} \otimes \rho^f_{(KC)^c_{i,\vec{j}} E \wedge \Omega_\mathcal{P} | i, \vec{j}, \Phi} \right\|_{\text{tr}} + \Pr[\Omega^c_\mathcal{P} \cap \Gamma^c_\mathcal{P} | i, \vec{j}, \Phi] \\
\leq& \left\| \rho^f_{KCc^c_{i,\vec{j}} E \wedge \Omega_\mathcal{P} \cap \Psi | i, \vec{j}, \Phi} - \tau_{K_{i,\vec{j}}} \otimes \rho^f_{(KC)^c_{i,\vec{j}} E \wedge \Omega_\mathcal{P} \cap \Psi | i, \vec{j}, \Phi} \right\|_{\text{tr}} + \Pr[\Omega^c_\mathcal{P} \cap \Gamma^c_\mathcal{P} \cap \Psi | i, \vec{j}, \Phi] \\
&+ \left\| \rho^f_{KCc^c_{i,\vec{j}} E \wedge \Omega_\mathcal{P} \cap \Psi^c | i, \vec{j}, \Phi} - \tau_{K_{i,\vec{j}}} \otimes \rho^f_{(KC)^c_{i,\vec{j}} E \wedge \Omega_\mathcal{P} \cap \Psi^c | i, \vec{j}, \Phi} \right\|_{\text{tr}} + \Pr[\Omega^c_\mathcal{P} \cap \Gamma^c_\mathcal{P} \cap \Psi^c | i, \vec{j}, \Phi] \\
=& \Pr[\Omega_\mathcal{P} \cap \Psi | i, \vec{j}, \Phi] \left\| \rho^f_{KCc^c_{i,\vec{j}} E | i, \vec{j}, \Phi, \Omega_\mathcal{P}, \Psi} - \tau_{K_{i,\vec{j}}} \otimes \rho^f_{(KC)^c_{i,\vec{j}} E | i, \vec{j}, \Phi, \Omega_\mathcal{P}, \Psi} \right\|_{\text{tr}} + \Pr[\Omega^c_\mathcal{P} \cap \Gamma^c_\mathcal{P} \cap \Psi | i, \vec{j}, \Phi] \\
&+ \Pr[\Omega_\mathcal{P} \cap \Psi^c | i, \vec{j}, \Phi] \left\| \rho^f_{KCc^c_{i,\vec{j}} E | i, \vec{j}, \Phi, \Omega_\mathcal{P}, \Psi^c} - \tau_{K_{i,\vec{j}}} \otimes \rho^f_{(KC)^c_{i,\vec{j}} E | i, \vec{j}, \Phi, \Omega_\mathcal{P}, \Psi^c} \right\|_{\text{tr}} + \Pr[\Omega^c_\mathcal{P} \cap \Gamma^c_\mathcal{P} \cap \Psi^c | i, \vec{j}, \Phi] \\
\leq& \Pr[\Omega_\mathcal{P} | i, \vec{j}, \Phi, \Psi] \left\| \rho^f_{KCc^c_{i,\vec{j}} E | i, \vec{j}, \Phi, \Omega_\mathcal{P}, \Psi} - \tau_{K_{i,\vec{j}}} \otimes \rho^f_{(KC)^c_{i,\vec{j}} E | i, \vec{j}, \Phi, \Omega_\mathcal{P}, \Psi} \right\|_{\text{tr}} + \Pr[\Omega^c_\mathcal{P} \cap \Gamma^c_\mathcal{P} | i, \vec{j}, \Phi, \Psi] \\
&+ \Pr[\Omega_\mathcal{P} \cap \Psi^c | i, \vec{j}, \Phi] + \Pr[\Omega^c_\mathcal{P} \cap \Gamma^c_\mathcal{P} \cap \Psi^c | i, \vec{j}, \Phi] \\
\leq& \Pr[\Omega_\mathcal{P} | i, \vec{j}, \Phi, \Psi] \left\| \rho^f_{KCc^c_{i,\vec{j}} E | i, \vec{j}, \Phi, \Omega_\mathcal{P}, \Psi} - \tau_{K_{i,\vec{j}}} \otimes \rho^f_{(KC)^c_{i,\vec{j}} E | i, \vec{j}, \Phi, \Omega_\mathcal{P}, \Psi} \right\|_{\text{tr}} + \Pr[\Omega^c_\mathcal{P} \cap \Gamma^c_\mathcal{P} | i, \vec{j}, \Phi, \Psi] + \Pr[\Psi^c | i, \vec{j}, \Phi] \\
\leq& \Pr[\Omega_\mathcal{P} | i, \vec{j}, \Phi, \Psi] \left\| \rho^f_{KCc^c_{i,\vec{j}} E | i, \vec{j}, \Phi, \Omega_\mathcal{P}, \Psi} - \tau_{K_{i,\vec{j}}} \otimes \rho^f_{(KC)^c_{i,\vec{j}} E | i, \vec{j}, \Phi, \Omega_\mathcal{P}, \Psi} \right\|_{\text{tr}} + \Pr[\Omega^c_\mathcal{P} \cap \Gamma^c_\mathcal{P} | i, \vec{j}, \Phi, \Psi] \\
&+ 2^{-r_V} + (n-1) 2^{-r_N} + 2(n-1)\varepsilon_{\text{enc}},
\end{aligned}
\tag{S57}
$$

where in the first inequality we used the convexity of the trace distance, in the second inequality we used the fact that the trace distance of two normalized states is bounded by one, and in the last inequality we used the failure probability of Veto ($2^{-r_V}$), of the Notification for every Bob ($(n-1)2^{-r_N}$) and of the decoding function $G$ ($2(n-1)\varepsilon_{\text{enc}}$). Note that since we want to derive a bound valid for every choice of sender and receivers, we maximize the number of Bobs to $m = n-1$. We now use Lemma 1 (applied to a fully-ACKA protocol conditioned on the event $\Psi$) in combination with the results of Lemmas 14 and 15 to bound the first two terms in the last expression:

$$
\text{Eq. (20)} \leq \varepsilon_{\text{EC}} + 2\varepsilon_x + \varepsilon_{\text{PA}} + 2^{-r_V} + (n-1)2^{-r_N} + +2(n-1)\varepsilon_{\text{enc}},
\tag{S58}
$$

which concludes the proof. $\square$

### C. Weak anonymity

As already discussed in Sec. II, our fully-ACKA protocol based on GHZ states (Protocol 2) does not satisfy the anonymity condition given in Eq. (B14). This is because the reduced state of a receiver could depend on the identities of the other participants and thus not satisfy anonymity as intended by Definition 4. This happens if some of the information collected by a receiver during the protocol –stored in their reduced state– depends on the identities of the other participants.

We provided two examples where this could happen, namely: the information on the sender's raw key (the string of $Z$ outcomes of the sender), which is recovered after EC, and the information on whether the protocol aborts or not. Indeed, exploiting multipartite entanglement for establishing a conference key while maintaining a certain robustness to noise, allows for asymmetries in the source state preparation. Such asymmetries can cause the two pieces of information mentioned above to depend on the specific set of participants. However, we stressed that only the combination of such information with a detailed knowledge of the shared multipartite states can lead to a leakage of the participants' identities[7].

For this, we introduced the notion of weak-anonymity in Definition 6. In the following we show that Protocol 2 satisfies the weak-anonymity definition.

---

[7] To exemplify strategies that would break anonymity if a receiver is aware of the details about the states prepared by Eve, consider the following:

1. Eve prepares, in a small fraction of the rounds, states with deterministic outcomes in the $Z$ basis. When the receiver reconstructs the sender's raw key, they can determine who is the sender by checking the sender's key bits corresponding to the rounds with deterministic $Z$ outcomes and comparing them with their knowledge of the states prepared by Eve.

2. Eve prepares $|\text{GHZ}_{n-1}\rangle_{t^c} \otimes |+\rangle_t$ in every round of the protocol. The state yields $Q_X^{\text{obs}} = 0$ so it passes the verification of secrecy (step 7) but it aborts in error correction if party $t$ is a participant. A receiver, based on the information on whether the error correction aborts or not, can deduce if party $t$ is a participant or not.

**Lemma 17.** The fully-ACKA protocol (Protocol 2) is $\varepsilon_{\text{wAN}}$-weak-anonymous according to Definition 6, with $\varepsilon_{\text{wAN}} = 2^{-(r_V-1)} + (n-1)(2^{-r_n} + 3\varepsilon_{\text{enc}}) + 2(\varepsilon_{\text{PA}} + 2\varepsilon_x)$.

*Proof.* In order to prove weak-anonymity of our fully-ACKA protocol (Protocol 2), we need show that it satisfies the anonymity condition with respect to non-participants and Eve, Eq. (B15), and the anonymity conditions with respect to honest but curious receivers, Eq. (B16) and Eq. (B17).

In the proof we will make use of the Quantum Leftover Hash Lemma.

**Proposition 2** (Quantum Leftover Hash Lemma [47].) Let $\sigma_{XD}$ be a sub-normalized quantum state classical on $X$. Let $\mathcal{F}$ denote a two-universal family of hash functions from $X = \{0,1\}^n$ to $C = \{0,1\}^s$. Then, for every $\varepsilon \in [0, \sqrt{\text{Tr}[\sigma_{XD}]})$, it holds:

$$\|\sigma_{CDF} - \tau_C \otimes \sigma_{DF}\|_{\text{tr}} \leq \frac{1}{2} 2^{-\frac{1}{2}(H^{\varepsilon}_{\min}(X|D)_{\sigma} - s)} + 2\varepsilon,$$
(S59)

where the register $F$ stores the randomly chosen hash function $f \in_R \mathcal{F}$, the state $\sigma_{CDF}$ is obtained from $\sigma_{XDF}$ by applying the random hash function to the register $X$: $\sigma_{CDF} = \sum_{x,f} \frac{1}{|\mathcal{F}|} p(x) |f(x)\rangle\langle f(x)| \otimes \sigma_{D|x} \otimes |f\rangle\langle f|$, and $\tau_C := \frac{1}{2^s} \sum_{c \in C} |c\rangle\langle c|$.

*Anonymity with respect to non-participants and Eve:* To prove condition Eq. (B15), we need to show that the reduced state of non-participants and Eve in the fully-ACKA protocol is close to a state which is anonymous according to Definition 3, namely, independent of the identities of honest participants. Conversely, no condition is imposed by Definition 3 in the case of dishonest participants, thus we can avoid analyzing this case.

As previously defined in the proof of Lemma 15, let $\Omega_{\text{S1}}$ be the event that the fully-ACKA protocol does not abort until step 7 is completed, and let $\Psi$ be the event in which Veto in step 7, Notification in TKD, and the decoding function $G$ in TKD and error correction are successful –i.e., Veto returns 1 when at least one party inputs 1, every party $s$ picked by Alice is notified in TKD, and $G(F(\vec{x}) \oplus \vec{b}) = \top$ for every $\vec{b} \neq 0$.

We first deal with the instances where only one party applied to be the sender. Then we can decompose the reduced state of non-participants and Eve in the fully-ACKA protocol as follows:

$$\begin{aligned}
\rho^f_{(PKC)^c_{A,\vec{B}}E|A,\vec{B}} = &\Pr[\Gamma|A,\vec{B}]\rho^f_{(PKC)^c_{A,\vec{B}}E|A,\vec{B},\Gamma} + \Pr[\Gamma^c \cap \Phi^c|A,\vec{B}]\rho^f_{(PKC)^c_{A,\vec{B}}E|A,\vec{B},\Gamma^c,\Phi^c} \\
&+ \Pr[\Phi \cap \Psi^c|A,\vec{B}]\rho^f_{(PKC)^c_{A,\vec{B}}E|A,\vec{B},\Phi,\Psi^c} \\
&+ \Pr[\Phi \cap \Psi \cap \Omega^c_{\text{S1}}|A,\vec{B}]\rho^f_{(PKC)^c_{A,\vec{B}}E|A,\vec{B},\Phi,\Psi,\Omega^c_{\text{S1}}} \\
&+ \Pr[\Phi \cap \Psi \cap \Omega_{\text{S1}}|A,\vec{B}]\rho^f_{(PKC)^c_{A,\vec{B}}E|A,\vec{B},\Phi,\Psi,\Omega_{\text{S1}}}.
\end{aligned}$$
(S60)

Now, we first note that all the probabilities appearing in the decomposition of the final state in (S60) are independent of the identity of the participants. Indeed, all the events contained in such probabilities are independent of the identities of the sender and the receivers. For this reason, we can decompose a generic anonymous state using the same probabilities as follows:

$$\begin{aligned}
\sigma^{\mathcal{D}}_{(PKC)^c_{A,\vec{B}}E|A,\vec{B}} = &\Pr[\Gamma|A,\vec{B}]\sigma^{\mathcal{D}}_{(PKC)^c_{A,\vec{B}}E|A,\vec{B},\Gamma} + \Pr[\Gamma^c \cap \Phi^c|A,\vec{B}]\sigma^{\mathcal{D}}_{(PKC)^c_{A,\vec{B}}E|A,\vec{B},\Gamma^c,\Phi^c} \\
&+ \Pr[\Phi \cap \Psi^c|A,\vec{B}]\sigma^{\mathcal{D}}_{(PKC)^c_{A,\vec{B}}E|A,\vec{B},\Phi,\Psi^c} \\
&+ \Pr[\Phi \cap \Psi \cap \Omega^c_{\text{S1}}|A,\vec{B}]\sigma^{\mathcal{D}}_{(PKC)^c_{A,\vec{B}}E|A,\vec{B},\Phi,\Psi,\Omega^c_{\text{S1}}} \\
&+ \Pr[\Phi \cap \Psi \cap \Omega_{\text{S1}}|A,\vec{B}]\sigma^{\mathcal{D}}_{(PKC)^c_{A,\vec{B}}E|A,\vec{B},\Phi,\Psi,\Omega_{\text{S1}}},
\end{aligned}$$
(S61)

for arbitrary states $\left\{\sigma^{\mathcal{D}}_{(PKC)^c_{A,\vec{B}}E|A,\vec{B},e}\right\}_e$, $e \in \{\Gamma, \Gamma^c, \Phi, \Phi^c, \Omega_{\text{S1}}, \Omega^c_{\text{S1}}, \Psi, \Psi^c\}$, such that $\sigma^{\mathcal{D}}_{(PKC)^c_{A,\vec{B}}E|A,\vec{B},e}$ satisfies the

anonymity definition for ACKA protocols (Definition 3). Thus, by the convexity of the trace distance, we can bound:

$$\left\| \rho^f_{(PKC)^c_{A,\vec{B}} E | A, \vec{B}} - \sigma^{\mathcal{D}}_{(PKC)^c_{A,\vec{B}} E | A, \vec{B}} \right\|_{\mathrm{tr}}$$

$$\leq \Pr[\Gamma | A, \vec{B}] \left\| \rho^f_{(PKC)^c_{A,\vec{B}} E | A, \vec{B}, \Gamma} - \sigma^{\mathcal{D}}_{(PKC)^c_{A,\vec{B}} E | A, \vec{B}, \Gamma} \right\|_{\mathrm{tr}} \tag{S62}$$

$$+ \Pr[\Gamma^c \cap \Phi^c | A, \vec{B}] \left\| \rho^f_{(PKC)^c_{A,\vec{B}} E | A, \vec{B}, \Gamma^c, \Phi^c} - \sigma^{\mathcal{D}}_{(PKC)^c_{A,\vec{B}} E | A, \vec{B}, \Gamma^c \cap \Phi^c} \right\|_{\mathrm{tr}} \tag{S63}$$

$$+ \Pr[\Phi \cap \Psi^c | A, \vec{B}] \left\| \rho^f_{(PKC)^c_{A,\vec{B}} E | A, \vec{B}, \Phi \cap \Psi^c} - \sigma^{\mathcal{D}}_{(PKC)^c_{A,\vec{B}} E | A, \vec{B}, \Phi, \Psi^c} \right\|_{\mathrm{tr}} \tag{S64}$$

$$+ \Pr[\Phi \cap \Psi \cap \Omega^c_{\mathrm{S1}} | A, \vec{B}] \left\| \rho^f_{(PKC)^c_{A,\vec{B}} E | A, \vec{B}, \Phi, \Psi, \Omega^c_{\mathrm{S1}}} - \sigma^{\mathcal{D}}_{(PKC)^c_{A,\vec{B}} E | A, \vec{B}, \Phi, \Psi, \Omega^c_{\mathrm{S1}}} \right\|_{\mathrm{tr}} \tag{S65}$$

$$+ \Pr[\Phi \cap \Psi \cap \Omega_{\mathrm{S1}} | A, \vec{B}] \left\| \rho^f_{(PKC)^c_{A,\vec{B}} E | A, \vec{B}, \Phi, \Psi, \Omega_{\mathrm{S1}}} - \sigma^{\mathcal{D}}_{(PKC)^c_{A,\vec{B}} E | A, \vec{B}, \Phi, \Psi, \Omega_{\mathrm{S1}}} \right\|_{\mathrm{tr}}. \tag{S66}$$

In order to bound the terms (S62)-(S66), we will either argue that the trace distance between the final state of fully-ACKA and the anonymous state is small, or we will show that the probability of a certain event is small and trivially bound the corresponding trace distance by 1.

In order to bound (S62), we use the fact that $\Gamma$ corresponds to the event where the fully-ACKA protocol aborts in step 1, when fully-ACKA-ID (Protocol 7) is executed. This event does not leak any information about the participants' identities thanks to the properties of the underlying Collision Detection, Parity and Veto protocols [20] that compose the fully-ACKA-ID protocol. Therefore we have that $\rho^f_{PKCE|A,\vec{B},\Gamma} = \sigma^{\mathcal{D}}_{PKCE|A,\vec{B},\Gamma}$ for some state $\sigma^{\mathcal{D}}_{PKCE|A,\vec{B},\Gamma}$ satisfying Definition 3 and therefore:

$$(S62) = 0. \tag{S67}$$

For the term (S63), we note that a failure of fully-ACKA-ID could imply, in particular, that a Bob is not aware that he is a participant. This leads to the leakage of the testing key in TKD (step 3 of fully-ACKA), since there are Parity rounds where Alice inputs the key $\vec{k}_l$ but the notified Bob –considering himself a non-participant– inputs $\vec{0}$, hence making the public Parity output coincide with $\vec{k}_l$. This opens an anonymity loophole[8]. Therefore, the trace distance in (S63) is upper bounded by 1 while we can bound the probability that this event happens. From the proof of Lemma 13 we have that

$$(S63) \leq \Pr[\Gamma^c \cap \Phi^c | A, \vec{B}] \leq 2^{-r_V} + (n-1)\varepsilon_{\mathrm{enc}}. \tag{S68}$$

The term (S64) corresponds to the case in which the protocol does not abort in step 1 and the identities are correctly assigned, but either the Notification in TKD, the Veto in step 7, or the decoding with $G$ (in TKD and EC) fail. Such failures can lead to anonymity loopholes. Indeed, if Notification fails, the key $\vec{k}_l$ can be leaked: the Bob who was supposed to be notified inputs $\vec{0}$ in the Parity protocols of TKD. And if the protocol does not abort in step 7, the leaked key can be used by dishonest parties to learn the participants' identities, as highlighted in the previous paragraph. Note, moreover, that the failure of Notification alone can lead to an anonymity issue[9], even when the protocol aborts in step 7. At the same time, a failure of Veto alone can lead to an identity leak[10]. Therefore, we trivially bound the trace distance of (S64) and bound the corresponding probability as following:

$$(S64) \leq \Pr[\Phi \cap \Psi^c | A, \vec{B}] \leq \Pr[\Psi^c | A, \vec{B}] \leq 2^{-r_V} + (n-1)2^{-r_N} + 2(n-1)\varepsilon_{\mathrm{enc}}, \tag{S69}$$

where for the last inequality we use the union bound and bound the failure of Veto by $2^{-r_V}$, of Notification of any Bob by $(n-1)2^{-r_N}$ and of the decoding function $G$ when used by Alice or the Bobs in TKD and error correction by $2(n-1)\varepsilon_{\mathrm{enc}}$.

---

[8] With the knowledge of the testing key, a dishonest party can measure in the $Z$ basis in the key generation rounds and recover Alice's raw key after error correction. Moreover they can also obtain the information on whether the protocol aborts after error correction. This information can leak the participants identities, if, for example, the eavesdropper employs the strategies exemplified in Footnote 7.

[9] Suppose that a dishonest party actively tampers with the notification of party $s$ in the Notification stage of TKD. In particular, the dishonest party does not input zero in the Parity protocols forming Notification when a specific party, say party $t$, is the last to broadcast. If the output $\vec{o}$ of the following Parity protocols is equal to $\vec{k}_l$ –which can be verified by the dishonest party by decoding the last $|\vec{r}_l|$ bits of $\vec{o}$ and not obtaining $\top$–, the dishonest party can guess the identity of party $t$ with a probability higher than the trivial one.

[10] This can happen e.g. if Eve distributes eigenstates of the $Z$ basis independently to every party, in each round, such that she knows beforehand the raw key of every party. These source states would certainly lead to a high test error rate $Q^{\mathrm{obs}}_X$ and would prompt Alice to input 1 in the Veto of step 7. However, if Veto fails to abort the protocol, the syndrome and hash computed by Alice from her raw key in error correction can help Eve to distinguish which party produced such strings, thus revealing the sender's identity.

The term (S65) corresponds to the case where identities are correctly assigned, the Notification in TKD, the Veto in step 7 and the decoding with $G$ work perfectly, but the protocol aborts in step 7, which means that Veto outputs 1. Since fully-ACKA-ID and Notification succeed, no information is leaked to the non-participants and Eve from steps 1 to 6. Now, the Veto in step 7 reveals information about the observed test error rate $Q_X^{\text{obs}}$ and the verification of correctness of the strings distributed in TKD (Protocol 8). From the fact that $Q_X^{\text{obs}}$ and the verification performed during TKD are invariant under permutation of the parties' roles[11], we deduce that the output of the Veto in step 7 does not reveal any information on the parties' identities. Therefore, under these circumstances, the fully-ACKA protocol behaves ideally and its conditional final state coincides with a state satisfying Definition 3:

$$\rho^f_{(PKC)^c_{A,\vec{B}}E|A,\vec{B},\Phi,\Psi,\Omega^c_{S1}} = \sigma^{\mathcal{D}}_{(PKC)^c_{A,\vec{B}}E|A,\vec{B},\Phi,\Psi,\Omega^c_{S1}}. \tag{S70}$$

Thus, it holds:

$$(S65) = 0. \tag{S71}$$

The term (S66) corresponds to the case where the protocol does not abort in step 7 and fully-ACKA-ID, Veto, Notification and decoding with $G$ are successful. Then, the protocol reaches step 8 –the error correction (EC)–, where it treats differently participants and non-participants, hence allowing for potential anonymity issues. Indeed, the error correction information publicly communicated during step 8 should allow the Bobs to retrieve Alice's raw key, but should not allow the non-participants and Eve to do so –otherwise an anonymity loophole arises, as discussed above.

In order to bound (S66), we show that the reduced state of the non-participants and Eve, despite including a copy of the error correction information, is close to a state that is independent of the identity of the remaining parties and that satisfies Definition 3.

To set things in a precise manner, we specify the different registers composing the total register $C^c_{A,\vec{B}}$, which contains the classical inputs and outputs of every non-participant at various steps of the protocol. Let $C'^c_{A\vec{B}}$ denote the classical inputs and outputs of the non-participants up to step 7 of fully-ACKA. Recall that the information $C^{\text{EC}}$ exchanged during error correction (step 8) is composed of some public information: the syndrome $\vec{y}$, the verification hash of Alice's raw key $\vec{h}_A$ and the hash functions $f_Y$ and $f_H$ chosen among 2-universal families, which are stored in the registers $C^Y, C^H, F^Y$ and $F^H$, respectively. Let us denote $C''$ the register containing the remaining information exchanged in the last steps of error correction (steps 6 to 9 in Protocol 10). There, the participants agree on whether to abort the protocol but this information is kept private and is only stored in their personal registers. Conversely, the inputs and outputs of the non-participants are either completely random or fixed, thus their registers $C''^c_{A,\vec{B}}$ are uncorrelated from everything else. Finally, the only information added in step 9 of fully-ACKA is the randomly chosen 2-universal hash function, stored in the public register $F^{\text{PA}}$. This is used by the participants to compute their secret conference key, hence it is uncorrelated from the registers of the non-participants. We thus have the following set of registers for the non-participants: $C^c_{A,\vec{B}} = C'^c_{A,\vec{B}}C^Y C^H F^Y F^H C''^c_{A,\vec{B}}F^{\text{PA}}$.

In view of the above considerations, we can recast the reduced state of non-participants and Eve appearing in the lhs of (S66) as following:

$$\rho^f_{(PKC)^c_{A,\vec{B}}E|A,\vec{B},\Phi,\Psi,\Omega_{S1}} = |\bot\rangle\langle\bot|_{P^c_{A,\vec{B}}} \otimes \rho^f_{C''^c_{A,\vec{B}}} \otimes \tau_{F^{\text{PA}}} \otimes \rho^f_{(KC)^c_{A,\vec{B}}C^Y C^H F^Y F^H E|A,\vec{B},\Phi,\Psi,\Omega_{S1}}, \tag{S72}$$

where we used the fact that, conditioned on the event $\Phi$, we have: $\rho^f_{P|A,\vec{B},\Phi} = \xi^{(A,\vec{B})}_P$, where $\xi^{(i,\vec{j})}_P$ is given in Eq. (B4). The anonymous state $\sigma^{\mathcal{D}}_{(PKC)^c_{A,\vec{B}}E|A,\vec{B},\Phi,\Psi,\Omega_{S1}}$ that we select in order to upper bound (S66) with a small quantity is defined starting from the final state of the protocol (S72). However, in the anonymous state we remove the error correction information $C^Y$ and $C^H$ that could cause identity leaks and replace it with uncorrelated random outputs. Thus, we define the following anonymous state:

$$\sigma^{\mathcal{D}}_{(PKC)^c_{A,\vec{B}}E|A,\vec{B},\Phi,\Psi,\Omega_{S1}} = |\bot\rangle\langle\bot|_{P^c_{A,\vec{B}}} \otimes \tau_{C^Y} \otimes \tau_{C^H} \otimes \tau_{F^Y} \otimes \tau_{F^H} \otimes \rho^f_{C''^c_{A,\vec{B}}} \otimes \tau_{F^{\text{PA}}} \otimes \rho^f_{(KC)^c_{A,\vec{B}}E|A,\vec{B},\Phi,\Psi,\Omega_{S1}}, \tag{S73}$$

and claim that it satisfies Definition 3. As a matter of fact, the last state remains unchanged from step 7 and is thus anonymous with the analogous argument used for (S70), while the other states are clearly independent of the participants' identities.

---

[11] In the TKD protocol (Protocol 8), any attempt of tampering with the strings distributed by Alice is detected either by the Bobs or by Alice, in the exact same way.

We employ (S72) and (S73) in (S66), together with the property that $\|A \otimes B_1 - A \otimes B_2\|_{\mathrm{tr}} = \|B_1 - B_2\|_{\mathrm{tr}}$ for any normalized quantum state $A$, to remove the registers $P^c_{A,\vec{B}}$, $F^{\mathrm{PA}}$ and $C''^c_{A,\vec{B}}$ from the trace distance:

$$(\mathrm{S66}) = \Pr[\Phi \cap \Psi \cap \Omega_{\mathrm{S1}}|A,\vec{B}] \left\| \rho^f_{(KC')^c_{A,\vec{B}}C^Y C^H F^Y F^H E|A,\vec{B},\Phi,\Psi,\Omega_{\mathrm{S1}}} - \tau_{C^Y} \otimes \tau_{C^H} \otimes \tau_{F^Y} \otimes \tau_{F^H} \otimes \rho^f_{(KC')^c_{A,\vec{B}}E|A,\vec{B},\Phi,\Psi,\Omega_{\mathrm{S1}}} \right\|_{\mathrm{tr}}$$

$$\leq \left\| \rho^f_{(KC')^c_{A,\vec{B}}C^Y C^H F^Y F^H E \wedge \Omega_{\mathrm{S1}}|A,\vec{B},\Phi,\Psi} - \tau_{C^Y} \otimes \tau_{C^H} \otimes \tau_{F^Y} \otimes \tau_{F^H} \otimes \rho^f_{(KC')^c_{A,\vec{B}}E \wedge \Omega_{\mathrm{S1}}|A,\vec{B},\Phi,\Psi} \right\|_{\mathrm{tr}}, \tag{S74}$$

where in the inequality we used $\Pr[\Phi \cap \Psi \cap \Omega_{\mathrm{S1}}|A,\vec{B}] \leq \Pr[\Omega_{\mathrm{S1}}|A,\vec{B},\Phi,\Psi]$ and the definition of sub-normalized states. We proceed by splitting the rhs of (S74) into two terms, thanks to the convexity of the trace distance:

$$(\mathrm{S66}) \leq \underbrace{\left\| \rho^f_{(KC')^c_{A,\vec{B}}C^Y C^H F^Y F^H E \wedge \Omega_{\mathrm{S1}}|A,\vec{B},\Phi,\Psi} - \tau_{C^H} \otimes \tau_{F^H} \otimes \rho^f_{(KC')^c_{A,\vec{B}}C^Y F^Y E \wedge \Omega_{\mathrm{S1}}|A,\vec{B},\Phi,\Psi} \right\|_{\mathrm{tr}}}_{(I)}$$

$$+ \underbrace{\left\| \rho^f_{(KC')^c_{A,\vec{B}}C^Y F^Y E \wedge \Omega_{\mathrm{S1}}|A,\vec{B},\Phi,\Psi} - \tau_{C^Y} \otimes \tau_{F^Y} \otimes \rho^f_{(KC')^c_{A,\vec{B}}E \wedge \Omega_{\mathrm{S1}}|A,\vec{B},\Phi,\Psi} \right\|_{\mathrm{tr}}}_{(II)}, \tag{S75}$$

where in the second term of the sum we traced out the states $\tau_{C^H} \otimes \tau_{F^H}$ thanks to the property $\|A \otimes B_1 - A \otimes B_2\|_{\mathrm{tr}} = \|B_1 - B_2\|_{\mathrm{tr}}$.

We now use Proposition 2 to bound the trace distances in (S75). In doing so, we keep in mind that the error correction information $C^Y$ and $C^H$ is generated from Alice's raw key $\vec{Z}^{\mathrm{KG}}_A$ and that the bit-lengths of the syndrome and of the hash, $\log_2 |C^Y|$ and $\log_2 |C^H|$, are given by:

$$\log_2 |C^Y| = L(1-p)h(Q_Z) \tag{S76}$$

$$\log_2 |C^H| = \log_2 \frac{2(n-1)}{\varepsilon_{\mathrm{EC}}}. \tag{S77}$$

By applying Proposition 2 with $\varepsilon = \varepsilon_x$, we obtain:

$$(I) \leq \frac{1}{2} 2^{-\frac{1}{2}\left( H^{\varepsilon_x}_{\min}(\vec{Z}^{\mathrm{KG}}_A|K^c_{A,\vec{B}}C'^c_{A,\vec{B}}C^Y F^Y E)_{\wedge \Omega_{\mathrm{S1}}|A,\vec{B},\Phi,\Psi} - \log_2 \frac{2(n-1)}{\varepsilon_{\mathrm{EC}}} \right)} + 2\varepsilon_x \tag{S78}$$

$$(II) \leq \frac{1}{2} 2^{-\frac{1}{2}\left( H^{\varepsilon_x}_{\min}(\vec{Z}^{\mathrm{KG}}_A|K^c_{A,\vec{B}}C'^c_{A,\vec{B}}E)_{\wedge \Omega_{\mathrm{S1}}|A,\vec{B},\Phi,\Psi} - L(1-p)h(Q_Z) \right)} + 2\varepsilon_x. \tag{S79}$$

We can further bound (S78) and (S79) by deriving lower bounds on the respective smooth min-entropies. For (S79) we can directly use (S53) from the secrecy proof of fully-ACKA and obtain:

$$H^{\varepsilon_x}_{\min}(\vec{Z}^{\mathrm{KG}}_A|K^c_{A,\vec{B}}C'^c_{A,\vec{B}}E)_{\wedge \Omega_{\mathrm{S1}}|A,\vec{B},\Phi,\Psi} \geq L(1-p)[1 - h(Q_X + \gamma(Q_X))]. \tag{S80}$$

For (S78) we can obtain the following lower bound:

$$H^{\varepsilon_x}_{\min}(\vec{Z}^{\mathrm{KG}}_A|K^c_{A,\vec{B}}C'^c_{A,\vec{B}}C^Y F^Y E)_{\wedge \Omega_{\mathrm{S1}}|A,\vec{B},\Phi,\Psi} \geq H^{\varepsilon_x}_{\min}(\vec{Z}^{\mathrm{KG}}_A|K^c_{A,\vec{B}}C'F^Y E)_{\wedge \Omega_{\mathrm{S1}}|A,\vec{B},\Phi,\Psi} - \log_2 |C^Y|$$

$$= H^{\varepsilon_x}_{\min}(\vec{Z}^{\mathrm{KG}}_A|K^c_{A,\vec{B}}C'E)_{\wedge \Omega_{\mathrm{S1}}|A,\vec{B},\Phi,\Psi} - \log_2 |C^Y|$$

$$\geq L(1-p)[1 - h(Q_X + \gamma(Q_X))] - L(1-p)h(Q_Z), \tag{S81}$$

where in the first inequality we use a chain rule for smooth entropies [50], in the equality we remove the register $F^Y$ since it is uncorrelated with the remaining systems, and for the last inequality we again use (S53) and (S76). By employing (S80) and (S81) in (S79) and (S78), respectively, we obtain:

$$(I) \leq \frac{1}{2} 2^{-\frac{1}{2}\left( L(1-p)[1-h(Q_X+\gamma(Q_X))-h(Q_Z)] - \log_2 \frac{2(n-1)}{\varepsilon_{\mathrm{EC}}} \right)} + 2\varepsilon_x \tag{S82}$$

$$(II) \leq \frac{1}{2} 2^{-\frac{1}{2}\left( L(1-p)[1-h(Q_X+\gamma(Q_X))-h(Q_Z)] \right)} + 2\varepsilon_x. \tag{S83}$$

By employing the expression Eq. (D2) for the final conference key length $\ell$ and by assuming that $\ell > 0$, we obtain:

$$(I) \leq \frac{1}{2} 2^{-\frac{1}{2}\left(\ell + 2\log_2\frac{1}{2\varepsilon_{\text{PA}}}\right)} + 2\varepsilon_x \leq \varepsilon_{\text{PA}} + 2\varepsilon_x \tag{S84}$$

$$(II) \leq \frac{1}{2} 2^{-\frac{1}{2}\left(\ell + 2\log_2\frac{1}{2\varepsilon_{\text{PA}}} + \log_2\frac{2(n-1)}{\varepsilon_{\text{EC}}}\right)} + 2\varepsilon_x \leq \varepsilon_{\text{PA}} + 2\varepsilon_x. \tag{S85}$$

We now combine the bounds (S84) and (S85) with (S75) to obtain the following bound on (S66):

$$(S66) \leq 2(2\varepsilon_x + \varepsilon_{\text{PA}}). \tag{S86}$$

By combining the upper bounds derived on the terms (S62)-(S66), we finally obtain the desired bound for the first condition Eq. (B15) of weak-anonymity:

$$\left\| \rho^f_{(PKC)^c_{A,\vec{B}}E|A,\vec{B}} - \sigma^{\mathcal{D}}_{(PKC)^c_{A,\vec{B}}E|A,\vec{B}} \right\|_{\text{tr}} \leq 2^{-(r_V-1)} + (n-1)(2^{-r_N} + 3\varepsilon_{\text{enc}}) + 2(\varepsilon_{\text{PA}} + 2\varepsilon_x). \tag{S87}$$

For the instances in which more than one party applies to be the sender, the fully-ACKA protocol aborts in step 1 (ID aborts) and does not reveal the participants' identities, except with probability bounded by $2^{-r_V}$. Thus we can obtain the following bound:

$$\left\| \rho^f_{(PKC)^c_{\vec{i},\vec{j}}E|\vec{i},\vec{j}} - \sigma^{\mathcal{D}}_{(PKC)^c_{\vec{i},\vec{j}}E|\vec{i},\vec{j}} \right\|_{\text{tr}} \leq 2^{-r_V}. \tag{S88}$$

By combining (S87) and (S88), we conclude that the fully-ACKA protool (Protocol 2) satisfies the first condition of weak-anonymity, Eq. (B15), with

$$\epsilon_{\text{npAN}} = 2^{-(r_V-1)} + (n-1)(2^{-r_N} + 3\varepsilon_{\text{enc}}) + 2(\varepsilon_{\text{PA}} + 2\varepsilon_x). \tag{S89}$$

*Anonymity with respect to honest but curious receivers:* At the end of the protocol, the receivers recover Alice's raw key and also know whether the protocol aborted or not after error correction. In general this information is sensitive and, together with a detailed knowledge of the source state, can lead to the leakage of the participants identities. However, in order to prove weak anonymity, we are interested in the case in which the only previous knowledge the receivers have about the source is that it generates a state that is invariant under permutation of the parties. Under this assumption, the extra knowledge acquired by the receivers during the execution of the protocol does not increase their ability to guess the participants. Therefore the fully-ACKA protocol (Protocol 2) satisfies the anonymity with respect to honest but curious receivers, with $\varepsilon_{\text{pAN}} = 0$.

In conclusion, the fully-ACKA protocol is $\varepsilon_{\text{wAN}}$-weak-anonymous, with $\varepsilon_{\text{wAN}} = 2^{-(r_V-1)} + (n-1)(2^{-r_N} + 3\varepsilon_{\text{enc}}) + 2(\varepsilon_{\text{PA}} + 2\varepsilon_x)$. This concludes the proof. $\qquad\square$

### D. Security

**Lemma 18.** The fully-ACKA protocol (Protocol 2) is $\varepsilon_{\text{tot}}$-secure according to Definition 5 (but with the anonymity condition, Eq. (B14), replaced by Definition 6) with $\varepsilon_{\text{tot}} = 2^{-(r_V-2)} + (n-1)(6\varepsilon_{\text{enc}} + 2^{-(r_N-1)}) + \varepsilon_{\text{EC}} + 6\varepsilon_x + 3\varepsilon_{\text{PA}}$.

*Proof.* The claim follows from combining the results of Lemmas 13, 16 and 17 with Definitions 5 and 6. $\qquad\square$

## VI. SECURITY PROOF OF bifully-ACKA

In this section we prove the security of the bifully-ACKA protocol (Protocol 12).

### A. Integrity

**Lemma 19.** The bifully-ACKA protocol (Protocol 12) is $\varepsilon_{\text{IN}}$-integrous with $\varepsilon_{\text{IN}} = 2^{-r_V} + (n-1)\varepsilon_{\text{enc}}$.

*Proof.* The bifully-ACKA protocol employs the same ID protocol of fully-ACKA (Protocol 7), thus the proof follows analogously. $\qquad\square$

## B. CKA-security

Similarly to the CKA-security proof of the fully-ACKA protocol, we separately address the role played by the Veto in step 4 and the Notification in step 3 of bifully-ACKA. As a matter of fact, if the Veto or the Notification fail, both correctness and secrecy are affected.

Concerning correctness, it can happen that $Bob_l$ is not notified in step 3 when Alice intended to and does not realize this[12]. Or it can happen or that he obtains $\top$ when decoding the conference key $F(\vec{k}_A)$ sent by Alice, but the Veto fails to abort the protocol in step 4. In both cases we have: $\vec{k}_{B_i} \neq \vec{k}_A$. Regarding secrecy, if a Bob is not notified in step 3, the output of the Parity protocols in step 3.3 coincides with Alice's input, $F(\vec{k}_A)$, thus publicly revealing the conference key. The protocol is not secret if Veto fails to abort the protocol in step 4 even if Bob inputs $v_{B_i} = 1$.

Let us denote by $\Psi$ the event in which the Veto in step 4 and the Notification in step 3 are successful –i.e., Veto returns 1 when at least one party inputs 1 and every party $s$ picked by Alice is notified in step 3. We now prove that the bifully-ACKA protocol is both correct and secret conditioned on the event $\Psi$. We then deal with the general case while proving CKA-security in Lemma 22.

**Lemma 20.** The bifully-ACKA protocol (Protocol 12) –conditioned on event $\Psi$– is $\varepsilon_{\text{cor}}$-correct, with $\varepsilon_{\text{cor}} = (n-1)\varepsilon_{\text{enc}}$.

*Proof.* According to the correctness definition (Definition 1), we must compute the probabilities $\Pr[\cup_{l=1}^m \vec{k}_A \neq \vec{k}_{B_l} \cap \Omega_{\mathcal{P}} | A, \vec{B}, \Phi, \Psi]$ and $\Pr[\Omega_{\mathcal{P}}^c \cap \Gamma_{\mathcal{P}}^c | A, \vec{B}, \Phi, \Psi]$, where we relabelled the sender $i$ and the receivers $\vec{j}$ as Alice ($i = A$) and Bobs ($\vec{j} = \vec{B}$) and where we conditioned on $\Psi$. The former is the probability that the conference key of some Bob differs from Alice's and the protocol did not abort for any participant. The latter is the probability that the protocol only aborts for a strict subset of participants.

Since the conference key is transmitted over error-free bipartite private channels, the only way in which the conference key of $Bob_l$ differs from Alice's is if the bits of the string $F(\vec{k}_A)$ transmitted by Alice are flipped. This happens if some party other than Alice and $Bob_l$ does not input $\vec{0}$ in the sequence of Parity protocols in step 3.3. However, the decoding function $G$ applied by $Bob_l$ detects any flip by returning $\top$, except for probability $\varepsilon_{\text{enc}}$. If a flip is detected, the protocol certainly aborts in step 4 (when we condition on $\Psi$).

Therefore, the event in which the protocol does not abort and $Bob_l$'s conference key differs from Alice's occurs if the decoding fails to detect a flip. Since this can happen to every Bob independently, by the union bound we can derive the following upper bound:

$$\Pr[\cup_{l=1}^m \vec{k}_A \neq \vec{k}_{B_l} \cap \Omega_{\mathcal{P}} | A, \vec{B}, \Phi, \Psi] \leq (n-1)\varepsilon_{\text{enc}}, \tag{S90}$$

where we chose the maximum number of Bobs, $m = n - 1$, so that the bound is independent of the choice of sender and receivers. Moreover, the bifully-ACKA protocol can only abort in step 4 and, if it does so, it aborts for every participant –indeed, for every party. Thus we have that:

$$\Pr[\Omega_{\mathcal{P}}^c \cap \Gamma_{\mathcal{P}}^c | A, \vec{B}, \Phi, \Psi] = 0. \tag{S91}$$

By combining (S90) and (S91) (which hold for any choice of sender and receivers) we obtain the Lemma's claim. This concludes the proof. $\qquad\square$

**Lemma 21.** The bifully-ACKA protocol (Protocol 12) –conditioned on event $\Psi$– is $\varepsilon_{\text{sec}}$-secret, with $\varepsilon_{\text{sec}} = 0$.

*Proof.* The only possibility for $\vec{k}_A$ to be leaked would be if a Bob receiving the key in step 3 would not input a random string while Alice inputs $F(\vec{k}_A)$ in the Parity protocols. Indeed in this case the output of the Parity protocols would coincide with $F(\vec{k}_A)$ and could be decoded back to $\vec{k}_A$.

However, this cannot happen since for the secrecy proof we must only consider the instances where the fully-ACKA-ID protocol is successful (event $\Phi$). This means that every Bob is aware of his identity (i.e. being a receiver) and inputs a random string when notified. Moreover, since we condition on the event $\Psi$, we are certain that every Bob is correctly notified. Finally, the conference key is uniformly distributed since Alice generates it in this way. Thus, the final state of the bifully-ACKA protocol is given by:

$$\rho^f_{K_{\vec{B}}^c C_{A,\vec{B}}^c E | A, \vec{B}, \Phi, \Omega_{\mathcal{P}}, \Psi} = \tau_{K_A} \otimes \rho^f_{(KC)_{A,\vec{B}}^c E | A, \vec{B}, \Phi, \Omega_{\mathcal{P}}, \Psi}, \tag{S92}$$

where Alice's key is factorized and uniformly distributed ($\tau_{K_A}$). By employing this equality in (S2), we conclude the proof. $\qquad\square$

---

[12] For instance, if Bob gets notified by a dishonest party in another round of step 3 and his decoding function $G$ does not return $\top$, see the discussion CKA-security proof of fully-ACKA.

**Lemma 22.** The bifully-ACKA protocol (Protocol 12) is $\varepsilon_{\text{CKA}}$-CKA-secure, with $\varepsilon_{\text{CKA}} = (n-1)\varepsilon_{\text{enc}} + 2^{-r_V} + (n-1)2^{-r_N}$.

*Proof.* Analogously to the proof of Lemma 16, we decompose the lhs of the CKA-security condition Eq. (B12) into two parts: one due to the event $\Psi$ and the other one due to $\Psi^c$:

$$Eq. \ (B12) \leq \Pr[\Omega_{\mathcal{P}}|i,\vec{j},\Phi,\Psi] \ \left\| \rho^f_{KC^c_{i,\vec{j}}E|i,\vec{j},\Phi,\Omega_{\mathcal{P}},\Psi} - \tau_{K_{i,\vec{j}}} \otimes \rho^f_{(KC)^c_{i,\vec{j}}E|i,\vec{j},\Phi,\Omega_{\mathcal{P}},\Psi} \right\|_{\text{tr}}$$
$$+ \Pr[\Omega^c_{\mathcal{P}} \cap \Gamma^c_{\mathcal{P}}|i,\vec{j},\Phi,\Psi] + 2^{-r_V} + (n-1)2^{-r_N}, \tag{S93}$$

where we used the failure probability of Veto or of Notification $(\Pr[\Psi^c] \leq 2^{-r_V} + (n-1)2^{-r_N})$ to bound the second part. We now use Lemma 1, in the case of the bifully-ACKA protocol conditioned on the event $\Psi$, in combination with the results of Lemmas 20 and 21 to bound the first two terms in the last expression. This concludes the proof. $\qquad\square$

### C. Anonymity

**Lemma 23.** The bifully-ACKA protocol (Protocol 12) is $\varepsilon_{\text{AN}}$-anonymous according to Eq. (B14), with $\varepsilon_{\text{AN}} = 2^{-r_V} + (n-1)(2^{-r_N} + \varepsilon_{\text{enc}})$.

*Proof.* Similarly to the anonymity proof of fully-ACKA (Lemma 17), we decompose the final state of bifully-ACKA into independent events and evaluate, for each of them, if the protocol could leak the participants' identities.

We first deal with the instances where only one party applies to be the sender. Let us denote by $\Psi'$ the event in which the Notification in step 3 is successful –i.e., every party $s$ picked by Alice is notified in step 3. Then we can decompose the final state of the bifully-ACKA protocol (Protocol 12) as follows:

$$\rho^f_{(PKC)^c_A E|A,\vec{B}} = \Pr[\Gamma|A,\vec{B}]\rho^f_{(PKC)^c_A E|A,\vec{B},\Gamma} + \Pr[\Gamma^c \cap \Phi^c|A,\vec{B}]\rho^f_{(PKC)^c_A E|A,\vec{B},\Gamma^c,\Phi^c}$$
$$+ \Pr[\Phi \cap \Psi'^c|A,\vec{B}]\rho^f_{(PKC)^c_A E|A,\vec{B},\Phi,\Psi'^c} + \Pr[\Phi \cap \Psi'|A,\vec{B}]\rho^f_{(PKC)^c_A E|A,\vec{B},\Phi,\Psi'}. \tag{S94}$$

Now, we first note that all the probabilities appearing in the decomposition of the final state in (S94) are independent of the identities of the participants. Indeed, by the description of the protocol, one can see that all the events contained in such probabilities are independent of the identities of the sender and the receivers. For this reason, we can decompose a generic state anonymous according to Definition 4 using the same probabilities, as follows:

$$\sigma^{\mathcal{D}}_{(PKC)^c_A E|A,\vec{B}} = \Pr[\Gamma|A,\vec{B}]\sigma^{\mathcal{D}}_{(PKC)^c_A E|A,\vec{B},\Gamma} + \Pr[\Gamma^c \cap \Phi^c|A,\vec{B}]\sigma^{\mathcal{D}}_{(PKC)^c_A E|A,\vec{B},\Gamma^c,\Phi^c}$$
$$+ \Pr[\Phi \cap \Psi'^c|A,\vec{B}]\sigma^{\mathcal{D}}_{(PKC)^c_A E|A,\vec{B},\Phi,\Psi'^c} + \Pr[\Phi \cap \Psi'|A,\vec{B}]\sigma^{\mathcal{D}}_{(PKC)^c_A E|A,\vec{B},\Phi,\Psi'} \tag{S95}$$

for arbitrary states $\left\{ \sigma^{\mathcal{D}}_{(PKC)^c_A E|A,\vec{B},e} \right\}_e$, $e \in \{\Gamma, \Gamma^c, \Phi, \Phi^c, \Psi', \Psi'^c\}$, such that $\sigma^{\mathcal{D}}_{(PKC)^c_A E|A,\vec{B},e}$ satisfies the anonymity definition for fully-ACKA protocols (Definition 4). Thus, by the convexity of the trace distance, we can bound:

$$\left\| \rho^f_{(PKC)^c_A E|A,\vec{B}} - \sigma^{\mathcal{D}}_{(PKC)^c_A E|A,\vec{B}} \right\|_{\text{tr}} \leq \Pr[\Gamma|A,\vec{B}] \left\| \rho^f_{(PKC)^c_A E|A,\vec{B},\Gamma} - \sigma^{\mathcal{D}}_{(PKC)^c_A E|A,\vec{B},\Gamma} \right\|_{\text{tr}} \tag{S96}$$
$$+ \Pr[\Gamma^c \cap \Phi^c|A,\vec{B}] \left\| \rho^f_{(PKC)^c_A E|A,\vec{B},\Gamma^c,\Phi^c} - \sigma^{\mathcal{D}}_{(PKC)^c_A E|A,\vec{B},\Gamma^c,\Phi^c} \right\|_{\text{tr}} \tag{S97}$$
$$+ \Pr[\Phi \cap \Psi'^c|A,\vec{B}] \left\| \rho^f_{(PKC)^c_A E|A,\vec{B},\Phi,\Psi'^c} - \sigma^{\mathcal{D}}_{(PKC)^c_A E|A,\vec{B},\Phi,\Psi'^c} \right\|_{\text{tr}} \tag{S98}$$
$$+ \Pr[\Phi \cap \Psi'|A,\vec{B}] \left\| \rho^f_{(PKC)^c_A E|A,\vec{B},\Phi,\Psi'} - \sigma^{\mathcal{D}}_{(PKC)^c_A E|A,\vec{B},\Phi,\Psi'} \right\|_{\text{tr}}. \tag{S99}$$

We now bound the terms (S96)-(S99), either by arguing that the trace distance between the final state of bifully-ACKA and the anonymous state is small, or by computing the probability of a certain event and trivially bounding the corresponding trace distance by 1.

In order to bound (S96), we use the fact that $\Gamma$ corresponds to the event where the bifully-ACKA protocol aborts in step 1, when fully-ACKA-ID (Protocol 7) is executed. As argued e.g. in the proof of Lemma 17, this event does not leak any information about the participants' identities thanks to the properties of the underlying Collision Detection, Parity and Veto protocols [20] that compose the fully-ACKA-ID protocol. Moreover, the private output of a Bob only

informs him about his role as a receiver but not about the identity of the remaining participants. Therefore, we have that $\rho^f_{(PKC)^c_A E|A,\vec{B},\Gamma} = \sigma^{\mathcal{D}}_{(PKC)^c_A E|A,\vec{B},\Gamma}$ for some state $\sigma^{\mathcal{D}}_{(PKC)^c_A E|A,\vec{B},\Gamma}$ satisfying Definition 4 and thus:

$$(S62) = 0. \tag{S100}$$

The term (S97) corresponds to a failure of the fully-ACKA-ID protocol, i.e. the identities are not correctly assigned as Alice planned, and this could lead to identity leakage. For example, it can happen that a dishonest party flips some bits destined to party $t$ in the fully-ACKA-ID protocol and the protocol does not abort. Suppose then that the dishonest party finds a round where the output $\vec{o}$ of the Parity protocol (in step 3.3 of fully-ACKA) does not decode to $\top$. The dishonest party can deduce, with high probability, that party $t$ was supposed to be a Bob but their flipping in fully-ACKA-ID did not allow him to acknowledge that, leading party $t$ to act as a non-participant in step 3.3, such that $\vec{o} = F(\vec{k}_A)$ in the round where party $t$ was supposed to obtain the conference key. From the integrity proof (Lemma 19) we have that:

$$(S97) \le \Pr[\Gamma^c \cap \Phi^c | A, \vec{B}] \le 2^{-r_V} + (n-1)\varepsilon_{\mathrm{enc}}. \tag{S101}$$

Term (S98) corresponds to the case in which the protocol does not abort in step 1, the identities are correctly assigned, but the Notification in step 3 fails. If the Notification in step 3 fails for at least one Bob and a dishonest party actively tampered with Notification, then we have an anonymity loophole. Indeed, suppose that a dishonest party tampers with the notification of a specific party $t$ in step 3.2 of bifully-ACKA. This means, in particular, that the dishonest party does not input zero in the Parity protocols forming the Notification protocol (step 2 in Protocol 8), when party $t$ is the last to broadcast. Then, the dishonest party decodes the output of the Parity protocol in step 3.3 of fully-ACKA. If the decoding function $G$ does not return $\top$, the dishonest party can guess the identity of party $t$ with a probability higher than the trivial one. Therefore, we trivially bound:

$$(S98) \le \Pr[\Phi \cap \Psi'^c | A, \vec{B}] \le \Pr[\Psi'^c | A, \vec{B}] \le (n-1)2^{-r_N}, \tag{S102}$$

where for the last inequality we use the union bound to obtain $(n-1)$ times the failure of Notification of a single Bob.

Finally conditioned on the events $\Phi$ and $\Psi'$, that is, fully-ACKA-ID and Notification in step 3 are successful, we have that no information about the participants' identities is leaked to non-participants and Eve. Indeed, the outcome of the Veto in step 4 is independent of the participants' identities. Moreover, the additional information obtained by a receiver during the protocol, namely the secret key, is randomly selected by Alice and therefore not correlated with the identity of the sender. Therefore, we have that: $\rho^f_{(PKC)^c_A E|A,\vec{B},\Phi,\Psi'} = \sigma^{\mathcal{D}}_{(PKC)^c_A E|A,\vec{B},\Phi,\Psi'}$ for some state $\sigma^{\mathcal{D}}_{(PKC)^c_A E|A,\vec{B},\Phi,\Psi'}$ satisfying Definition 4, and thus:

$$(S99) = 0. \tag{S103}$$

Combining the bounds on the terms (S96)-(S99), we obtain:

$$\left\| \rho^f_{(PKC)^c_A E|A,\vec{B}} - \sigma^{\mathcal{D}}_{(PKC)^c_A E|A,\vec{B}} \right\|_{\mathrm{tr}} \le 2^{-r_V} + (n-1)(2^{-r_N} + \varepsilon_{\mathrm{enc}}). \tag{S104}$$

For the instances in which more than one party applies to be the sender, the fully-ACKA protocol aborts in step 1 (ID aborts) and does not reveal the participants' identities, except with probability bounded by $2^{-r_V}$. Thus we can obtain the following bound:

$$\left\| \rho^f_{(PKC)^c_{\vec{t}} E|\vec{t},\vec{j}} - \sigma^{\mathcal{D}}_{(PKC)^c_{\vec{t}} E|\vec{t},\vec{j}} \right\|_{\mathrm{tr}} \le 2^{-r_V}. \tag{S105}$$

By combining (S104) and (S105), we deduce that the bifully-ACKA protocol is $\epsilon_{\mathrm{AN}}$-anonymous, with:

$$\epsilon_{\mathrm{AN}} = 2^{-r_V} + (n-1)(2^{-r_N} + \varepsilon_{\mathrm{enc}}) \tag{S106}$$

$$\square$$

## D. Security

**Lemma 24.** The bifully-ACKA protocol (Protocol 12) is $\varepsilon_{\mathrm{tot}}$-secure according to Definition 5, with $\varepsilon_{\mathrm{tot}} = 3 \cdot 2^{-r_V} + (n-1)(2^{-(r_N-1)} + 3\varepsilon_{\mathrm{enc}})$.

*Proof.* The claim follows from combining the results of Lemmas 19, 22, and 23 with Definition 5. $\square$

## VII. CONFERENCE KEY RATE OPTIMIZATION

In this work we compare the protocols in terms of their conference key rate, that is, the number of fresh secret bits shared by Alice and the intended Bobs per network use. We define one network use to be the (attempted) distribution of one GHZ state or multiple Bell pairs from the quantum server to the parties. Recall that, in each quantum channel linking the server to a party, only one photon is allowed to travel per network use.

### A. Protocols' required resources

In order to compute the conference key rates of our protocols, we need to derive the number of network uses required to establish the conference key. To this aim, here we summarize the number of GHZ states and uses of bipartite private channels (which are implemented with Bell pairs) required by each protocol.

### ACKA (Protocol 1)

The total number of bipartite private channel uses in the ACKA protocol, $g$, is:

$$
\begin{aligned}
g = {} & n^2(n-1)(3r_V + |F(\vec{d_t})|) && \text{ACKA-ID (Protocol 6)} \\
& + Lh(p)n(n-1) && \text{Testing key anonymous broadcast (step 5)} \\
& + Lp\,n(n-1) && \text{Parameter estimation (step 6)}
\end{aligned}
\tag{S107}
$$

where $|F(\vec{d_t})|$ is given in Eq. (C6). The total number of $n$-party GHZ states is $L$.

### bACKA (Protocol 11)

The total number of bipartite private channel uses in the bACKA protocol, $g_b$, is:

$$
\begin{aligned}
g_b = {} & n^2(n-1)(3r_V + |F(\vec{d_t})|) && \text{ACKA-ID (Protocol 6)} \\
& + n(n-1)L_b && \text{Conference key distribution (step 3)}
\end{aligned}
\tag{S108}
$$

### fully-ACKA (Protocol 2)

The total number of bipartite private channel uses in the fully-ACKA protocol, $g_f$, is:

$$
\begin{aligned}
g_f = {} & n^2(n-1)(3r_V + |F(d_t)|) && \text{fully-ACKA-ID (Protocol 7)} \\
& + n^2(n-1)^2 r_N \; + \; n(n-1)^2|\vec{k_l}| && \text{TKD (Protocol 8)} \\
& + Lh(p)n(n-1) && \text{Testing key broadcast (step 5)} \\
& + Lp\,n(n-1) && \text{Parameter estimation (step 6)} \\
& + n^2(n-1)r_V && \text{Veto (step 7)} \\
& + \left[ L(1-p)h(Q_Z) + \log_2\frac{n-1}{\varepsilon_{\text{EC}}} + 1 + 2\log_2\frac{1}{\varepsilon_{\text{enc}}} \right] n(n-1) && \text{fully-ACKA-EC (Protocol 10)}
\end{aligned}
\tag{S109}
$$

where $|F(d_t)|$ and $|\vec{k_l}|$ are given in Eq. (C7) and Eq. (C8), respectively. The total number of $n$-party GHZ states is $L$.

### bifully-ACKA (Protocol 12)

The total number of bipartite private channel uses in the bifully-ACKA protocol, $g_{bf}$, is:

$$
\begin{aligned}
g_{bf} = {} & n^2(n-1)(3r_V + |F(d_t)|) && \text{fully-ACKA-ID (Protocol 7)} \\
& + n^2(n-1)^2 r_N \; + \; n(n-1)^2\left[ L_b + 2\left( \log_2 L_b + \log_2\frac{1}{\varepsilon_{\text{enc}}} \right) \right] && \text{Conference key distribution (step 3)} \\
& + n^2(n-1)r_V && \text{Veto (step 4)}
\end{aligned}
\tag{S110}
$$

### B. Conference key rates

The conference key rate of the bACKA and bifully-ACKA protocol reads:

$$
r_b = \frac{L_b}{L_{b\text{tot}}} \quad , \quad r_{bf} = \frac{L_b}{L_{bf\text{tot}}},
\tag{S111}
$$

respectively, where $L_b$ is the length of the conference key transmitted by Alice and $L_{b\text{tot}}$ ($L_{bf\text{tot}}$) is the total number of network uses.

Recall that bACKA requires $g_b$ uses of bipartite private channels in order to deliver an $L_b$-bit conference key. Each secret bit transmitted in a bipartite private channel requires the prior distribution of a two-qubit Bell state to the two parties, which corresponds to one network use. However, since the Bell states prepared and sent by the quantum server may be subjected to errors in the state preparation and loss in the quantum channels, the effective number of Bell states is greater than $g_b$. In particular, we label $Q_{Xb}$ and $Q_{Zb}$ the QBERs in the $X$ and $Z$ bases affecting the two qubits of a distributed Bell state. Then, according to our model, the number of Bell pairs required to transmit $g_b$ secret bits in the bipartite private channels is given by $\eta^{-2}[1 - h(Q_{Xb}) - h(Q_{Zb})]^{-1}g_b$, i.e. it increases by a factor equal to the inverse of the asymptotic secret key rate[13] of the BB84 protocol [4,6], when implemented with photon pairs.

Finally, since we allow the source to distribute multiple Bell pairs in a single network use, the total number of network uses is smaller than the total number of required Bell pairs. Indeed, we can distribute Bell states belonging to different pairs of parties in parallel, so as to count as a single network use. For instance, in a network of five parties we can distribute two Bell pairs in a single network use. In general, this reduces the total number of network uses in an $n$-party network by a factor $\lfloor n/2 \rfloor^{-1}$. Thus, the total number of network uses for the bACKA protocol is:

$$L_{b\text{tot}} = \frac{g_b}{\left\lfloor \frac{n}{2} \right\rfloor \eta^2 [1 - h(Q_{Xb}) - h(Q_{Zb})]}. \tag{S112}$$

where $g_b$ is given in (S108), while $\eta, Q_{Xb}$ and $Q_{Zb}$ parametrize the losses and errors affecting the Bell states. For the same argument, the total number of network uses required by bifully-ACKA is given by:

$$L_{bf\text{tot}} = \frac{g_{bf}}{\left\lfloor \frac{n}{2} \right\rfloor \eta^2 [1 - h(Q_{Xb}) - h(Q_{Zb})]}. \tag{S113}$$

where $g_{bf}$ is provided in (S110).

In a similar fashion, the conference key rate of the ACKA and fully-ACKA protocol is given by:

$$r = \frac{\ell_{\text{net}}}{L_{\text{tot}}} \quad , \quad r_f = \frac{\ell}{L_{f\text{tot}}}, \tag{S114}$$

respectively, where $\ell_{\text{net}}$ is the net[14] secret conference key length of the ACKA protocol given by Eq. (1), $\ell$ is the key length of fully-ACKA and is given by Eq. (2), and $L_{\text{tot}}$ ($L_{f\text{tot}}$) is the total number of network uses.

The ACKA protocol requires $g$ bipartite private channel uses, which correspond, after accounting for losses and errors in the Bell states, to the following number of network uses:

$$\frac{g}{\left\lfloor \frac{n}{2} \right\rfloor \eta^2 [1 - h(Q_{Xb}) - h(Q_{Zb})]}. \tag{S115}$$

The ACKA protocol also requires $L$ successful distributions of an $n$-partite quantum state. The state ideally is $|\text{GHZ}_n\rangle$, but the protocol tolerates errors reflected in the QBERs $Q_X$ and $Q_Z$ appearing in the key length $\ell_{\text{net}}$ given in Eq. (1). The number of network uses needed to successfully transmit $L$ $n$-partite states is, on average, $L/\eta^n$. By summing the total number of network uses required by ACKA, we obtain:

$$L_{\text{tot}} = \frac{g}{\left\lfloor \frac{n}{2} \right\rfloor \eta^2 [1 - h(Q_{Xb}) - h(Q_{Zb})]} + \frac{L}{\eta^n}, \tag{S116}$$

where $g$ is given by (S107) and $L$ is an input parameter. Analogously, the total number of network uses of the fully-ACKA protocol reads

$$L_{f\text{tot}} = \frac{g_f}{\left\lfloor \frac{n}{2} \right\rfloor \eta^2 [1 - h(Q_{Xb}) - h(Q_{Zb})]} + \frac{L}{\eta^n}, \tag{S117}$$

where $g_f$ is given in (S109).

---

[13] We remark that the asymptotic key rate provides an upper bound on the actual BB84 key rate achievable with a finite number of rounds, which implies that the number of Bell pairs estimated above is a lower bound on the actual value. This means, in particular, that the conference key rates of bACKA and bifully-ACKA plotted in Fig. 2 do not present finite-key effects. In the last section of this document, we investigate the finite-key effects affecting bACKA and bifully-ACKA by replacing the BB84 asymptotic key rate with its counterpart valid in the finite-key regime.

[14] Recall that the ACKA protocol consumes bits from a pre-established conference key, thus we consider the net amount of fresh secret key bits it produces.

## C. Numerical optimization

In Fig. 2 we compare the conference key rates (S114) of ACKA and fully-ACKA with the corresponding key rates (S111) of bACKA and bifully-ACKA. The key rates are numerically optimized over the protocols' parameters as a function of the total number of network uses, while having fixed the parameters of our error model ($\eta$, $f_G$) and the security parameter (see Definition 5) to $\varepsilon_{\text{tot}} \leq 10^{-8}$.

The optimized key rate of the ACKA (fully-ACKA) protocol is obtained by solving the following optimization problem:

$$\max_{\{L, p, \varepsilon_{\text{enc}}, \varepsilon_x, \varepsilon_{\text{EC}}, \varepsilon_{\text{PA}}, r_V, (r_N)\}} \frac{\ell_{\text{net}}}{L_{\text{tot}}} \quad \left(\frac{\ell}{L_{f\text{tot}}}\right)$$
$$\text{subject to } \varepsilon_{\text{tot}} \leq 10^{-8} \,; \, L_{\text{tot}} = l \quad (L_{f\text{tot}} = l) \tag{S118}$$

where $L_{\text{tot}}$ and $L_{f\text{tot}}$ are given by (S116) and (S117), respectively, $\varepsilon_{\text{tot}}$ is provided in Theorem 1, and $l$ is varied in the interval $l \in [10^5, 10^{10}]$. Similarly, the optimized key rate of the bACKA (bifully-ACKA) protocol is obtained by solving:

$$\max_{\{L_b, \varepsilon_{\text{enc}}, r_V, (r_N)\}} \frac{L_b}{L_{b\text{tot}}} \quad \left(\frac{L_b}{L_{bf\text{tot}}}\right)$$
$$\text{subject to } \varepsilon_{\text{tot}} \leq 10^{-8} \,; \, L_{b\text{tot}} = l \quad (L_{bf\text{tot}} = l), \tag{S119}$$

where $L_{b\text{tot}}$ and $L_{bf\text{tot}}$ are given in (S112) and (S113), respectively, $\varepsilon_{\text{tot}}$ is provided in Theorem 2, and $l$ is varied in the interval $l \in [10^5, 10^{10}]$.

In the optimizations, we set the CNOT gate failure probability to $f_G = 0.02$ (which is realistic for certain quantum computer architectures [39]) and to a slightly more optimistic value of $f_G = 0.01$, while the transmittance $\eta$ of each channel party-server is calculated by assuming the channel to be an ultra-low-loss fiber with optical attenuation $\gamma = 0.17$ dB/km and length $d$, so that $\eta = 10^{-\gamma d/10}$. We set $d$ to $d = 2$ km and $d = 10$ km to simulate an intra-building and intra-city scenario. The QBERs appearing in the optimized key rates are a function of $f_G$ and are given by (S135), (S136) and (S137).

## VIII. NOISE MODEL FOR STATE PREPARATION

The error rates affecting the Bell pairs ($Q_{Xb}$ and $Q_{Zb}$) and GHZ states ($Q_X$ and $Q_Z$) can be derived as a result of an error model for the state preparation and state distribution. In our model we focus on the errors introduced by the quantum server when preparing the GHZ state and the Bell state, while we assume that fibers used to distribute the state do not introduce further noise, apart from photon loss.

We assume that the server prepares the two states with the same procedure and we describe the procedure for the GHZ state:

$$|\text{GHZ}_n\rangle = \frac{1}{\sqrt{2}} \left( |0\rangle^{\otimes n} + |1\rangle^{\otimes n} \right). \tag{S120}$$

The procedure to prepare the Bell state is recovered for $n = 2$.

The server initializes $n$ qubits in the state $|+\rangle_1 |0\rangle^{\otimes n-1}$. The server then applies a CNOT gate to every pair of qubits $(1, t)$, for $t = 2, \ldots, n$. Ideally, the CNOT gate is a two-qubit gate represented by the unitary operator:

$$C_{1,t} = |0\rangle\langle 0|_1 \otimes \mathbb{1}^{(t)} + |1\rangle\langle 1|_1 \otimes X^{(t)}, \tag{S121}$$

where $X^{(t)}$ is the first Pauli matrix acting on the $t$-th qubit. The GHZ state is then obtained as indicated:

$$\circ_{t=2}^n C_{1,t} |+\rangle_1 |0\rangle^{\otimes n-1} = |\text{GHZ}_n\rangle. \tag{S122}$$

However, we assume a more realistic CNOT gate which might depolarize the qubits on which it acts. This implies that the CNOT is no longer described by a unitary operation, but rather by a quantum map $\mathcal{C}_{1,t}$. The map is given by the unitary CNOT gate followed by a local depolarization on each involved qubit. The local depolarization replaces the qubit with the maximally mixed state with probability $f_Q$. The CNOT map can be written as follows:

$$\mathcal{C}_{1,t}(\tau) = (1 - f_Q)^2 C_{1,t} \tau C_{1,t}^\dagger + f_Q(1 - f_Q) \left( \frac{\mathbb{1}^{(1)}}{2} \otimes \text{Tr}_1[C_{1,t} \tau C_{1,t}^\dagger] + \frac{\mathbb{1}^{(t)}}{2} \otimes \text{Tr}_t[C_{1,t} \tau C_{1,t}^\dagger] \right)$$
$$+ f_Q^2 \frac{\mathbb{1}^{(1)}}{2} \otimes \frac{\mathbb{1}^{(i)}}{2} \otimes \text{Tr}_{1,t}[\tau]. \tag{S123}$$

Then, the multipartite state prepared by the quantum server reads:

$$\rho_{1,\ldots,n} = \circ_{t=2}^{n} \mathcal{C}_{1,t}(|+\rangle\langle+|_1 \otimes |0\rangle\langle0|^{\otimes n-1}). \tag{S124}$$

In order to compute the relevant QBERs affecting the GHZ and Bell states, we first compute the QBER $Q_Z^{(1,t)}$ affecting qubits 1 and $t$ in the $Z$ basis. The qubits 1 and $t$ become correlated in the $Z$ basis when $\mathcal{C}_{1,t}$ is applied and no qubit gets depolarized, i.e. with probability $(1 - f_Q)^2$. The CNOT maps applied previously do not influence this fact, while those applied later do. In fact, the two qubits remain correlated in the $Z$ basis when the subsequent maps $\mathcal{C}_{1,t'}$ (with $t' > t$) do not depolarize qubit 1. This happens with probability $(1 - f_Q)^{n-t}$. Thus, the probability that qubits 1 and $t$ are perfectly correlated in the prepared state $\rho_{1,\ldots,n}$ is given by $(1 - f_Q)^{2+n-t}$. Note that when the two qubits are not perfectly correlated –i.e. with complementary probability–, they are mutually random in the $Z$ basis, hence their QBER $Q_Z^{(1,t)}$ would be 1/2. Therefore, the QBER $Q_Z^{(1,t)}$ computed on the state (S124) prepared by the server reads:

$$Q_Z^{(1,t)} = \frac{1}{2}\left[1 - (1 - f_Q)^{2+n-t}\right]. \tag{S125}$$

Now we compute the QBER $Q_Z^{(q,t)}$ affecting qubits $q$ and $t$ in the $Z$ basis, for $2 \leq q < t$. We must do so since the server does not know which party is Alice, and therefore cannot consistently send her qubit 1. Recall that for the ACKA fully-ACKA protocols we need the largest QBER in the $Z$ basis between Alice and any Bob. Similarly to the computation of $Q_Z^{(1,t)}$, qubits $q$ and $t$ are correlated in the prepared state (S124) when the following chain of events take place: with probability $(1 - f_Q)^2$ the map $\mathcal{C}_{1,q}$ works ideally; with probability $(1 - f_Q)^{t-q-1}$ qubit 1 is not depolarized by the following maps $\mathcal{C}_{1,t'}$ ($i < t' < t$); with probability $(1 - f_Q)$ the map $\mathcal{C}_{1,t}$ does not depolarize qubit $t$. Putting everything together we obtain the following QBER affecting qubits $q$ and $t$ in the $Z$ basis:

$$Q_Z^{(q,t)} = \frac{1}{2}\left[1 - (1 - f_Q)^{2+t-q}\right]. \tag{S126}$$

Having computed the $Z$ basis QBERs between any pair of parties, we should then take the worst-case QBER as the value for $Q_Z$ in the ACKA (fully-ACKA) protocol:

$$\max_{q<t} Q_Z^{(q,t)} = \frac{1}{2}\left[1 - (1 - f_Q)^n\right]. \tag{S127}$$

Now, note that we can actually obtain a smaller value for the worst-case QBER than the one above. Indeed, by assuming that the quantum server applies the gates in (S124) in a uniformly random order, we can recompute the QBERs in (S125) and (S126) and obtain a lower value than the one in (S127).

In order to recompute QBER $Q_Z^{(1,t)}$, we observe that qubits 1 and $t$ are correlated in the $Z$ basis when $\mathcal{C}_{1,t}$ is applied ideally and qubit 1 does not get depolarized by the following maps. Given the random position of map $\mathcal{C}_{1,t}$, the probability that qubits 1 is not depolarized averages to:

$$\frac{1}{n-1}\sum_{k=0}^{n-2}(1 - f_Q)^k. \tag{S128}$$

Then the QBER $Q_Z^{(1,t)}$, for randomized CNOT gates, reads:

$$Q_Z^{(1,t)} = \frac{1}{2}\left[1 - \frac{(1 - f_Q)^2}{n-1}\sum_{k=0}^{n-2}(1 - f_Q)^k\right]. \tag{S129}$$

With a similar argument, one can compute the QBER $Q_Z^{(q,t)}$ (for $2 \leq q < t$) for randomized CNOT gates and obtain:

$$Q_Z^{(q,t)} = \frac{1}{2}\left[1 - \frac{(1 - f_Q)^3}{n-2}\sum_{k=0}^{n-3}(1 - f_Q)^k\right]. \tag{S130}$$

We now take the worst-case QBER between (S129) and (S130) as the largest QBER in the $Z$ basis between Alice and any Bob ($Q_Z$). One can easily verify that the QBER in (S130) is the largest, thus we obtain the QBER $Q_Z$ for our error model:

$$Q_Z = \frac{1}{2}\left[1 - \frac{(1 - f_Q)^3}{n-2}\sum_{k=0}^{n-3}(1 - f_Q)^k\right]. \tag{S131}$$

The other relevant QBER in ACKA and fully-ACKA is $Q_X = \Pr[\prod_{t=1}^{n} X_t = -1]$, which is easily computed from $\langle X^{\otimes n} \rangle$ evaluated on the prepared state (S124), as follows [8]:

$$Q_X = \frac{1 - \langle X^{\otimes n} \rangle_\rho}{2}.$$  (S132)

By writing (S124) in explicit form, one can see that every term contains an identity operator whose contribution to $\langle X^{\otimes n} \rangle_\rho$ is null due to $X$ being traceless. The only term in $\rho$ yielding a non-zero contribution to $\langle X^{\otimes n} \rangle_\rho$ is $(1 - f_Q)^{2n-2} |\text{GHZ}_n\rangle\langle\text{GHZ}_n|$. Thus we obtain:

$$Q_X = \frac{1}{2} \left[ 1 - (1 - f_Q)^{2n-2} \right].$$  (S133)

In order to enhance the readability of the QBERs in our model, we define $f_G$ to be the failure probability of the CNOT gate, that is, the probability that the CNOT gate was not implemented ideally:

$$f_G = 1 - (1 - f_Q)^2.$$  (S134)

Then we can recast the QBERs of the GHZ state as follows:

$$Q_Z = \frac{1}{2} \left[ 1 - \frac{(1 - f_G)^{3/2}}{n - 2} \sum_{k=0}^{n-3} (1 - f_G)^{k/2} \right]$$  (S135)

$$Q_X = \frac{1}{2} \left[ 1 - (1 - f_G)^{n-1} \right].$$  (S136)

Finally, the QBERs of the Bell state are easily recovered from the ones of the GHZ state by setting[15] $n = 2$:

$$Q_{Zb} = Q_{Xb} = \frac{f_G}{2}.$$  (S137)

## IX. FULL FINITE-KEY ANALYSIS OF bACKA AND bifully-ACKA

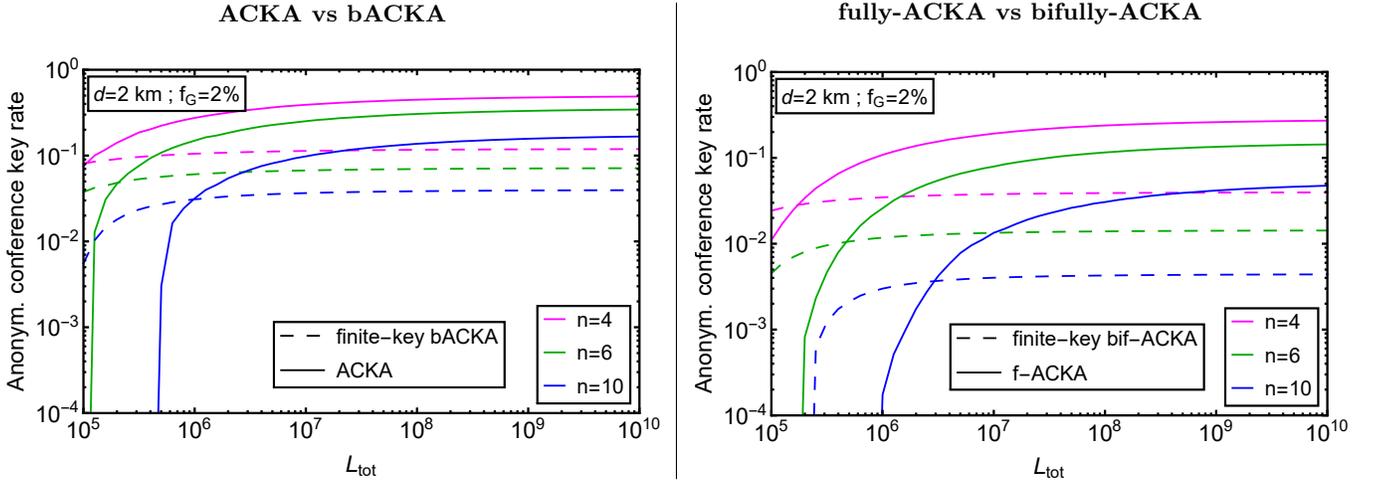

FIG. S1. Finite-key conference key rates of our anonymous protocols as a function of the total number of network uses, for different numbers of parties. We use the same parameters of the top plots in Fig. 2. The difference is that the bACKA and bifully-ACKA rates in this figure fully account for the finite-key effects affecting the BB84 protocols used to implement the bipartite private channels. Nevertheless, the finite-key effects are more prominent on the ACKA and fully-ACKA rates due to the smaller number of GHZ states that are effectively distributed to the parties, compared to the number of Bell pairs in bACKA and bifully-ACKA.

---

[15] Note that to recover the QBER $Q_{Zb}$ we need to set $n = 2$ in (S129).

In Sec. III we observe that the protocols based on GHZ states require a higher number of network uses to yield a non-zero key rate, compared to the protocols based on Bell states. As argued, this fact occurs even if we perform a full finite-key analysis for the bACKA and bifully-ACKA rates.

This is shown in Fig. S1, where we replot the top graphs in Fig. 2 but include the finite-key effects in the bipartite private channels. Specifically, we view the transmission of secret bits between every pair of parties in the network as a standalone BB84 protocol with finite-key effects. In practice, this amounts to replacing the BB84 asymptotic key rate used in (S112) with an independent finite-key rate for every pair of parties. From Fig. S1 we clearly see that the finite-key effects are less detrimental for the bACKA (bifully-ACKA) rate, compared to the ACKA (fully-ACKA) rate. This is due to the fact that the effective number of Bell pairs that two parties have at their disposal to perform the BB84 protocol and establish the private channel is larger than the number of GHZ states used by ACKA (fully-ACKA), for a given value of $L_{\text{tot}}$. Indeed, each channel use can distribute multiple Bell pairs simultaneously and the probability of photon loss is lower for Bell pairs than for GHZ states.

## X.   MAXIMUM NUMBER OF NETWORK USERS WITH GHZ-STATE ADVANTAGE

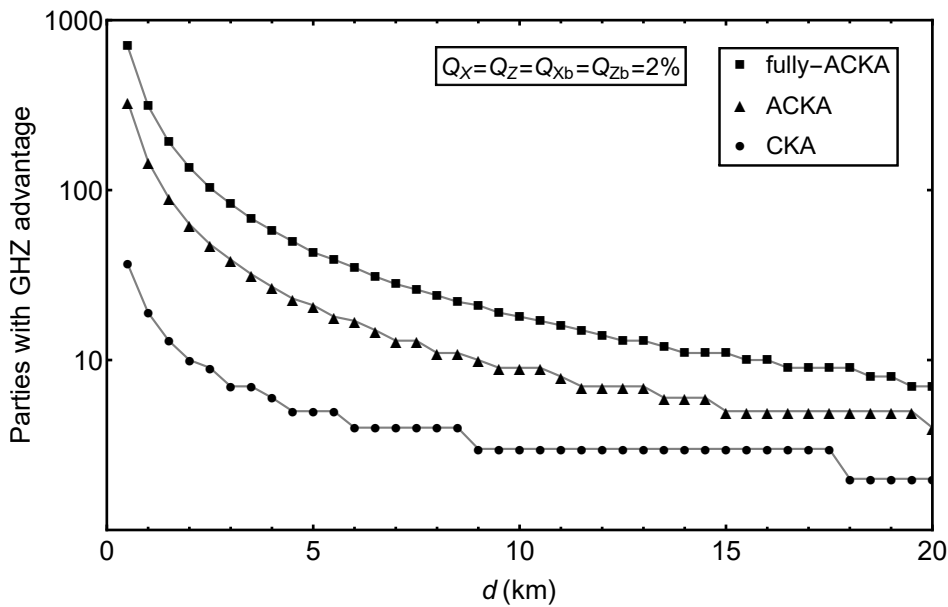

FIG. S2. Maximum number of parties ($n$) in the network, such that the GHZ-based protocols (ACKA and fully-ACKA) yield larger asymptotic secret conference key rates than the corresponding protocols based on Bell pairs (bACKA and bifully-ACKA). We also include the case of standard CKA. The error rates $Q_Z, Q_X, Q_{Zb}$ and $Q_{Xb}$ are fixed to 2%.

In Sec. III we observe that the advantage provided by GHZ states for the tasks of ACKA and fully-ACKA is limited to a maximum number of parties $n$ in the network. Above this threshold, the exponential suppression due to photon loss ($\eta^{n-2}$) dominates and the ratios of asymptotic conference key rates become smaller than 1. However, the maximum $n$ is heavily dependent on the distance $d$ party-server, since $\eta = 10^{-0.17d/10}$.

In Fig. S2 we plot the maximum number of network parties such that the ratios between the asymptotic rates in Eqs. (3) and (4), (5) and (6), and (7) and (8) are greater than 1, as a function of the distance $d$. We observe that for applications where the distances between the parties are of few kilometers (e.g. within a building or neighbourhood), the use of GHZ states is advantageous in networks of more than a hundred parties. If we increase the distance up to $d = 10$ km, the advantage is retained in networks with a number of parties between ten and a hundred.

As a final remark, note that the detrimental effect of photon loss when distributing GHZ states, instead of Bell pairs, can be removed in networks where GHZ states are distributed in a more sophisticated manner. For example, the central source could be distributing one qubit of a Bell pair to each party, while keeping the other qubit for itself. Each party heralds the arrival of their qubit and the source only needs to resend the Bell pairs that were not successfully transmitted. Then, a Bell-state measurement on the $n$ qubits at the source projects the qubits of the $n$ parties in a GHZ-like state.

Alternatively, employing W states [42] in place of GHZ states to perform anonymous CKA yields rates that scale

with $\eta$ instead of $\eta^n$, i.e., only one out of $n$ photons needs to be successfully transmitted [10]. This would entirely remove the photon-loss drawback of the anonymous protocols based on multipartite entanglement.


\fi

\end{document}